\documentclass[12pt,preprint]{aastex}

\newcommand{\jcomp}{ $Journal$ $of$ $Computational$ $Physics$ }
\newcommand{\SJSci}{ $SIAM$ $J.$ $Scientific$ $Computing$ }
\newcommand{\jsci} {$J.$ $Scientific$ $Computing$ }

\begin{document}

\title{A New Multidimensional Hydrodynamics code based 
on Semidiscrete Central and WENO schemes.}

\author{Tanvir Rahman \altaffilmark{1}, R. B. Moore\altaffilmark{1}}

\altaffiltext{1}{Department of Physics, Rutherford Physics Building, 
McGill University, 3600 University Street, 
Montreal, QC H3A 2T8, Canada.  For comments, please 
contact tanvir@physics.mcgill.ca}

\begin{abstract}

We present a new multidimensional classical hydrodynamics code
based on Semidiscrete Central Godunov-type schemes and 
high order Weighted Essentially Non-oscillatory (WENO) data reconstruction.  
This approach is a lot simpler and easier to implement than other 
Riemann solver based methods.  The algorithm incorporates 
elements of the Piecewise Parabolic Method (PPM) in the reconstruction 
schemes to ensure robustness and applications of high order 
reconstruction schemes.  A number of one and 
two dimensional benchmark tests have been carried out 
to verify the code.  The tests show that this 
new algorithm and code is comparable in accuracy, efficiency 
and robustness to others.  

\end{abstract}

\section{Introduction} 
\label{SecIntro}
Gas dynamics and their simulations are of considerable interest in many
areas of astrophysics.  Computational astrophysics is now one of 
the main branches of theoretical astrophysics that provide 
invaluable tools 
for studying complex astrophysical phenomena.  The areas in which
computational tools have proved absolutely necessary in 
astrophysics have mainly involved fluid/gas flows and N-body simulations.   
Such flows are described by nonlinear
equations that can be expressed as hyperbolic conservation laws and may
contain shock waves as solutions.  These equations 
cannot be treated analytically in multi-dimensions and hence
one must rely on numerical approximations to
study them.  Using standard finite difference techniques to
handle shock waves, discontinuities etc.,  usually lead to spurious
oscillations and instabilities in the solution that render 
these approaches useless for tackling most 
astrophysical scenarios of interest.  
Therefore non-standard approaches are
necessary to deal with these problems.  High resolution shock capturing
schemes (HRSC) are a class of numerical methods devoted specifically to
this purpose.  The key feature of such methods is their ability to
accurately approximate the solution in smooth regions while also 
handling shock waves and discontinuities away from them.  Due to its 
wide ranging applications HRSC scheme research is one of the most
active areas of research in applied mathematics.  
For an introduction and a pedagogical review 
of HRSC schemes we refer the reader
to \cite{leveque1}, \cite{toro}, \cite{leveque2} 
and references therein.

Computational astrophysics has benefited immensely from 
HRSC research
based on which numerous codes have been developed to study
astrophysical fluid dynamics.  Among the many HRSC schemes
that have been developed for applications
in astrophysics one of the most significant is the  
Piecewise Parabolic method (PPM) of 
Colella $\&$ Woodward (1985-1,2).  Several
multidimensional, multipurpose codes have been developed based 
on the PPM technique and it remains the most often-applied HRSC 
approach in computational astrophysics (for Eulerian grid based schemes).  
Some of the recent astrophysical legacy codes based on the PPM approach 
are the ZEUS (\cite{ZEUS}), the PROMETHEUS (\cite{prometheus}) and 
more recently the FLASH (\cite{FLASH1,FLASH2}) codes.  Some of these codes 
have been in development for years and are designed to 
incorporate new features within their structure as progress is made in
Applied Mathematics/Numerical analysis.  
Based on the experience gained from the development of 
these codes, the computational astrophysics community has 
learned that the path from the inception of a particular 
HRSC scheme to its robust 
application (for multidisciplinary use) usually requires years of
development.  The issues borne out by 
multidisciplinary applications are used to fine tune algorithms and 
make them as robust and user friendly as possible.  
Therefore most of the legacy codes mentioned above were 
designed using HRSC schemes whose inception preceded the 
codes by years and despite the flexibility of 
incorporating newer techniques, many recent advances in HRSC research 
have not been incorporated into these codes.  
For example, research in 
areas such as Essentially Non Oscillatory 
data reconstructions (ENO) (\cite{HEOC}) methods and 
Central Godunov type schemes (\cite{NT}) have made good progress and
are much simpler than conventional shock capturing schemes.  But to 
date they have not seen widespread applications in 
computational astrophysics.   As astrophysical simulations 
become more complex and computationally demanding,  
we need to study the suitability of these newer,  
simpler algorithms in computational astrophysics.  
With these issues in mind the purpose of this work is to take the
first few steps and lay the foundation toward the 
development of an efficient, robust,
multi-purpose astrophysical hydrodynamics code 
using a new HRSC scheme that is simpler and 
relatively inexpensive (computationally) but is as 
accurate as some of the legacy codes.  

Before we introduce the algorithm and the code, we 
begin by reviewing the main areas of research 
in HRSC schemes that are essential to 
understanding the code presented here.  In principle, HRSC 
schemes deal with discontinuities by high order 
numerical approximations 
away from shock waves and low order 
approximations around them.  
Accuracy and stability are the two main issues to consider
when developing HRSC algorithms.  These issues are 
addressed by two main areas of research.  
These are, the HRSC formulations used to advance 
the solution in space and time 
and the non-oscillatory data reconstruction technique,  which
ensures that the algorithm avoids spurious oscillations when 
interpolating the data.  Each of these are introduced in turn.  

First,  the formulation of HRSC schemes is considered.  
In general, two main approaches have been used for 
formulating HRSC schemes.  These are the $Upwind$ 
(\cite{Harten, vanleer}) and the $Central$ $Godunov$ 
type schemes (\cite{lax54,friedrichs54}).  
Their main difference is that in the central 
approach the solution is 
advanced on a staggered grid.  This difference has 
far reaching consequences with respect to the  
simplicity, efficiency and accuracy of the 
respective methods.  Most 
hydrodynamic codes, and in particular the
multi-purpose legacy codes mentioned above in 
computational astrophysics are almost exclusively 
based on the upwind approach.  The main reason for
this is that upwind schemes are generally 
less dissipative than central schemes.  And until 
recently, progress has been slow in developing 
high order, less dissipative central schemes.

Although robust, upwind methods are
generally computationally expensive and complicated to implement.
This is because upwind methods require computations of 
the eigenvalues and eigenvectors of the Jacobian of the flux 
matrix as well as flux splitting and other complicated 
computations for advancing the solution.  In contrast, the 
central approach is much simpler.  Unlike their 
upwind counterparts, they are computationally less 
expensive and do not require the computations of 
eigenvectors, eigenvalues, flux splitting etc.  Mainly 
because of this, much effort 
has been made recently on the development
of central schemes that are as accurate as their upwind counterparts.  
Some recent advances in central-type HRSC schemes 
include  their high order extensions, 
semidiscrete, genuinely multidimensional 
formulations, unstructured grid formulations etc.  Among these,  
the most significant are the semidiscrete 
formulations (\cite{KT1}) of central 
schemes which are much less dissipative than all other previous 
central approaches.  
Despite this progress,  surprisingly few computations have been done 
using the central approach in computational astrophysics.  In fact, no
detailed study has been done to test their suitability for 
astrophysical simulations besides some very recent ones that 
are mentioned below and which were done concurrently with this work.

The other main aspect of HRSC research is
the development of non-oscillatory data reconstruction techniques.  Data
reconstruction is an integral aspect of any HRSC scheme and it involves
the interpolation of a given set of data that may contain
discontinuities over the computational domain.
Two of the main techniques for non-oscillatory data reconstructions 
include the PPM method of \cite{WC2} and 
ENO approach of \cite{vanleer,OT,Harten,HEOC}.  
Recently, a newer approach known as the Weighted Essentially 
Non-Oscillatory (WENO) (\cite{LOC})
reconstructions scheme, that can be considered an extension of
the ENO approach, have also been developed.  
Each of these reconstruction methods have their own
advantages and disadvantages and they have 
all been shown to perform well for both upwind 
and central schemes.  Even though the PPM method has been used 
extensively in many legacy codes, the relatively newer WENO 
schemes have had a comparatively smaller number 
of applications despite the fact that they are 
generally more accurate than other reconstructions schemes.  
Also,  they admit arbitrarily high order formulations 
which can be useful for computations requiring 
high level of accuracy.

Rapid developments in central schemes research 
and WENO data reconstruction methods are duly 
attracting the attention of 
the computational astrophysics community.  Based on 
progress made in the areas mentioned above, both 
semidiscrete central and WENO schemes are beginning to be 
applied.  Among these applications the following are noteworthy;
Balsara D. (2001), has used the WENO and the upwind approach 
extensively for multidimensional
Magnetohydrodynamics simulations.  Recently 
Feng $\&$ Shu (2004), have applied the WENO technique 
for cosmological simulations.  Del Zanna $\&$ Bucciantini (2002),
and Anninos $\&$ Fragile (2003), have developed 
central-type relativistic hydrodynamic codes and applied it to the 
study of pulsar bow shock structures (\cite{pulsar}) 
and accretion disks close to black holes (\cite{accretion}), respectively.
Lucas-Serano et al. (2004), have investigated the 
suitability of using the semidiscrete central 
schemes in relativistic hydrodynamics.  
For multidisciplinary applications, extensions of the 
central schemes of Kurganov $\&$ Tadmore (2000),  
is also being considered for integration into the FLASH (\cite{FLASH1, 
FLASH2}) code.  One interesting aspect about all these works is that 
none of them combine the central approach with the 
WENO data reconstruction technique that admits high order formulations.  
For this reason, we have considered a different 
formulation of the central approach for 
solving the multidimensional Euler equations and 
coupled it with a reconstruction scheme that is a 
combination of the WENO and the PPM reconstruction techniques.  
Specifically, we consider the central 
semidiscrete formulation of Kurganov $\&$ Levy (2000), (KL) 
in which the central semidiscrete scheme is coupled 
to a 3$^{rd}$ order WENO reconstruction scheme for
solving hyperbolic conservation laws.  KL have 
demonstrated the compatibility of combining 
the semidiscrete central scheme with a 3$^{rd}$ order WENO reconstruction 
algorithm for solving general hyperbolic equations.  This was a step 
forward in HRSC research as it opened the 
possibility of combining arbitrarily high order accurate 
WENO reconstruction schemes
within the semidiscrete central paradigm.  
Following them,  we propose here 
a new algorithm for data reconstruction 
that can be used with the central scheme of KL for robust applications 
in computational astrophysics.  Essentially, 
the simplicity of central schemes, the accuracy of 
WENO reconstruction method and the robustness of the PPM 
algorithm have been combined in this new algorithm.  
This algorithm has been tested by an extensive 
collection of one and two dimensional problems with the Euler 
equations using both 3$^{rd}$ and 4$^{th}$ order reconstructions.  
Many of the tests presented here are a first using 
the semidiscrete central WENO approach.  In particular, 
the two dimensional Riemann Problems 
presented in this work are the first such set of 
computations using the WENO reconstruction scheme.  
In addition to testing the code, these 
computations address some other issues with respect to the robust 
application of WENO schemes for shock capturing schemes 
that will be discussed later.  Building on the work presented here,  
we have also developed a multidimensional 
relativistic hydrodynamics code that will be presented 
in a forthcoming paper.  

The outline of this paper is as follows.  Sec. 2 presents a
brief review of the semidiscrete central scheme and the WENO
data reconstruction methods.  The hydrodynamics 
algorithm is then presented in Sec. 3.  
Sec. 4 presents the tests of the algorithm for the Euler equations.  
Sec. 4.1 presents the one dimensional test and Sec. 4.2 
presents the two dimensional tests.  
Some concluding remarks are in Sec. 5.

\section{Semidiscrete Central and WENO schemes: A Review}
\label{section:review}


This section begins with a brief overview of Godunov type
central schemes and the motivations behind their semidiscrete
formulation.  WENO data reconstruction schemes are also discussed and 
some of their advantages over other non-oscillatory
reconstruction schemes are highlighted.   This is followed 
by a step by step account of the
particular central scheme used in our algorithm.  This follows a
description of the WENO reconstruction scheme used in the code.

The first central scheme was developed by Lax (1954), followed by 
Frieidrichs $\&$ Lax (1971) (LxF).   These schemes were both first order  
schemes.  They were extended to second order 
by Nessyahu $\&$ Tadmore (1990) (NT).  Since then, there has been a number of
formulations of the central approach which can be considered
extensions of the LxF and NT schemes.  These newer formulations explored
a number of numerical approaches and applications that improved upon their
predecessors.  These include high order 
extensions in multidimensions 
(\cite{LT,GT,LPR1,LPR2,KP1,KNP,KL}), genuinely multidimensional 
formulation (that include fluxes from diagonal directions 
in multidimensions) (\cite{KP1,KNP}) and recently, 
the formulation of central schemes on unstructured grids (\cite{KP2}).
 As mentioned above,  the main advantage of 
central schemes over their upwind counterparts is
 their simplicity.  These schemes do not require the use of
computationally expensive Riemann solvers to advance the
solution.  However, the trade-off for their simplicity is accuracy.  In
general, central schemes are more dissipative than upwind schemes.  To
address this issue, Kurganov $\&$ Tadmore (2000) (KT), developed a newer
formulation of central schemes known as 
the semidiscrete central schemes.  Numerical tests by KT showed that for 
central schemes of a given order, semidiscrete central ones were 
far less dissipative than their predecessors.  Hence, these 
schemes retained the advantages of
the central formulation while enhancing its performance.  
Semidiscrete schemes are different from
their predecessors mainly in two ways.  First, the solution is 
advanced on a non-staggered grid.  This means that after advancing 
the solution, it need not be projected back to the original grid.  
Second, semidiscrete schemes are more accurate 
because local speed of propagation of information are taken into account 
in their formulation. The success of the KT scheme 
precipitated a flurry of research in semidiscrete schemes which 
included their extensions to higher orders and their 
multidimensional formulations (\cite{KP1,KNP,KL}) among others 
that will be mentioned later.  

We turn now to the WENO data reconstruction methods.  
WENO schemes were first proposed by 
Liu et al. (1994), as a natural extension to ENO schemes.  The main 
advantage of WENO over ENO schemes is their arbitrary 
high order formulations.  They are also more accurate then
their ENO counterparts.  An interesting feature of WENO 
schemes is that they show ``super convergent'' behavior not observed in
other non-oscillatory data reconstruction methods.  Because of these,  
WENO schemes were incorporated with central schemes by Levy et al.  
(\cite{LPR3,LPR1,LPR2,LPR4}) who proposed a new class of
HRSC schemes called the Central Weighted 
Essentially Non-Oscillatory (CWENO) method.  
In order to take advantage
of the characteristics of semidiscrete schemes mentioned earlier 
and WENO reconstruction methods, Kurganov $\&$ Levy (2000) (KL) 
combined the semidiscrete central and WENO methods 
and proposed a new HRSC scheme.  This work proved the compatibility 
of WENO within the central framework.  The work 
presented here is based on the scheme by KL.  In some respects, it 
can even be considered as an extension of the scheme by KL.   

The next section (Sec. 2.1) provides a step by step description of the
development of central Godunov type schemes and presents the scheme by KL
that is used in this work.  For the sake of simplicity, a one 
dimensional scalar hyperbolic conservation 
law will be considered.  
The extension to multidimensions and that of a system of 
equations can be done using standard techniques such as 
dimensional splitting (\cite{toro}) and other methods.  
The description of KL will be followed by that of the 3$^{rd}$ and 
4$^{th}$ order WENO reconstruction scheme used in this work (Sec. 2.2).

\subsection{Semidiscrete Central Scheme}
Consider the following equation along with the given
initial condition 
\begin{eqnarray}
u_{t} + \frac {\partial f(u)} {\partial x}   = 0 \quad \nonumber \\ 
u(x,t=t^n)=u^n(x) \quad .
\label{F1}
\end{eqnarray}

Our objective is to numerically advance the solution of this equation from
$t=t^n$ to $t=t^{n+1}$.  In order to discretize the problem the following 
notations are defined. Let $x_j:= j \Delta x$,  $x_{j \pm \frac
{1}{2}} : = (j \pm \frac {1} {2} ) \Delta x$ and  $t^{n}:= n \Delta
t$, where $\Delta x$ and $\Delta t$ are unit
intervals in space and time respectively.  Also define the interval
$I_j := [x_{j-1/2},x_{j+1/2}]$ and $u_{j}^n$ $:=$ 
$\{u(x,t=t^n);$ $x \in I_j \}$.  
The problem may now be 
summarized as follows; given $u(x,t=t^n)= \{u_{j}^n\}$, 
we would like to find $ \{u_j^{n+1} \}$.  Alternatively, for 
finite volume methods that HRSC schemes are,  cell averages 
instead of point values are updated. Defining
\begin{eqnarray}
\overline {u} (x) := \frac {1} {\Delta x} \int_{I(x)} u(\zeta,t) d
\zeta \quad , \nonumber \\ 
& I(x) = \Big \{ \zeta : \vert \zeta - x \vert < \frac {\Delta x}
{2} \Big \} \quad ,
\end{eqnarray}
and integrating in space, Eq. 1 becomes 
\begin{eqnarray}
\overline u_t(x, t) = \frac {1} {\Delta x} \Big \{ f (  u ( x +
\frac {\Delta x}{2}, t ) ) - 
f (  u ( x - \frac {\Delta x}{2}, t ) ) \Big \} \quad .
\end{eqnarray}
Now integrating in time from $t=t^n$ to $t=t^{n+1}$ gives,
\begin{eqnarray}
\overline u(x, t + \Delta t) - \overline u(x, t) =
\frac {1} {\Delta x} \left[   
\int_{t^n}^{t^{n+1}}f (  u ( x +
\frac {\Delta x}{2}, \tau ) )  d \tau -
\int_{t^n}^{t^{n+1}}f (  u (x - \frac {\Delta x}{2}, \tau ) )  d \tau
\right] \quad .
\end{eqnarray}
This equivalent formulation is the starting point for the construction
of Godunov-type schemes for numerically approximating hyperbolic
conservation laws.  In Eq. 4, the solution is evolved 
in terms of sliding averages.  Setting $x=x_j$ leads to a formulation 
that is known as the upwind scheme.  The upwind scheme requires the
evaluation of the flux integrals on cell boundaries, where the data
could be discontinuous.  This is customarily done by using Riemann solvers.  
On the other hand setting $x=x_{j + 1/2}$ leads to the central scheme
formulation.  Under this scheme, the flux integrals are evaluated at
the center of the cell where the data is continuous and finite 
speed of propagation of information guarantees that 
Riemann solvers are not needed.  
Approximations of Godunov type central schemes generally 
involve three main steps; reconstruction, evolution and projection.  
The next few paragraphs describe these steps.

First consider the reconstruction step.  In this step, 
given $\{\overline {u}_j^n\}$, an $n th.$ order, non-oscillatory 
piecewise polynomial interpolation $\{p_j^n(x)\}$ of the data is
constructed over the computational domain.  The $\{p_j(x)\}$'s
are polynomials that can only be discontinuous at cell interfaces (if
the data contains discontinuities) and
they are determined from two main constraints.  
These are, the conservation of cell averages
\begin{eqnarray}
\int_{x_{j-1/2}}^{x_{j+1/2}} 
p_j^n(\zeta) \quad {\rm d} \zeta = 
\overline u_j^n & \quad \forall \quad j \quad ,
\end{eqnarray}
and accuracy requirements
\begin{eqnarray}
u(x,t^n)= \sum_{j} p_j^n(x,t^n) \chi_j(x) + O(\Delta x^r) \quad ,
\end{eqnarray}
where $\chi_j$ is the characteristic function of each cell defined as
\begin{eqnarray}
\chi_j (x) =  \begin{array}{cc}
1 & {\rm if} \qquad x \in \quad I_j   \\
0   &\quad {\rm otherwise} \quad . \nonumber
\end{array}
\end{eqnarray}
The simplest interpolation is of course the piecewise
constant case,  i.e., $p_j^n(x,t^n)=
\overline u_j^n$, which leads to the 
central LxF scheme.  Higher order polynomials 
lead to better approximations but more
oscillations near discontinuities.  The general strategy used 
to manage these oscillations is to lower the order of the interpolation
near discontinuities.  This is known as 
non-oscillatory reconstruction and ENO and WENO schemes 
mentioned above are examples of such interpolation.

Once the data reconstruction is done the RHS of Eq. 4 can be
computed to advance the solution in time.  Using this reconstruction, 
we may write 
$u(x,t^n) \approx \sum_j p_j^{n} (x) \chi_j$.  
Substituting this in Eq. 4 and setting $x=x_{j+1/2}$ leads to the 
the following reformulation of Eq. 4   
\begin {eqnarray}
\overline u_{j+ \frac {1}{2}}^{n+1} = \frac {1}{\Delta x} 
\left[ \int_{x_j}^{x_{j+\frac{1}{2}}} p_j^n(x) dx +
 \int_{x_{j+\frac{1}{2}}}^{x_{j+1}} p_{j+1}^n(x) dx \right] - \nonumber \\ 
\frac{\lambda}{\Delta t} 
\left[ \int_{t^n}^{t^{n+1}} f(u(x_{j+1},t)) dt -
\int_{t^n}^{t^{n+1}} f(u(x_{j},t)) dt \right] \quad ,
\end{eqnarray}
where  $\lambda = \frac {\Delta t} {\Delta x}$.  The first two integrals in 
in Eq. 7 can be computed exactly given an appropriate piece-wise
polynomial interpolation.  The flux terms can be 
approximated by quadrature rules of the
appropriate order.   The function values needed in
the quadrature formula can be computed 
by using Taylor expansion or the appropriate 
Runge-Kutta method.  For example, using the 
second order reconstruction of Nessyahu and Tadmore (1990) (NT),
\begin {eqnarray}
p_j^n(x)= \overline u_j^n + s_j^n (x-x_j) \quad ,
\end{eqnarray}
and using the midpoint rule for flux evaluation results in the
NT staggered scheme,
\begin {eqnarray}
\overline u_{j+ \frac {1}{2}}^{n+1} = 
\frac{1}{2} (\overline u_{j}^{n} + \overline u_{j+1}^{n}) -
\frac{1}{8} (\overline s_{j}^{n} - \overline s_{j+1}^{n}) - \nonumber \\
\lambda \left[ f(\overline u_{j+1}^{n+1/2})-
f(\overline u_{j}^{n+1/2}) \right] \quad ,
\end{eqnarray}
where $s_j^n$ is constructed using $minmod$ limiters to
minimize oscillations.  For example,
\begin{eqnarray}
s_j^n = {\rm minmod} (\frac {\overline u_j^n - \overline u_{j-1}^n}{\Delta x},
\frac {\overline u_{j+1}^n - \overline u_{j}^n}{\Delta x}) \quad ,
\end{eqnarray}
where the $minmod$ function is defined as,
\begin{eqnarray}
{\rm minmod} (a,b) := \frac { {\rm sgn(a)} +{\rm  sgn(b)} }{2} 
{\rm min} (|a|,|b|) \quad .
\end{eqnarray}
Eq. 9 is the second order NT scheme and Eqns. 8, 10 are 
an example of a second order ENO data reconstruction method.  All modern 
central schemes can be thought of as extensions of 
the scheme by NT including the KL scheme presented below.  
As mentioned earlier, 
numerical experiments with the NT scheme
show this approach to be numerically dissipative and in order 
to address this issue,  Kurganov $\&$ Tadmore (2000) (KT) derived the 
semidiscrete version of the central scheme.  The derivation 
of this approach is complicated and for details
we refer the reader to \cite{KT1}.  The semidiscrete scheme 
that we have used is a further extension of the KT scheme 
and uses a 3$^{rd}$ order WENO data reconstruction method.  It 
was presented in KL and is given by
\begin{eqnarray}
\frac {d}{dt} \overline u_j (t) = - \frac {H_{i+\frac {1}{2}} (t) -
H_{i+\frac {1}{2}} (t) } {\Delta x} \quad ,
\label{updata}
\end{eqnarray}
where the flux $H_{i+\frac {1}{2}}$ is given by
\begin{eqnarray}
H_{i+\frac {1}{2}} (t) := \frac { f(u^+_{i+\frac {1}{2}} (t)) + 
 f(u^-_{i+\frac {1}{2}} (t)) } {2} - \nonumber \\ 
\frac {a_{i+\frac {1}{2}} (t)} {2}
\left[ u^+_{i+\frac {1}{2}} - u^-_{i+\frac {1}{2}} \right].
\label{flux}
\end{eqnarray}
where $u^{+}_{i+\frac {1}{2}}$, $a_{i+\frac {1}{2}}$ are
given by,
\begin{eqnarray}
a_{i+\frac {1}{2}} := \max \{ \rho ( \frac {\partial f}{\partial u}
(u_{j+1/2}^{-}) ), \rho ( \frac {\partial f}{\partial u}
(u_{j+1/2}^{+}) ) \}
\end{eqnarray}
\begin{eqnarray}
u^{+}_{i+\frac {1}{2}} = P(x_{j+1/2}) \quad .
\end{eqnarray}
In Eqns. 13 and 14, $a_{i+\frac {1}{2}}$ is the speed of propagation of
$u$ at the interface of a cell that is determined from 
the spectral radius of the Jacobian of the flux $f$.  Using Eq. 12, 
the solution can be
updated by any high order Total Variation Diminishing (TVD) 
Runge-Kutta type ODE solver.  

The extension of the scheme presented above 
to multi-dimensions can be
done as follows.  Eq. 12 in multidimensions contains flux contributions 
from every dimension that can be determined as in the one
dimensional case.  This leads to an unsplit scheme, which 
is what is used in our algorithm.  
\subsection{WENO Reconstruction Scheme}
 
Since its inception, WENO (\cite{LOC}) data reconstruction schemes 
have been improved and incorporated into a number 
of HRSC schemes and have also been applied to a number of
problems in computational astrophysics.  WENO schemes posses most of 
the advantages of the ENO methods and some significant others that ENO 
schemes do not have.  The main advantage is their arbitrarily high order 
formulations.  In this work we have implemented a 
3$^{rd}$ and a 4$^{th}$ order WENO reconstruction
schemes given by \cite{KL} and \cite{LPR2,LPR3}, respectively.  
The following reproduces the details 
of the 4$^{th}$ order scheme presented in \cite{LPR3}.
Whenever necessary, the modifications needed for
the 3$^{rd}$ order reconstruction scheme have also been highlighted.  
For simplicity, only the 
one dimensional reconstruction scheme is described here.  
Multidimensional extensions of the scheme presented 
here can be done using dimensional splitting in a 
straight forward manner.  

Our task is as follows; given $\{\overline u_j^n\}$, we would 
like to construct a 4$^{th}$ order piecewise parabolic 
interpolation of the data
$u^n(x)$ over the computational domain.  Described below is the
reconstruction procedure for a single interval $I_j$.  The same
procedure can be extended to interpolate the data
 over the entire domain.  The main
idea of the WENO approach is as follows.  Let $R_j(x)$ be the
non-oscillatory interpolant over the cell $I_j$.  It is given by a
weighted convex combination of interpolation 
polynomials $p_k(x)$ (where k = j, j+1, j-1), constructed
over different stencils.  This weighted combination is used to ensure
that the interpolant receives the most significant contribution from
the smoothest stencil, thereby preserving the non-oscillatory character
of the interpolation.  Therefore, $R_j(x)$ is given by 
\begin{eqnarray}
R_j(x) = w_{j-1}^j p_{j-1}^j(x) + w_{j+1}^j p_{j+1}^j(x) 
+ w_j^j p_{j}(x) \quad .
\label{p_j}
\end{eqnarray}
The $w_{k}^j$'s are weights corresponding to each polynomial 
and are subject to the normalization constraint
\begin{eqnarray}
\sum w_{k}^j = 1 \quad .
\end{eqnarray}
The computations of the weights will be described shortly.  The 
polynomials $p_{k}(x)$'s are
constructed by using different stencils around the point 
$x_{k}$ where k = $\{j, j+1, j-1\}$.  For example,
$p_{j-1}(x)$ is the polynomial based on the left stencil
$\{x_{j-3},x_{j-2},x_{j-1},x_j,x_{j+1}\}$.  Other polynomials are
constructed similarly.   The coefficients of the
polynomials $p_{k}(x)$ are fixed by satisfying the following
conservation and accuracy requirements,
\begin{eqnarray}
\frac {1}{h} \int p_j(x) dx = \overline {u}_k, \qquad  k=j-1,j,j+1,
\end{eqnarray}
\begin{eqnarray}
\frac {1}{2h} \int p_{j}(x,t^n) \quad dx = 
\frac {1}{2h} \int u(x,t^n) \quad dx + O(h^s) \quad .
\end{eqnarray}

\noindent Therefore writing the polynomials as,
\begin{eqnarray}
p_j(x)= \tilde {u}_{k} +  \tilde {u}_{k}' (x-x_k) + \nonumber \\
\frac {1}{2} \tilde u_{k}'' (x-x_k)^2, \qquad k= j-1, j, j+1
\end{eqnarray}
and applying conditions set by Eqns. 18 and 19 gives,
\begin{eqnarray}
\tilde u_{k}'' = \frac {\overline {u}_{k+1}-2\overline {u}_{k} +
\overline {u}_{k-1}} {h^2} \quad ,  
\end{eqnarray}
\begin{eqnarray}
\tilde u_{k}' = \frac {\overline
{u}_{k+1}-\overline {u}_{k-1}} {2h} \quad ,
\end{eqnarray}
\begin{eqnarray}
\tilde u_{k}= \overline
{u}_{k} - \frac {h^2}{24} \tilde u_{k}'' \quad .
\end{eqnarray}

Note: In case of 3$^{rd}$ order reconstruction \cite{KL}, the
polynomials $p_{j-1}^j(x)$ and $p_{j+1}^j(x)$ are piecewise linear.  The
method for computing their coefficients is the same as that shown above.

The weights $w_k^j$ $(k=j-1,j,j+1)$ are given by 
(for details, see \cite{JS}),
\begin{eqnarray}
w_{k}^j = \frac {\alpha_k^j} { \alpha_{j-1}^j +\alpha_{j}^j +
\alpha_{j+1}^j} \quad ,
\end{eqnarray}
\noindent where
\begin{eqnarray}
 \alpha_{k}^j = \frac {C_k}{(\epsilon + IS_k^j)^p},  
\quad C_k > 0 \quad .
\end{eqnarray}
The constants $C_k$ are known as $optimal$ $weights$ and their evaluation
is described in \cite{SO}.   The parameter $IS_k^j$ is
used to compute the smoothness of the various 
stencils and is defined below.  The parameter 
$\epsilon$ is needed to prevent the
denominator of $\alpha_{k}^j$'s from going to zero.  From Eqns. 24, 25
above we note that the parameters $C_k$, $p$ and
$\epsilon$ are the only free parameters in 
WENO reconstruction schemes.  
They must be set a priori depending on the 
application being considered.  However numerical 
tests have shown that
some specific choice seems to work well for a number of test problems.  
For example when computing  point values, any symmetric
combination of $C_k$ gives the desired order of
accuracy.  The parameter $p$ is empirically
chosen and is set to 2 for 3$^{rd}$ order schemes and 3 for 4$^{th}$
order schemes.  The parameter $\epsilon$ is usually set to
$10^{-6}$.  For our computations we have kept the values corresponding to
the 3$^{rd}$ and 4$^{th}$ order schemes fixed.  This was done 
deliberately to test the robustness of our algorithm.

\noindent The smoothness indicators $IS_k^j$ are defined by,
\begin{eqnarray}
IS_k^j = \sum_{l} \int h^{2l-1} (p_k^l)^2 \quad dx, 
\quad k= j-1,j,j+1 \quad .
\end{eqnarray}
These represent the $L^2$ norms of the first
and second derivatives, where $p_k^{(l)}$ denotes the $l^{th.}$ derivative
of $p_k^j(x)$.  For our 4$^{th}$ order reconstruction, the smoothness
indicators are given by
\begin{eqnarray}
IS_{j-1}^j = \frac {13}{12} ( \overline u_{j-2} - 2   \overline u_{j-1} +
 \overline u_{j} )^2 +  \frac {1}{4} ( \overline u_{j-2} - 4 \overline
u_{j-1} + 3 \overline u_{j} )^2 \quad , \nonumber \\
IS_{j}^j = \frac {13}{12} ( \overline u_{j-1} - 2   \overline u_{j} +
 \overline u_{j+1} )^2 +  \frac {1}{4} ( \overline u_{j-1} - \overline
u_{j+1})^2 \quad , \nonumber \\
IS_{j+1}^j = \frac {13}{12} ( \overline u_{j} - 2   \overline u_{j+1} +
 \overline u_{j+2} )^2 + \nonumber \\
\frac {1}{4} ( 3 \overline u_{j} - 4 \overline
u_{j+1} + \overline u_{j+1} )^2 \quad .
\end{eqnarray} 

For a system of equations, some modifications are necessary for
computing the smoothness indicators $IS_{k}$.  Even though computations
can be done component wise, best results are obtained by using universal
smoothness indicators \cite{KL}.  They are given by,

\begin{eqnarray}
IS_k = \frac {1}{d} \sum_{r=1}^d \frac {1}{ \parallel \overline{u}_r
\parallel_2} ( \sum_{l=1}^2 \int_{x_{j-1/2}}^{x_{j+1/2}} h^{2l-1} 
(p_{k,r}^l )^2 \quad dx ) \quad , \nonumber \\
k \in {j-1,j,j+1} \quad .
\end{eqnarray}
Where $d$ is the number of equations.  The scaling factor $\parallel
\overline{u}_r \parallel_2$ is defined as the $L^2$ norm of the cell
averages of the $r^{th}$ component of $u$ defined by,
\begin{equation}
\parallel \overline{u}_r \parallel_2 = 
( \sum_{j} | {\overline u}_{j,r}|^2 h )^{1/2}.
\end{equation}
This completes the description of the 4$^{th}$.  order 
WENO reconstruction scheme.  We now present some tests 
of the scheme used in our hydrodynamics code.  
Consider the function $f(x)$ $=\sin (x)$, where $x \in
[0, 2\pi]$.  Tables 1 and 2 below shows the $L^1$ and $L^{\infty}$
errors for reconstructing $f(x)$ using the schemes
described above.  The free parameters of 
WENO reconstructions ($C_i$'s,
$\epsilon$ and $p$) are
the same as those from \cite{KL}, \cite{LPR3}.  There are 
several interesting features worth noting from these
results.  For the 4$^{th}$ order reconstruction (Table 2), 
the order remains constant around four as 
the mesh spacing is decreased, as is
expected.  However for the 3$^{rd}$ order scheme (Table 1), 
non-linear behavior can be noticed in the errors.  In fact, the
reconstruction becomes better than order three and shows
the so called ``super convergent'' behavior noted by \cite{LOC} when
they introduced WENO reconstructions.  
This non-linear super convergence will also be seen  
in some of the tests of our hydrodynamic codes.  

\section{The Multidimensional Hydrodynamics Algorithm}

In principle, the combination of the WENO reconstruction scheme with the
central semidiscrete scheme is adequate for solving general hyperbolic
conservations laws.  However, if one demands a robust scheme for general
applications, then the 
scheme above would require modifications as the 
standard WENO prescriptions given above is still too 
oscillatory for cases in which shock waves are very strong.  How the WENO
schemes could itself be modified to handle such strong shocks is an 
interesting research project in numerical analysis.  However, given our 
objective of building a robust hydrodynamic code, this 
is beyond the scope of this work.  Instead,  to ensure robustness 
for our applications,  we have considered certain 
elements of the PPM scheme and incorporated these 
into our data reconstructions scheme.  
The features of the PPM scheme 
that makes the algorithm presented here robust are 
its contact steepening, flattening 
and monotonicity preserving algorithms.  These steps are outlined 
in detail in Colella $\&$ Woodward (1985-2).  We have 
incorporated them into our 
algorithm without any modification.  We discovered that when semidiscrete 
central WENO schemes are combined with these extra, it is 
robust with respect to a large number of benchmark 
tests (see Sec. 4).  Our hydrodynamic algorithm 
can be summarized as follows,

{\bf step 1}: \textit {Given ${\overline u}_{j}^n$, use 
the $nth.$ order WENO
reconstruction algorithm to construct ${p_j(x,t^n)}$.  Use the
$p_j(x,t^n)$'s to compute $u_{j}^+$, $u_{j}^-$ }.  

{\bf step 2}: \textit{ Apply the steepening (only to the density 
$\rho$), flattening and monotonicity
preserving algorithms to $u_{j}^+$, $u_{j}^-$ (Eq. 15) }.  

{\bf step 3}: \textit{ Update ${\overline u}_{j}^n$ to  ${\overline
u}_{j}^{n+1}$ using the scheme described in Sec 3.1 (Eq. 12)}.  

In step 3 of the algorithm above, we have used a 
total Variation diminishing (TVD) multi-step Runge-Kutta (RK) 
ODE solver \cite{roe} that we give below.  
\begin {eqnarray}
U^{(1)} = U^n + \Delta T L (U^n) \nonumber \\
U^{(2)} = \frac {3}{4} U^n + \frac {1}{4}U^{(1)} + \frac {3}{4} \Delta t
L(U^{(1)}) \nonumber \\
U^{n+1} = \frac {1}{3} U^n + \frac {2}{3} U^{(2)} +  \frac {2}{3} \Delta
t L(U^{(2)}).  
\end{eqnarray}
All our tests have done using this RK scheme.    

\section{Tests of the Hydrodynamics Code}

The standard approach to testing any HRSC scheme consists of
simple advection tests; shock capturing using Berger's equation
followed by more complex tests.  For the
semidiscrete CWENO scheme used here, 
advection and Berger's equation related tests have
already been published \cite{KL}, and while our codes were 
being developed, we have also reproduced them 
(without the PPM type modifications
mentioned in the last section).  However, we do not present 
the results here.  Since our primary
interest is in gas dynamics, we present test results 
of our hydrodynamic code.  

In two dimensions, the Euler equations of hydrodynamics 
are given by
\begin{displaymath}
\frac {\partial}{\partial t} \left( \begin{array}{ll} \rho \\
\rho u \\
\rho v \\
E \end{array} \right) + \frac {\partial}{\partial x} \left(
\begin{array}{ll} \rho u \\
\rho u^2 + p\\
\rho uv \\
u (E+p) \end{array} \right) + \frac {\partial}{\partial y} \left(
\begin{array}{ll} \rho v \\
\rho uv \\
\rho v^2 +p \\
v (E+p) \end{array} \right) = 0 \quad .
\end{displaymath}
The equations above are closed by an equation of state.  For ideal
gases this is given by $p= (\gamma -1) (E- \rho/2 (u^2+v^2))$.  
Here $\rho$, $u$, $v$, $p$ and $E$ are
the density, the $x$ and $y$ velocities, the pressure and the total
energy respectively.  For most of our tests, we 
have used an ideal gas equation of state for which
the adiabatic index, $\gamma$ $=$ $5/3$.  
The tests chosen for the code follow from
those chosen to test the ZEUS and FLASH codes 
(\cite{ZEUS,FLASH1,FLASH2}).  In one dimension,
these include several advection tests and some standard shock tube
problems.  In 2D, they include an exhaustive list of 
Riemann problems and several blast wave tests.

\subsection{One Dimensional Tests}
\label{section:results2D}
As already mentioned, following Stone $\&$ Norman (1992), 
Fryxell et al. (2000) and Calder et al. (2002), several 
advection tests were carried out followed by some standard shock tube
tests.  The results are discussed below and compared to
previously published results.  

\subsubsection{Advection Tests}
\label{section:results2D}
Advection problems were used to check the ability of our scheme to
transport and  maintain the shape of a density pulse.  The tests
presented here were first suggested by \cite{BB,forester}
and considered by both \cite{ZEUS} and \cite{FLASH1}.  First, we
considered the simple advection of a rectangular and
a Gaussian pulse.  For example, 
advecting a rectangular pulse tests the codes 
ability to lead and trail 
contact discontinuities while advecting a Gaussian pulse 
tests its ability to handle narrow flow features.
The pulse profile we use for these two test 
profiles is given by
\begin{equation}
\rho(s)=\rho_{1} \phi (s/w) + \rho_{0} (1 - \phi (s/w)) \quad ,
\end{equation}
where $s$ is the distance from of a point from the pulse midplane and $W$
is the characteristic width of the pulse.  For the square and Gaussian
pulse, the $\phi(x)$ as,
\begin{displaymath}
 \phi (x) = \left \{ \begin {array} {ll}
1 & \textrm{if $\mid x \mid < 1$}\\
0 & \textrm{if $\mid x \mid > 1$}\\
\exp (-x^2) & \textrm{if pulse is Gaussian}\\
\end{array} \right \} \quad .
\end{displaymath}
With this definition, the Euler equation reduces to a simple
advection equation.

We begin with the rectangular profile.  The advection of 
the rectangular pulse were followed to time $t=0.2$ and 
the positions of both the advected and the analytic solutions 
are shown in Figs. 1 and 2 for n=80, 160, 320 and 640.  
The solution has been
plotted using both the 3$^{rd.}$ and 4$^{th.}$ order 
reconstruction schemes and the analytic solutions.  Several 
features can be noted from these plots.  First, there is the convergence 
of the solution with increasing
resolution.  Second, that the 4$^{th}$ order reconstruction gives
better results than the 3$^{rd}$ order scheme for the 
same grid spacings.  Finally, comparing our
results to those of \cite{ZEUS}, \cite{FLASH1}, 
there is good qualitative agreement.  

Continuing with the advection of a
Gaussian pulse, we set the width of the pulse to 
$w=.015625$ and advected the pulse to time $t=0.2$.
The results are shown in Figs. 3 and 4.
Once again, as with the case of the rectangular 
pulse, there is convergence to the analytical solution 
as $n$ increases.  As expected, the 
4$^{th.}$  order scheme converges 
faster than the 3$^{rd.}$ order scheme.  
We have tabulated the $L1$-error norms of the 
solution in Table 3 for the 3$^{rd.}$  and 4$^{th.}$  order 
reconstructions.  The results 
show some interesting behavior similar
to tests done on this problem by both 
\cite{FLASH1} and \cite{ZEUS}.   
In general we would expect the schemes to be of 
high order accuracy away from shocks and discontinuities 
and of first order accuracy close to discontinuities.  
When a narrow Gaussian profile is 
discretized it behaves as neither a 
discontinuity nor a smooth function.  Hence with decreasing 
mesh spacing the order of convergence is fractional and
increasing before it starts to decrease for 
the 4th  order reconstruction.

The next consideration  was the propagation of a sinusoidal sound
wave consisting of a density and a pressure wave 
perturbation propagating at the speed of sound, $C_s$.  The initial 
conditions are given in Eq. 32 and periodic
boundary conditions were applied while propagating 
the perturbation.
\begin{eqnarray}   
\rho = \rho_0 + \epsilon \rho_0 \cos (kx) \quad , \nonumber \\
p = p_0 + C_s^2 (\rho-\rho_0) \quad , \nonumber \\
v=C_s \frac {\rho-\rho_0}{\rho_0} \quad .
\end{eqnarray}
The background density $\rho_0=3.0$ and pressure $p_0=50.0$, 
$\epsilon$ is set to $10^{-6}$ and $k$ is the wave
number.  Shown in Table 4, are the $L_1$ norms 
of density error of our solutions for the 
3$^{rd.}$ and 4$^{th.}$ order reconstructions.

\subsubsection{Shock Tube Tests}

Shock tube tests are used to test a codes ability to 
capture shock waves.  In one
dimension, shock tube tests can be described as follows. 
A one dimensional domain of
length $l$ is divided into two halves and its initial thermodynamic
states specified.  The thermodynamic state of the tube is then advanced
in time.  The initial configurations usually give rise to shock waves,
contact discontinuities and rare faction waves whose 
amplitudes and position can be determined analytically.  
Any reliable HRSC scheme would be able
to capture these shock waves.  A collection of standard shock tube tests
have been designed to verify various qualities of any given scheme.  We
begin here by presenting some of these results in one dimension.  
The flattening and monotonicity preserving 
sub-steps involve setting some free parameters \cite{WC2}.  For each
test done here, we have indicated the values of these parameters in 
Table 5  Outflow boundary conditions were used for all 
our tests unless otherwise specified.

{\bf Test 1:} We begin with the \textit{Sod shock tube} test.  
The initial
condition for this test is given by; 
$\rho_{left}=1.0$, $v_{left}=0$, $P_{left}=1.0$,
$\rho_{right}=.125$, $v_{left}=0$, $P_{right}=0.1$.  With 
this initial condition, the following happens.  
The left pressure being greater
than the right one results in a shock wave
that will propagate rightward.  In addition, the central contact
discontinuity that is visible in the density plot 
propagates rightward, while
a rarefaction wave propagates left from its the origin.  
The results are
shown in Fig. 5 where both the numerical and 
analytical solutions are
presented for the 3$^{rd}$  and 4$^{th}$  order reconstructions.  
We note excellent
agreement between the analytic and numerical approximations.  To
complement these results, we have also shown 
the $L1$ error versus grid spacing of
our solutions for both the 3$^{rd}$ and 4$^{th}$ order 
reconstructions in Table 6.  The
results satisfy the expected first order convergence rate for both
reconstructions.  Finally, comparing the results 
to \cite{ZEUS}, \cite{FLASH1} and
the analytic solution, we note excellent agreement.  

{\bf Test 2:} The next test is 
\textit{Lax's Problem} (\cite{lax54}).  The
thermodynamics state is given by, $\rho_{left}=0.445$, $u_{left}=0.698$,
$P_{left}=3.528$, $\rho_{right}=0.5$, $u_{right}=0$, and 
$P_{right}=0.571$.  The density, velocity and pressure 
profiles are shown in Fig. 6 for both the 3$^{rd.}$ 
and 4$^{th.}$ order reconstructions.  Note that the 4$^{th.}$ order 
results are marginally better than the
3$^{rd.}$ order scheme around the shock and 
contact discontinuities.  These
results also show good qualitative agreements 
with those given by \cite{suresh}.

{\bf Test 3:} The next test is \textit{Shu's Problem} (\cite{SHU90}).  This
problem is designed to test the ability of the scheme to resolve both a
discontinuity as well as an oscillatory solution.  The initial 
state is, $\rho_{left}=3.857143$, $u_{left}=2.629369$, $P_{left}=10.3333$,
$\rho_{right}=1.0+ 0.2 \sin (5 \pi x)$, $u_{right}=0.0$,
$P_{right}=1.0$.  The left and right sides are defined as left$:$ 
$-1 < $ x $< -0.8$ and right$:$ $-0.8 <$ x $< 1.0$.  Our results are shown
in Fig. 7.  There is a noticeable difference between the 3$^{rd.}$ and
4$^{th.}$ order reconstructions in this case. We find the 4$^{th.}$ order solution
to be more oscillatory than  that using the 3$^{rd.}$ order results.  
Qualitatively our results match those obtained by \cite{suresh}.

{\bf Test 4:} The next test is \textit{Sod Strong
Shock Problem} (\cite{FLASH1}).  This test is designed to capture
stronger shocks than any of the above, and 
hence, is quite challenging.  The
initial state is, $\rho_{left}=10.0$, $P_{left}=100.0$,
$\rho_{right}=1.0$ and $P_{right}=1.0$.  The results are shown in 
Fig. 8.  There is excellent agreements between our results and 
the analytic solution.  Also, a direct comparison
between our scheme and that of \cite{FLASH1} shows good 
qualitative agreement.  

{\bf Test 5:} The next test is the interaction between two blast waves
described by \cite{WC1}.  The initial state consists of
three constant states on the domain $x$ $\in$ $[0,1]$; $\rho_{[0,.1]}=1.0$,
$u_{[0,.1]}=1.0$, $P_{[0,.1]}=1000.0$, $\rho_{[.1,.9]}=1.0$,
$u_{[.1,9]}=0.0$, $P_{[.1,9]}=.01$, $\rho_{[.9,1.0]}=1.0$,
$u_{[0.9,1.0]}=0.0$ and $P_{[0.9,1.0]}=100.0$.  Solid reflective boundary
conditions are used on the computational domain.  This is one of the most
demanding tests for an HRSC code.  The expected solution structure is as
follows; shocks are driven into the middle part of the grid while
rarefaction waves propagate toward the outer boundaries.  By the time the
shocks collide, the rarefaction waves have caught up to them, making 
their post shock structure complex.  For more details of the complexities
involved with this test, see \cite{FLASH1}.  Our results are shown
in Figs. 9, 10 and 11.  
The density and velocity profiles are shown from
$t=0.026$ (before the shock waves begin to interact) to $t=0.038$ (after
they have stopped interacting).  These results are compared directly to
those of \cite{FLASH1} and \cite{ZEUS} and show excellent 
qualitative agreement with them.

\subsection{2D Riemann Problems using WENO}

As was the case in one dimension, the most elementary two dimensional tests
of HRSC schemes are 2D shock tube (2D Riemann problem) tests.
It can be described as follows. A 
square computational domain is divided into
four quadrants and thermodynamics states specified.  The initial
data are constant in each quadrant and restricted so that only one
elementary wave, a one dimensional shock, a one dimensional rarefaction
wave or a two dimensional contact discontinuity
appears at each interface.  According to Lax $\&$ Liu (1998), 
the total number of
genuinely different configurations for polytropic gases in 2D shock
tube tests is nineteen.  Lax $\&$ Liu (1998) solved for 
all nineteen configurations
to demonstrate the utility of their so-called $positive$ $Scheme$ 
(an HRSC scheme).  Kurganov $\&$ Tadmore (2002) perform
exactly the same calculations to test their HRSC scheme, that was based
on ENO reconstructions and a genuinely multidimensional CENO 
approach.  Following these two works, we have performed similar
computations to test our scheme.  Kurganov $\&$ Tadmore (2002) 
have demonstrated that the central scheme in 
combination with an ENO reconstruction scheme does
indeed satisfactorily solve the 2D Riemann problems 
of Lax $\&$ Liu (1998).  However, they comment
that because WENO reconstruction is  based on 
smoothness indicators, a priori
information of the solution structure is necessary for solving 
the variety of 2D Riemann problems they considered.  We
demonstrate here that a fixed set of parameters for computing the
smoothness indicators solves the 2D Riemann problems
without a priori knowledge of the solution structure.  
To do this, we turned off the steepening,
flattening and monotonicity preserving component 
of our algorithm for these tests.  This also
allows us to make a direct qualitative comparison between 
our results and those of Lax $\&$ Liu (1998) 
and Kurganov $\&$ Tadmore (2002).  Hence we have tackled two issues; 
the WENO related issue just mentioned above
as well as testing our algorithm.  To describe the initial conditions
for our 2D Riemann problems and the initial patterns expected from
them, we define the following notations;

\noindent $R^{\rightarrow}_{lr} :$ Forward rarefaction wave
\newline
$R^{\leftarrow}_{lr} :$ Backward rarefaction wave
\newline
$S^{\rightarrow}_{lr} :$ Forward shock wave
\newline
$S^{\leftarrow}_{lr} :$ Backward shock wave
\newline
$J^{-}_{lr} :$ Positive Slip line
\newline
$J^{+}_{lr} :$ Negative Slip line.

Table 7 gives the 
wave patterns expected for
each of the nineteen configurations.  In it, the subscripts represent the 
wave pattern expected between the quadrants, e.g., $J_{21}^{+}$ 
means that between the second and the first quadrant, a negative slip 
line will results from the initial conditions at these two quadrants.  
The convention used to label the quadrants are as follows.  
Considering a square, 
the North-East quadrant is labelled 1,  North-East quadrant is labelled 2,  
South-West quadrant is labelled 3,  South-East quadrant is labelled 4.     
Tables 8-9 presents the
initial conditions that give rise to the various wave patterns 
which are shown in table 7.  
For details of the thermodynamics conditions that give rise to the various
wave patterns,  see Lax $\&$ Liu (1998).  
Each of our computations were done using $n=400$ and the adiabaticity 
constant $\gamma$ $=$ $1.4$.  The time to
integration is case dependent.  An unsplit algorithm is used to 
advance the solution in time.  We also used
the 4$^{th}$ order, dimensionally split WENO reconstruction in 
all our calculations in this and the next section.  
We expect to obtain comparable results with 
3$^{rd}$ order reconstructions.

Our results are shown in Figs. 12-16.
By direct qualitative comparison with Lax $\&$ Liu (1998) and
Kurganov $\&$ Tadmore (2002), it is noted that all features 
of every configurations obtained by the previous studies 
have been recovered.

\subsection{Some more 2D test Problems}
Next we present some other standard tests for 
multidimensional HRSC schemes.  Following 
Fryxell et al.  (2000), and Stone $\&$ Norman (1992),
we consider an explosion problem,  
a two dimensional Sod Shock
problem and a Sedov blast wave problem.  
The purpose of these tests is to
verify the robustness of the algorithm 
for stronger shock waves than the previous tests.  

{\bf Test 1: The 2D explosion problem.} The following 
initial condition sets off a spherically symmetric explosion;
$(\rho,u,v,P)$ $=$ $(1.0,0.0,0.0,1.0)$ if $x^2 +y^2 < .2^2$, 
else $(\rho,u,v,P)$ $=$ $(.125,0.0,0.0,0.1)$.    
The computational domain is a square of length 2 units and the initial
high density and pressure region is cantered around the origin.  We have
shown the density and pressure profiles at time t$=$.25 for n=200 in
Fig. 17.  The profile shown is along the x-axis of the explosion.  The
solid line represents the profile obtained from a one dimensional
analytical computation.  Note that our 2D explosion 
results are in good agreement with the one dimensional calculations.

{\bf Test 2: The 2D Sod shock problem.} The 
thermodynamic state is the
same as the 1D Sod shock problem 
except the membrane separating the two
regions is chosen to be along the diagonal of the square 
region.  Shown in
Fig. 18 are the density, velocity and pressure profiles along the
diagonal of the domain.  Similar to the one 
dimensional test, there is 
good agreement between the analytical and numerical approximations.

{\bf Test 3: The Sedov Blast wave problem.} This problem is the 
self-similar evolution of a cylindrical blast wave 
from a delta-function initial
pressure perturbation in an otherwise homogeneous medium.  The initial
conditions are exactly those considered by \cite{FLASH2} in their
test.  We consider a small region of radius $\delta r$ at the canter of
the grid.  The pressure inside this region is given by
\begin{equation}
p_0 = \frac {3(\gamma-1) \epsilon}{3 \pi \delta r^2} \quad .
\end{equation}
The ambient pressure is set to $10^{-5}$ and the density 
is set to $\rho$ $=$ $1.0$ throughout the domain.  The gas is
assumed to be stationary at time t=0 and we have taken 
$\delta r$ $=$ $3.5*mesh spacing$.  For analytical solutions to the
problem we refer to \cite{FLASH2}.  In Fig. 19, we have shown the
density, pressure and velocity profiles at 
time $t$ $=$ $.05$ units.  In the same plot we have indicated the
analytical solutions as well.  We note good agreement between the
analytical and computed results.

\section{Conclusions and future work}
\label{section:summary}

We have tested a new dimensionally unsplit multidimensional 
hydrodynamics code using a robust,
multidimensional HRSC scheme based on
central semidiscrete Godunov type schemes,  
WENO data reconstruction algorithms and the PPM method.  
To our knowledge, this is the first
multi-purpose hydrodynamics code based 
on this approach that has been 
developed for computational astrophysics.  To ensure 
robustness, we have modified the
standard WENO schemes and added elements 
of the PPM reconstruction scheme.  
Our new algorithm and code is 
tested by a collection of standard one 
and two dimensional tests.  Whenever
possible,  the results have been compared to the literature and 
analytic solutions.  Overall, both our one and two dimensional codes
perform well without having to fine tune some of the user supplied 
input parameters for a wide range of tests.  From the results 
we may conclude that the
algorithm proposed here performs comparably to 
other HRSC schemes.  This success in implementing WENO 
based codes efficiently takes us a step closer 
towards using arbitrarily high order  data reconstruction 
schemes in computational astrophysics.  

In this context, the present work should be considered 
as taking the first few steps toward the development of a 
robust, multipurpose, multidimensional HRSC scheme for
computational astrophysics.  The reliability and applicability of a given
algorithm is best tested by applying it to a variety of problems.  This
usually exposes potential weaknesses of the algorithm that can then be
rectified.  It is our intention to apply this code to a number of
astrophysical problems.  At present, we are considering a
multidimensional study of pulsar bow shock 
structure simulations.  We are also planning to study gravitational 
waveforms emitted by collapsing stars.  To extend the code's capabilities,
several extensions are planned for the future.  Key among them are an 
extension to three dimensions, addition of 
adaptive capabilities and application of the 
algorithm to Magnetohydrodynamics.  
In addition, we are also planning to implement the algorithm
using MPI for parallel architectures.  As mentioned before, there have been
a number of other advances of the central semidiscrete schemes that is
worth investigating along the lines of this algorithm.  
These include the genuinely multidimensional
formulation of the scheme and the scheme on 
an unstructured grid, etc.  It is
clear that there is room for a good deal of work using such
algorithms.  We look forward to making progress in the future.

\begin{acknowledgements}

The authors would like to thank Martin Gander for providing 
valuable advice, insight  and review of some of the results 
during various stages of this work.  
T.R. would like to thank Jose font,  Andrew MacFadyen and 
Chris Fragile for useful discussions and suggestions.  
Thanks also to Gil Holder for taking interest in this work.  
T.R. would also like to 
thank Steve Liebling at the C.W. Post campus of Long Island University for 
hospitality during part of this work.  This work was supported 
by the National Sciences and Engineering Research Council (NSERC) of Canada. 

\end{acknowledgements}

\begin{center}
\begin{table}[!h]
\centering

\begin{tabular}{|c|cccc|}
\hline
N & $L^1$-error & Rate & $L^{\infty}$-error & Rate \\
\hline
40 &  0.00189 & - & 0.001020 & - \\
80 & 0.0002815 & 2.76 & 0.000254 & 2.00 \\
160 & 3.73E-05 & 2.90 & 6.17E-05 & 2.00 \\
320 & 2.93E-06 & 3.7 & 6.92E-06 & 3.2 \\
640 & 9.87E-08 & 4.9 & 1.49E-07 & 5.53 \\
1280 & 2.71E-09 & 5.2  & 2.38E-09 & 6.0 \\
\hline
\end{tabular}

\caption{$L^1$ and $L^{\infty}$ errors for 3$^{rd.}$  order reconstruction (Sec. 3.2)}
\end{table}
\end{center}

\begin{center}
\begin{table}[!h]
\centering

\begin{tabular}{|c|cccc|}
\hline
N & $L^1$-error & Rate & $L^{\infty}$-error & Rate \\
\hline
40 & 2.64E-05 & - & 7.80E-06 & - \\
80 & 1.83E-06 & 3.86  & 4.79E-07 & 4.06 \\
160 & 1.17E-07 & 3.96 & 2.98E-08 & 4.0 \\
320 & 7.41E-09 & 4.03 & 1.86E-09 & 4.03 \\
640 & 4.64E-10 & 4.00 & 1.16E-10 & 4.00 \\
1280 & 2.90E-11 & 4.00 & 7.41E-12 & 4.00 \\
\hline
\end{tabular}

\caption{$L^1$ and $L^{\infty}$ errors for 4$^{th}$ order reconstruction (Sec. 3.2)}
\end{table}
\end{center}

\begin{center}
\begin{table}[!h]
\centering

\begin{tabular}{|c|cccc|}
\hline
N & $L^1$-error (3$^{rd.}$  order) & Rate &$L^1$-error (4$^{th.}$  order) & Rate  \\
\hline
40 & 3.87E-2  & - &3.38E-2 & -\\
80 & 2.67E-2 & .53 &1.61E-2 & 1.07 \\
160 & 1.63E-2 & .74 & 5.19E-3  & 1.63  \\
320 & 7.12E-3  & 1.19 & 5.04E-4 & 3.37 \\
640 & 2.09E-3  & 1.77 & 4.27E-05  & 3.57 \\
1280 & 5.92E-4 & 1.82 & 9.29E-06 & 2.20 \\
2560 & 1.38E-4 & 2.11 & 1.83E-06 & 2.35 \\
\hline
\end{tabular}

\caption{$L^1$ errors for the advection of a narrow Gaussian pulse
by 3$^{rd.}$  and 4$^{th.}$  order WENO reconstruction.}
\end{table}
\end{center}

\begin{center}
\begin{table}[!h]
\centering

\begin{tabular}{|c|cccc|}
\hline
N & $L^1$-error (3$^{rd.}$  order) & Rate &$L^1$-error (4$^{th.}$  order) & Rate  \\
\hline
40 & 6.59E-07 & - & 7.56E-08  & -\\
80 & 8.22E-08 & 3.01 & 9.07E-09 & 3.06 \\
160 & 1.02E-08  & 2.93 & 1.19E-09& 2.94 \\
320 & 1.28E-09 & 3.00 & 1.76E-10 & 2.76 \\
\hline
\end{tabular}

\caption{$L^1$ errors for the advection of a sinusoidal
perturbation by 3$^{rd.}$  and 4$^{th.}$  order WENO reconstruction.}
\end{table}
\end{center}

\begin{center}
\begin{table}[!h]
\centering

\begin{tabular}{cccccccc}
\hline
Test & $K_0$ & $\eta^{(1)}$ & $\eta^{(2)}$ & $\epsilon^{(1)}$ & 
$\omega^{(1)}$ & $\omega^{(2)}$ & $\epsilon^{(2)}$ \\
 \hline
Test 1 &  0.10 & 20.0  & 0.05 & 0.1 & 0.52 & 10.0 & 0.1 \\
Test 2 &  0.10 & 20.0 & 0.05 & 0.1 & 0.52 & 10.0 & 0.1 \\
Test 3 &  0.10 & 20.0 & 0.05 & 0.1 & 0.52 & 10.0 & 0.1 \\
Test 4 &  0.10 & 20.0  & 0.05 & 0.01 & 0.52 & 10.0 & 0.33 \\
Test 5 &  0.10 & 20.0  & 0.05  & 0.01 & 0.52 & 10.0 & 0.33 \\
\hline
\end{tabular}

\label{tabla:ppm_param}
\caption{Values of the monotonicity, flattening and contact steepening parameters 
used for one dimensional shock tube tests presented in Sec. 5.1}
\end{table}
\end{center}

\begin{center}
\begin{table}[!h]
\centering

\begin{tabular}{|c|cccc|}
\hline
N & $L^1$-error (3$^{rd.}$  order) & Rate &$L^1$-error (4$^{th.}$  order) & Rate  \\
\hline
40 &  4.82E-1 & - & 1.75E-1 & - \\
80 & 2.45E-1 & 0.97 & 9.99E-02 & 0.82  \\
160 & 1.20E-01 & 1.03 & 5.01E-02 & 0.99 \\
320 & 6.18E-02 & 0.97 & 2.47E-02 & 1.02 \\
640 & 3.18E-02 & 0.97 & 1.25E-02 & 0.99 \\
\hline
\end{tabular}

\caption{$L^1$-errors in density for the Sod shock tube test for
3$^{rd.}$  and 4$^{th.}$  order WENO reconstructions.}
\end{table}
\end{center}

\begin{center}
\begin{table}[h]
\centering

\begin{tabular}{cccccccc}
\hline
Test & $K_0$ & $\eta^{(1)}$ & $\eta^{(2)}$ & $\epsilon^{(1)}$ & $\omega^{(1)}$ & 
$\omega^{(2)}$ & $\epsilon^{(2)}$ \\
 \hline
Test 1 &  .10 & 20.0  & 0.05 & 0.1 & 0.52 & 10.0 & 0.33 \\
Test 2 &  .10 & 20.0 & 0.05 & 0.1 & 0.52 & 10.0 & 0.33 \\
Test 3 &  .10 & 20.0 & 0.05 & 0.01 & 0.52 & 10.0 & 0.33 \\
\hline

\end{tabular}

\label{tabla:ppm_param}
\caption{Values of the monotonicity, flattening and contact steepening parameters 
used for 2D tests from Sec. 5.3}
\end{table}
\end{center}

\begin{center}
\begin{table}[h]
\centering

\begin{tabular}{ccccc}
\hline
Configuration & $C_{12}$ & $C_{32}$ & $C_{34}$ & $C_{41}$ \\
\hline
Config. 1  & $R_{21}^{\rightarrow}$  & $R_{32}^{\rightarrow}$ &
 $R_{34}^{\rightarrow}$ & $R_{41}^{\rightarrow}$ \\
Config. 2  & $R_{21}^{\rightarrow}$  & $R_{32}^{\leftarrow}$ &
 $R_{34}^{\leftarrow}$ & $R_{41}^{\rightarrow}$ \\
Config. 3  & $S_{21}^{\leftarrow}$  & $S_{32}^{\leftarrow}$ &
 $S_{34}^{\leftarrow}$ & $S_{41}^{\leftarrow}$ \\
Config. 4  & $S_{21}{\leftarrow}$  & $S_{32}{\rightarrow}$ &
 $S_{34}^{\rightarrow}$ & $R_{41}^{\leftarrow}$ \\
Config. 5  & $J_{21}^{-}$  & $J_{32}^{-}$ &
 $J_{34}^{-}$ & $J_{41}^{-}$ \\
Config. 6  & $J_{21}^{-}$  & $J_{32}^{+}$ &
 $J_{34}^{-}$ & $J_{41}^{+}$ \\
Config. 7  & $R_{21}^{\rightarrow}$  & $J_{32}^{-}$ &
 $J_{34}^{-}$ & $R_{41}^{\rightarrow}$ \\
Config. 8  & $R_{21}^{\leftarrow}$  & $J_{32}^{-}$ &
 $J_{34}^{-}$ & $R_{41}^{\leftarrow}$ \\
Config. 9  & $J_{21}^{+}$  & $R_{32}^{\leftarrow}$ &
 $J_{34}^{-}$ & $S_{41}^{\leftarrow}$ \\
Config. 10  & $J_{21}^{-}$  & $R_{32}^{\rightarrow}$ &
 $J_{34}^{\rightarrow}$ & $R_{41}^{\rightarrow}$ \\
Config. 11  & $S_{21}^{\leftarrow}$  & $J_{32}^{+}$ &
 $J_{34}^{+}$ & $R_{41}^{\leftarrow}$ \\
Config. 12  & $S_{21}^{\rightarrow}$  & $J_{32}^{+}$ &
 $J_{34}^{+}$ & $R_{41}^{\rightarrow}$ \\
Config. 13  & $J_{21}^{-}$  & $S_{32}^{\leftarrow}$ &
 $J_{-}^{\rightarrow}$ & $S_{41}^{\leftarrow}$ \\
Config. 14  & $J_{21}^{+}$  & $R_{32}^{\rightarrow}$ &
 $J_{+}^{\rightarrow}$ & $R_{41}^{\rightarrow}$ \\
Config. 15  & $R_{21}^{\rightarrow}$  & $J_{32}^{-}$ &
 $S_{34}^{\leftarrow}$ & $R_{41}^{\leftarrow}$ \\
Config. 16  & $R_{21}^{\leftarrow}$  & $J_{32}^{-}$ &
 $R_{34}^{+}$ & $S_{41}^{\rightarrow}$ \\
Config. 17  & $J_{21}^{-}$  & $S_{32}^{\leftarrow}$ &
 $J_{34}^{-}$ & $R_{41}^{\rightarrow}$ \\
Config. 18  & $J_{21}^{+}$  & $S_{32}^{\leftarrow}$ &
 $J_{34}^{+}$ & $R_{41}^{\rightarrow}$ \\
Config. 19  & $J_{21}^{+}$  & $S_{32}^{\leftarrow}$ &
 $J_{34}^{-}$ & $R_{41}^{\rightarrow}$ \\
\hline
\end{tabular}

\label{tabla:2Dinitconfig}
\caption{Expected wave Patterns for 2D Riemann Problem tests}
\end{table}
\end{center}

\begin{center}
\begin{table}[h]
\centering

\begin{tabular}{ccccc}
\hline
Configuration & $\rho$ & $u_x$ & $u_y$ & P \\
 \hline
(config. 1) Quad. 1  &  1.0 & 0. & 0. & 1.0 \\
Quad. 2 &  .5197 & -.7259 & 0. & .4 \\
Quad. 3 &  .1072 & -.7259 & -1.4045 & .0439 \\
Quad. 4 &  .2579 & 0. & -1.4045 & .15 \\
 \hline
(config. 2) Quad. 1 &  1.0 & 0. & 0. & 1.0 \\
Quad. 2 &  .4 & -.7259 & 0. & .4 \\
Quad. 3 &  1. & -.7259 & -.7259 & 1. \\
Quad. 4 &  .5197 & 0. & -.7259  & .4 \\
 \hline
(config. 3) Quad. 1 &  1.5 & 0. & 0. & 1.5 \\
Quad. 2 &  .5323 & 1.206 & 0. & .3 \\
Quad. 3 &  .138 & 1.206 & 1.206 & .029 \\
Quad. 4 &  .5323 & 0. & 1.206 & .3 \\
 \hline
(config. 4) Quad. 1 &  1.1 & 0. & 0. & 1.1 \\
Quad. 2 &  .5065 & .8939 & 0. & .35 \\
Quad. 3 &  1.1 & .8939 & .8939 & 1.1 \\
Quad. 4 &  .5065 & 0. & .8939 & .35 \\
 \hline
(config. 5) Quad. 1 &  1. & -.75 & -.5 & 1. \\
Quad. 2 &  2. & -.75 & .5 & 1. \\
Quad. 3 &  1. & .75 & .5 & 1. \\
Quad. 4 &  3. & .75 & -.5 & 1. \\
\hline
(config. 6) Quad. 1 &  1. & .75 & -.5 & 1. \\
Quad. 2 &  2. & .75 & .5 & 1. \\
Quad. 3 &  1. & -.75 & .5 & 1. \\
Quad. 4 &  3. & -.75 & -.5 & 1. \\
\hline
(config. 7) Quad. 1 &  1. & .1 & .1 & 1. \\
Quad. 2 &  .5197 & -.6259 & .1 & .4 \\
Quad. 3 &  .8 & .1 & .1 & .4 \\
Quad. 4 &  .5197 & .1 & -.6259 & .4 \\
\hline
\end{tabular}

\label{tabla:2Dinitconfig}
\caption{Initial conditions for 2D Riemann Problem tests}
\end{table}
\end{center}

\begin{center}
\begin{table}[h]
\centering

\begin{tabular}{ccccc}
\hline
Configuration & $\rho$ & $u_x$ & $u_y$ & P \\
 \hline
(config. 8) Quad. 1 &  .5197 & .1 & .1 & .4 \\
Quad. 2 &  1. & -.6259 & .1 & 1. \\
Quad. 3 &  .8 & .1 & .1 & 1. \\
Quad. 4 &  1. & .1 & -.6259 & 1. \\
\hline
(config. 9) Quad. 1 &  2. & 0. & -.5606 & 1. \\
Quad. 2 &  1. & 0. & -1.2172 & 8. \\
Quad. 3 &  .4736 & 0. & 1.2172 & 2.6667 \\
Quad. 4 &  .9474 & 0. & 1.1606 & 2.6667 \\
\hline
(config. 10) Quad. 1  &  1. & 0. & .4297 & 1. \\
Quad. 2 &  .5 & 0. & .6076 & 1. \\
Quad. 3 &  .2281 & 0. & -.6076 & .3333 \\
Quad. 4 &  .4562 & 0. & -.4297 & .3333 \\
 \hline
(config. 11) Quad. 1 &  1. & .1 & 0. & 1. \\
Quad. 2 &  .5313 & .8276 & 0. & .4 \\
Quad. 3 &  .8 & .1 & 0. & .4 \\
Quad. 4 &  .5313 & .1 & .7276 & .4 \\
\hline
(config. 12) Quad. 1 &  .5313 & 0. & 0. & .4 \\
Quad. 2 &  1. & 7276. & 0. & 1. \\
Quad. 3 &  .8 & 0. & 0. & 1. \\
Quad. 4 &  1. & 0. & .7276 & 1. \\
\hline
(config. 13) Quad. 1 &  1. & 0. & -.3 & 1. \\
Quad. 2 &  2. & 0. & .3 & 1. \\
Quad. 3 &  1.0625 & 0. & .8145 & .4 \\
Quad. 4 &  .5313 & 0. & .4276 & .4 \\
\hline
(config. 14) Quad. 1  &  1. & 0. & .3 & 1. \\
Quad. 2 &  2. & 0. & -.3 & 1. \\
Quad. 3 &  1.039 & 0. & -.8133 & .4 \\
Quad. 4 &  .5197 & 0. & -.4259 & .4 \\
\hline
\end{tabular}

\label{tabla:2Dinitconfig}
\caption{Initial conditions for 2D Riemann Problem tests (continued)}
\end{table}
\end{center}

\begin{center}
\begin{table}[h]
\centering

\begin{tabular}{ccccc}
\hline
Configuration & $\rho$ & $u_x$ & $u_y$ & P \\
 \hline
(config. 15) Quad. 1 &  1. & .1 & -.3 & 1. \\
Quad. 2 &  .5197 & -.6259 & -.3 & .4 \\
Quad. 3 &  .8 & .1 & -.3 & .4 \\
Quad. 4 &  .5313 & .1 & .4276 & .4 \\
\hline
(config. 16) Quad. 1 &  .5313 & .1 & .1 & .4 \\
Quad. 2 &  1.0222 & -.6179 & .1 & 1. \\
Quad. 3 &  .8 & .1 & .1 & 1. \\
Quad. 4 &  1. & .1 & .8276 & 1. \\
\hline
(config. 17) Quad. 1  &  1. & 0. & -.4 & 1. \\
Quad. 2 &  2. & 0. & -.3 & 1. \\
Quad. 3 &  1.0625 & 0. & .2145 & .4 \\
Quad. 4 &  .5197 & 0. & -1.1259 & .4 \\
\hline
(config. 18) Quad. 1 &  1. & 0. & 1. & 1. \\
Quad. 2 &  2. & 0. & -.3 & 1. \\
Quad. 3 &  1.0625 & 0. & .2145 & .4 \\
Quad. 4 &  .5197 & 0. & .2741 & .4 \\
\hline
(config. 19) Quad. 1 &  1. & 0. & .3 & 1. \\
Quad. 2 &  2. & 0. & -.3 & 1. \\
Quad. 3 &  1.0625 & 0. & .2145 & .4 \\
Quad. 4 &  .5197 & 0. & -.4259 & .4 \\
\hline
\end{tabular}

\label{tabla:2Dinitconfig}
\caption{Initial conditions for 2D Riemann Problem tests (continued)}
\end{table}
\end{center}

\begin{center}
\begin{figure}[h]
\centering

  \includegraphics[angle=-90,width=0.8\textwidth]{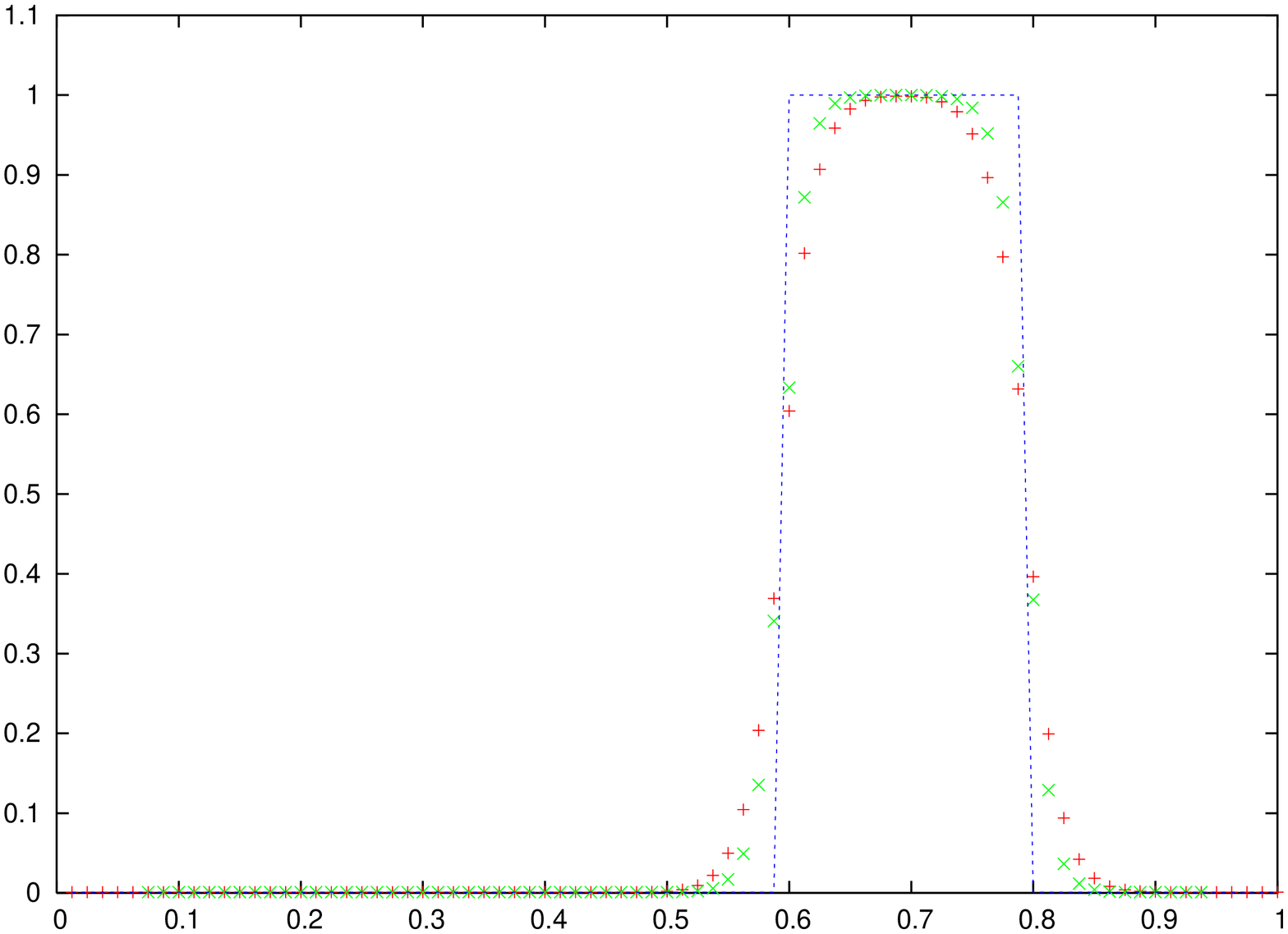}
  \includegraphics[angle=-90,width=0.8\textwidth]{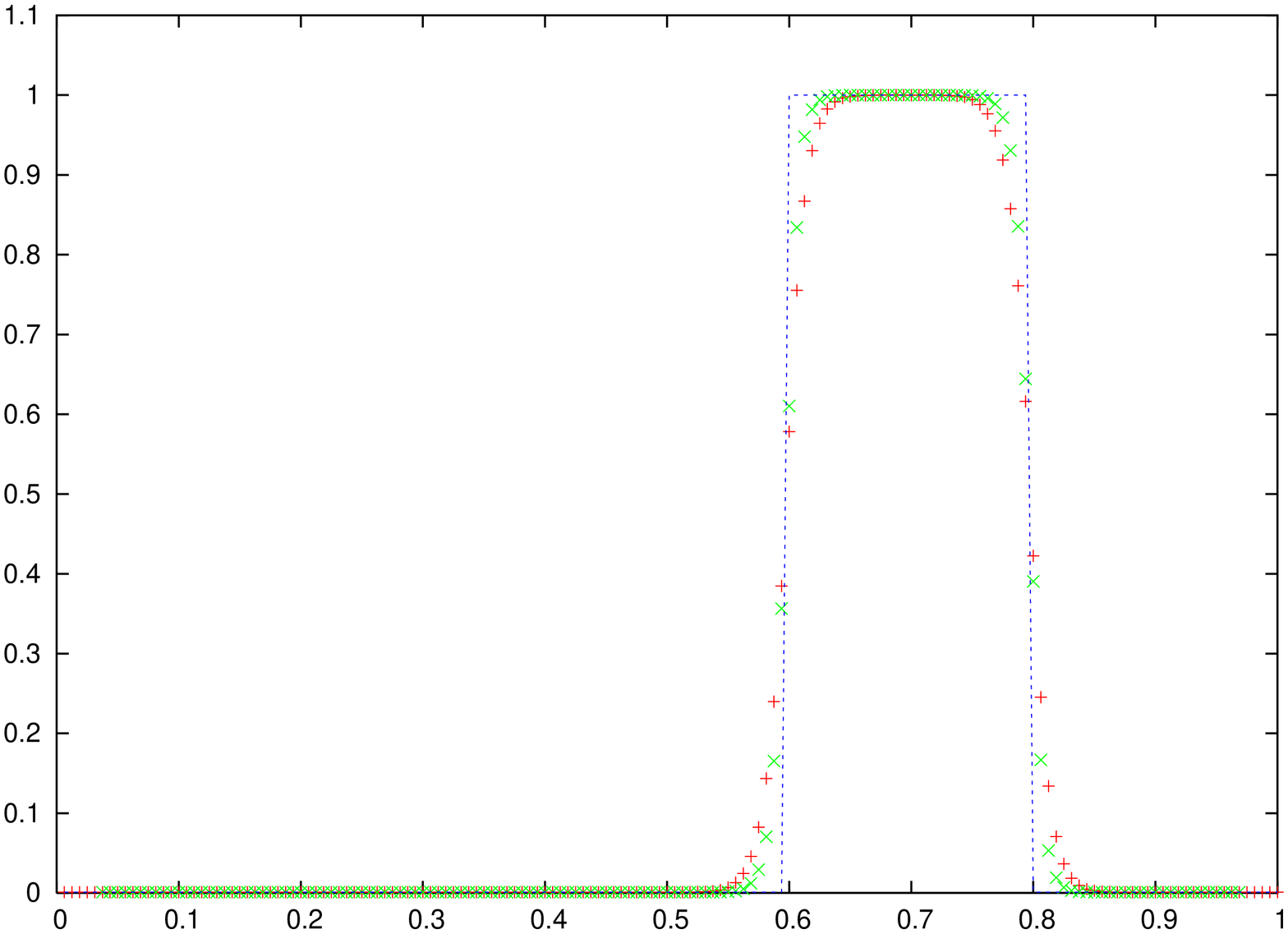}
\label{adv1}
\caption{Advection of a rectangular pulse in Eq. 31. 
Shown here are the advected pulse at time t=.2 for
n=80 (top) and n=160 (bottom). The blue dotted line represents the exact
solution. ``Green'' represents 4$^{th.}$ order WENO reconstruction and ``Red''
represents 3$^{rd.}$ order WENO reconstruction}
\end{figure}
\end{center}

\begin{figure}[h]
\centering
  \includegraphics[angle=-90,width=0.8\textwidth]{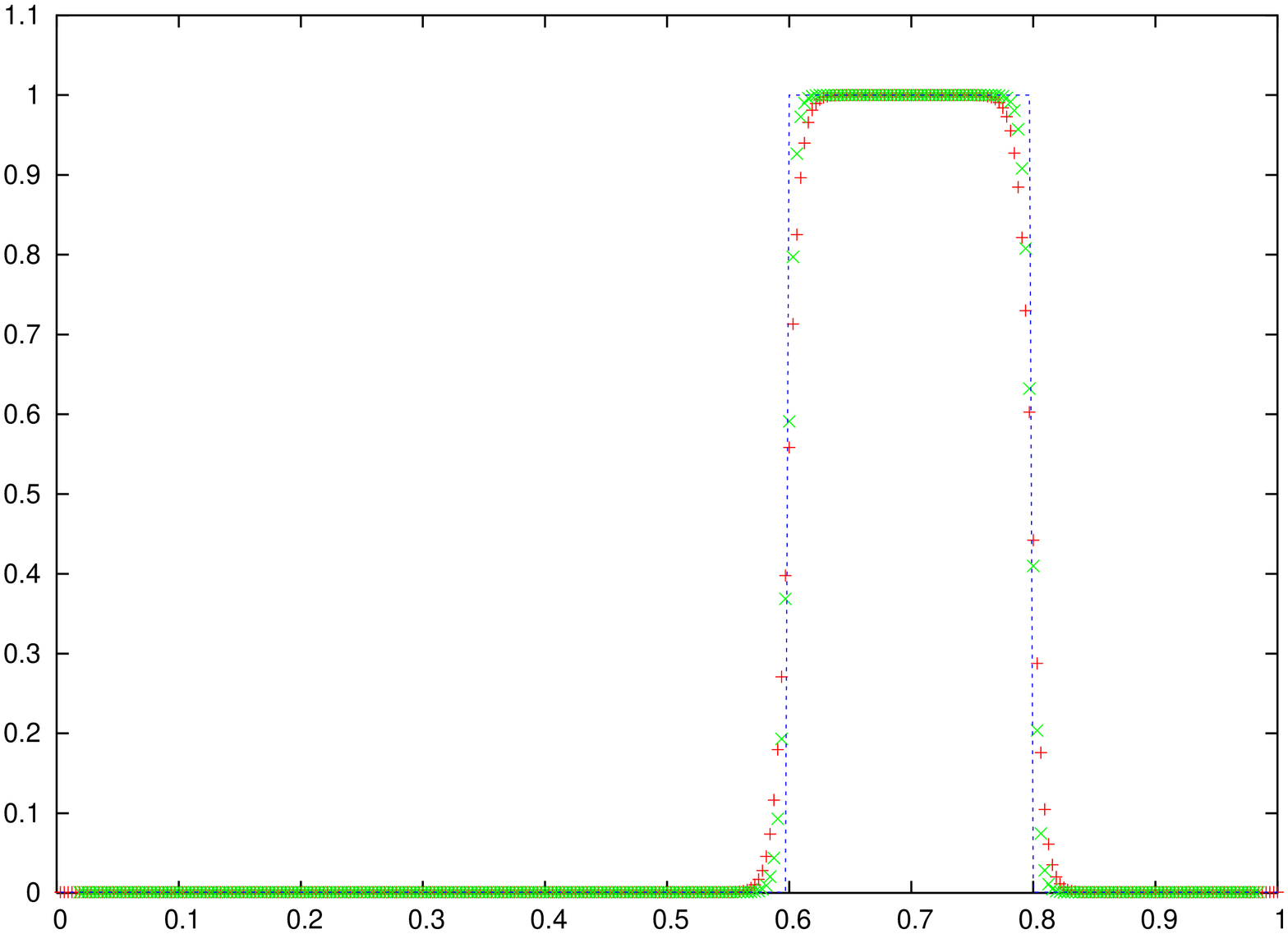}
  \includegraphics[angle=-90,width=0.8\textwidth]{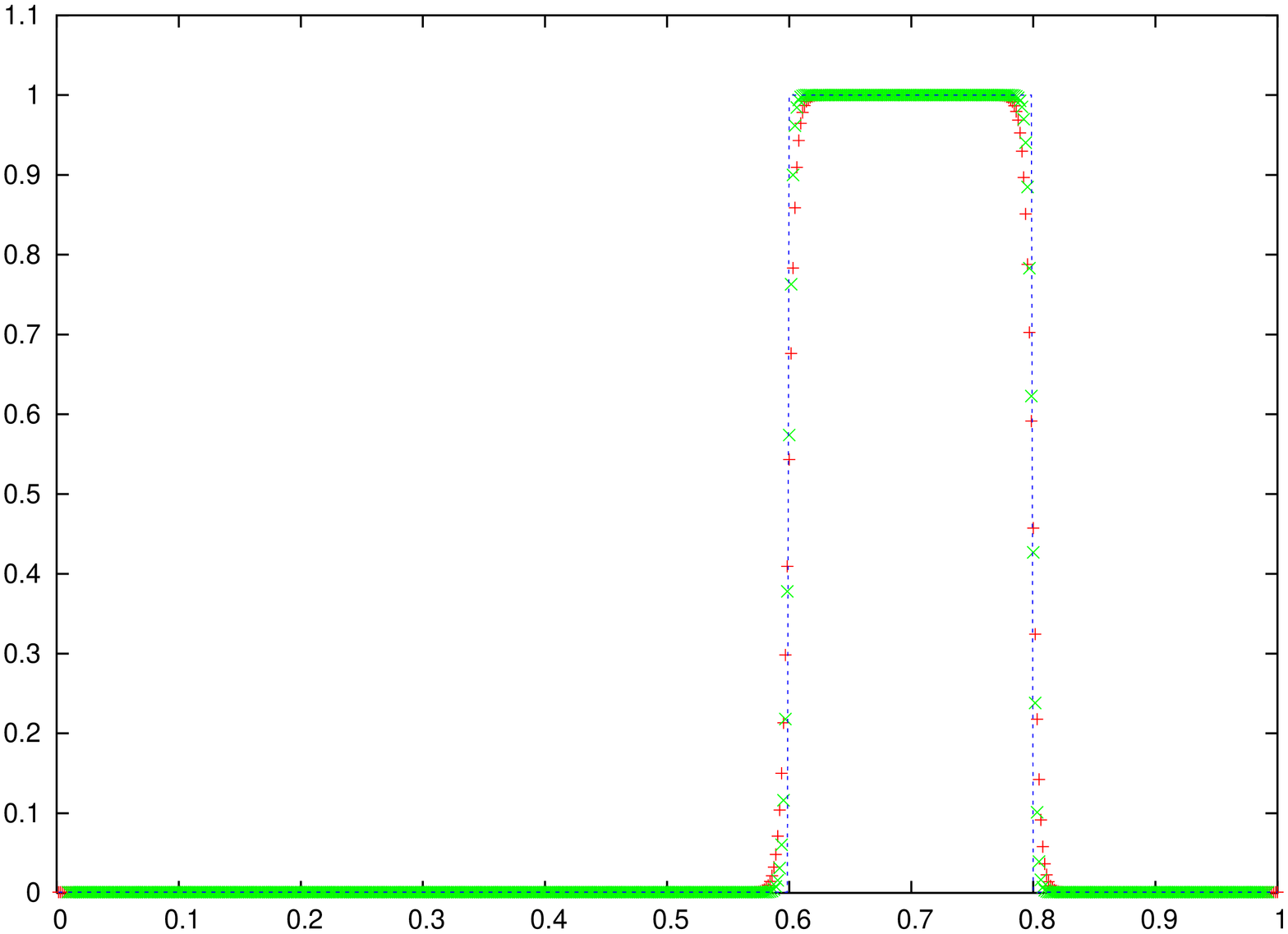}
\label{adv2}
\caption{Same as Fig. 2, n=320(top), n=640(bottom)}
\end{figure}

\begin{figure}[h]
\centering
  \includegraphics[angle=-90,width=0.8\textwidth]{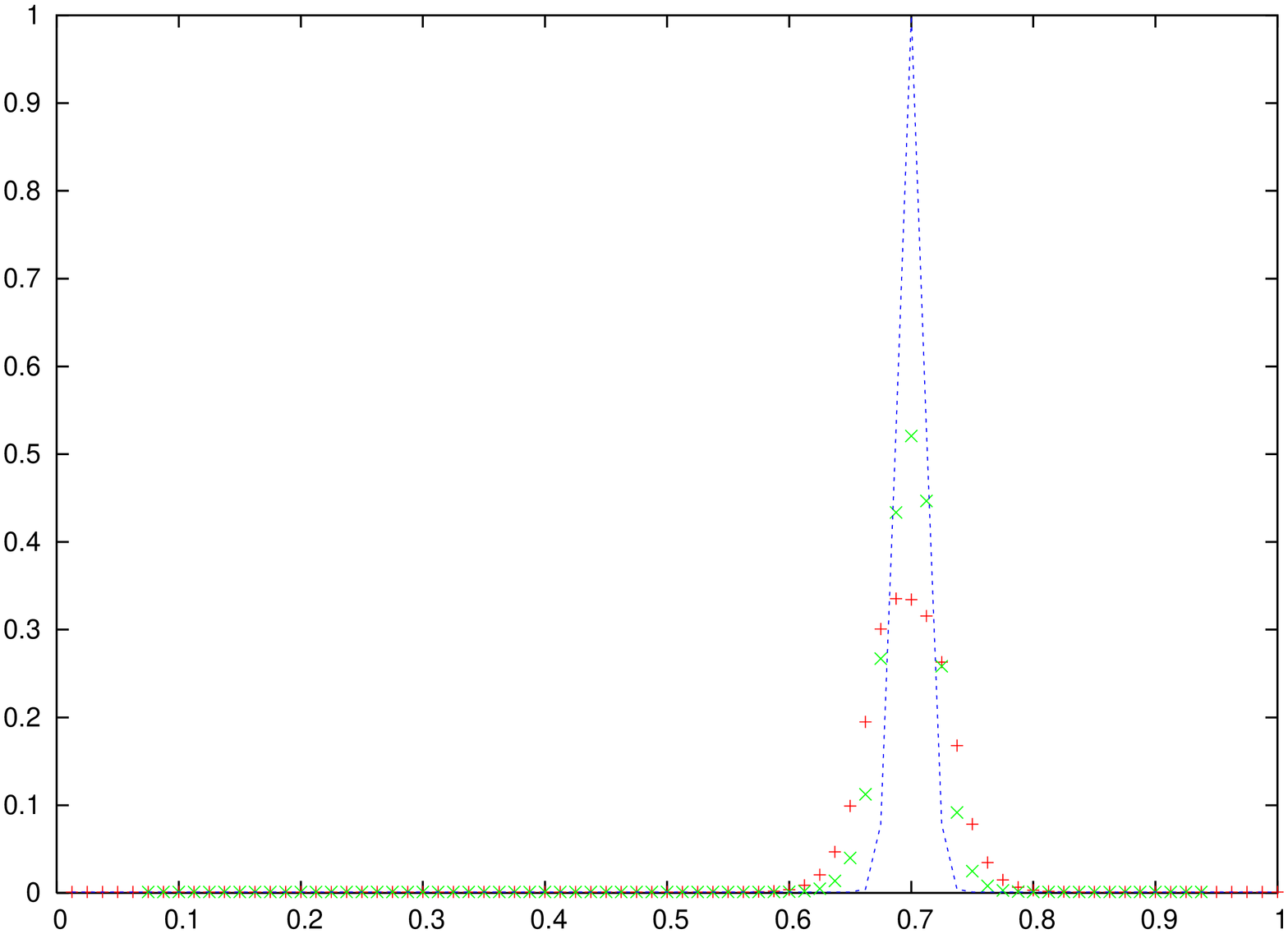}
  \includegraphics[angle=-90,width=0.8\textwidth]{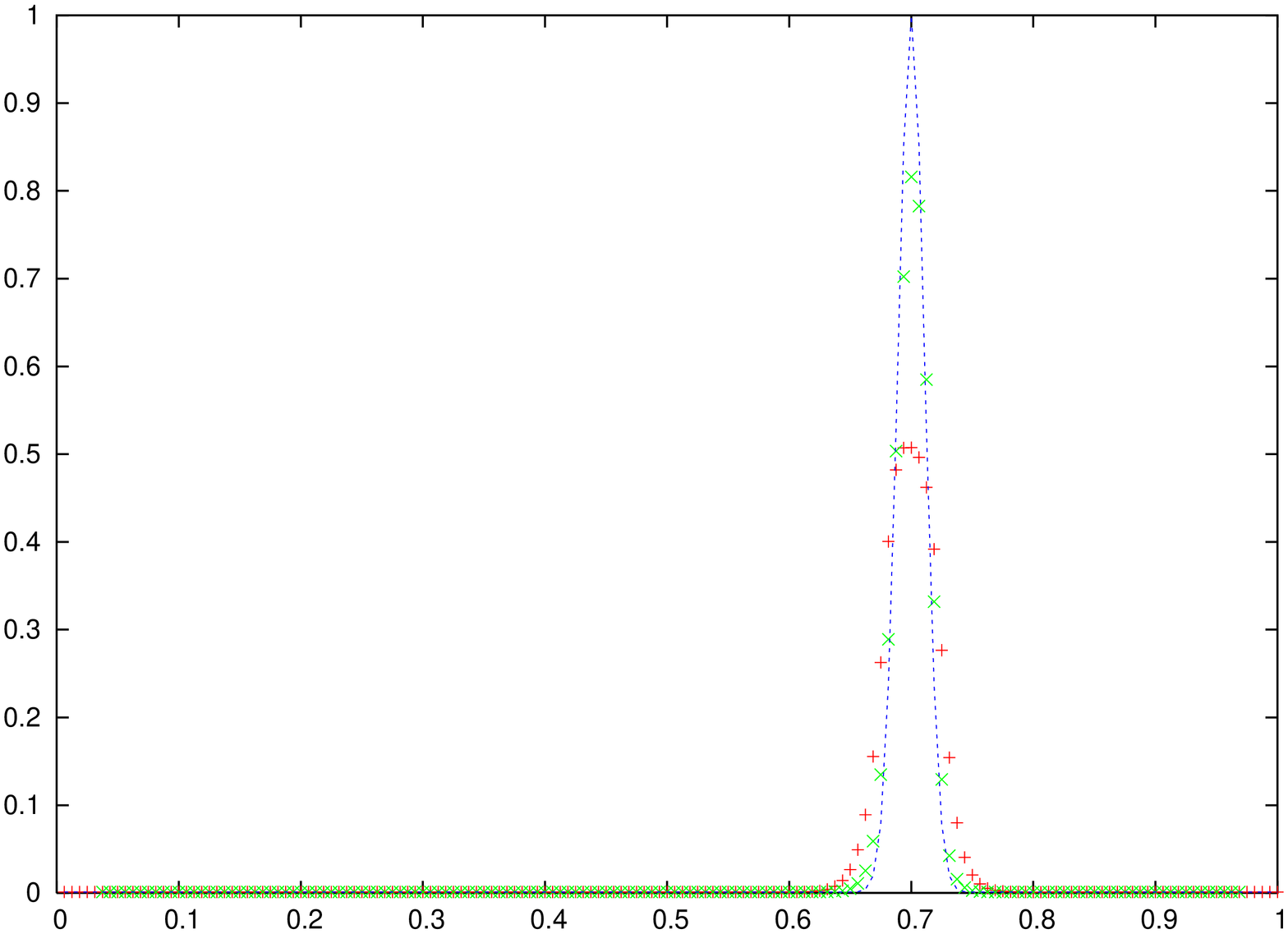}
\label{adv3}
\caption{Advection of a Gaussian pulse in Eq. 31. 
Shown here are the advected pulse at t=0.2 for
n=80(top), and n=160(bottom).  The blue dotted line represents the exact
solution. ``Green'' represents 4$^{th.}$ order WENO reconstruction and ``Red''
represents 3$^{rd.}$ order WENO reconstruction}
\end{figure}

\begin{figure}[h]
\centering
  \includegraphics[angle=-90,width=0.8\textwidth]{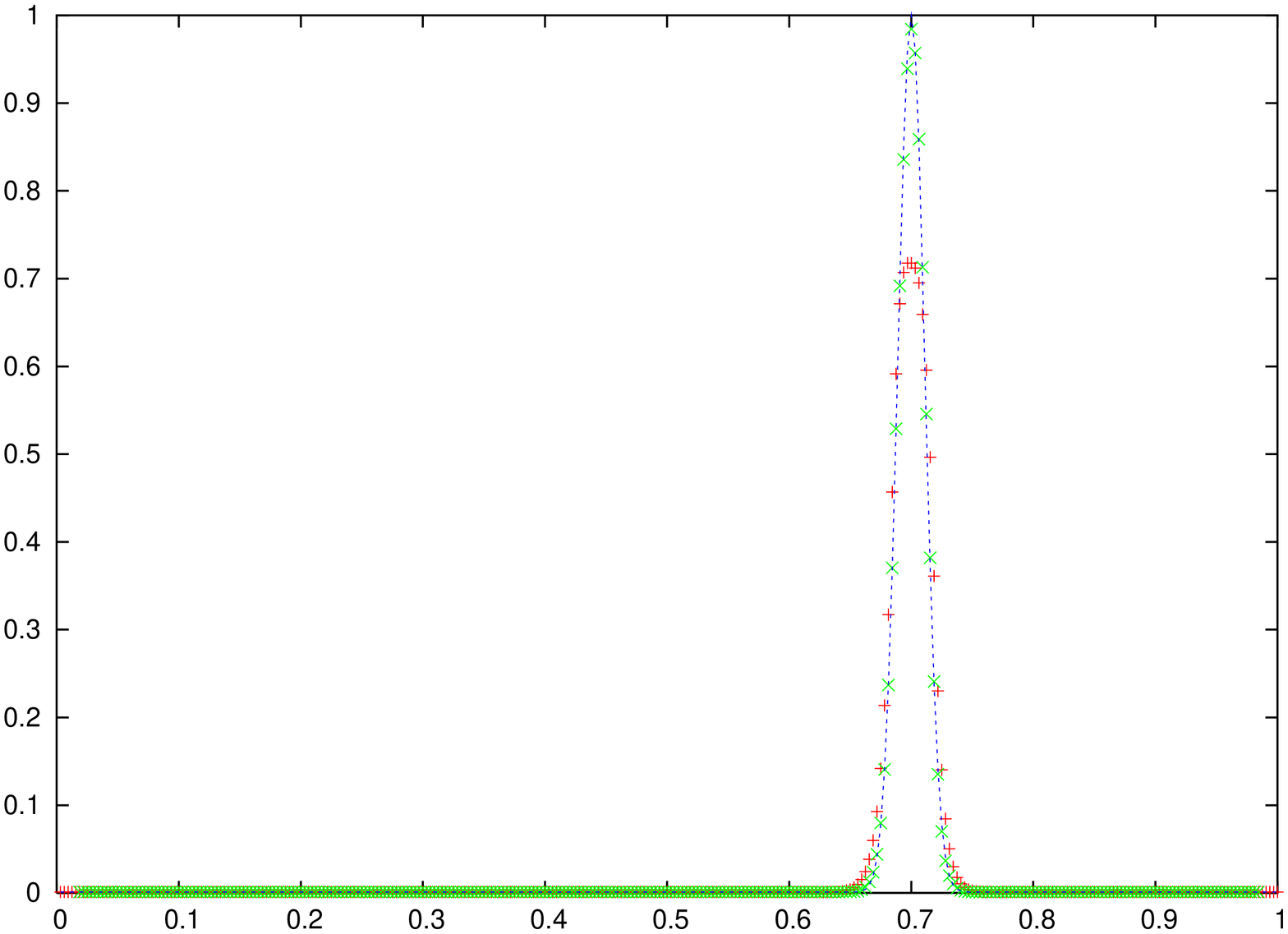}
  \includegraphics[angle=-90,width=0.8\textwidth]{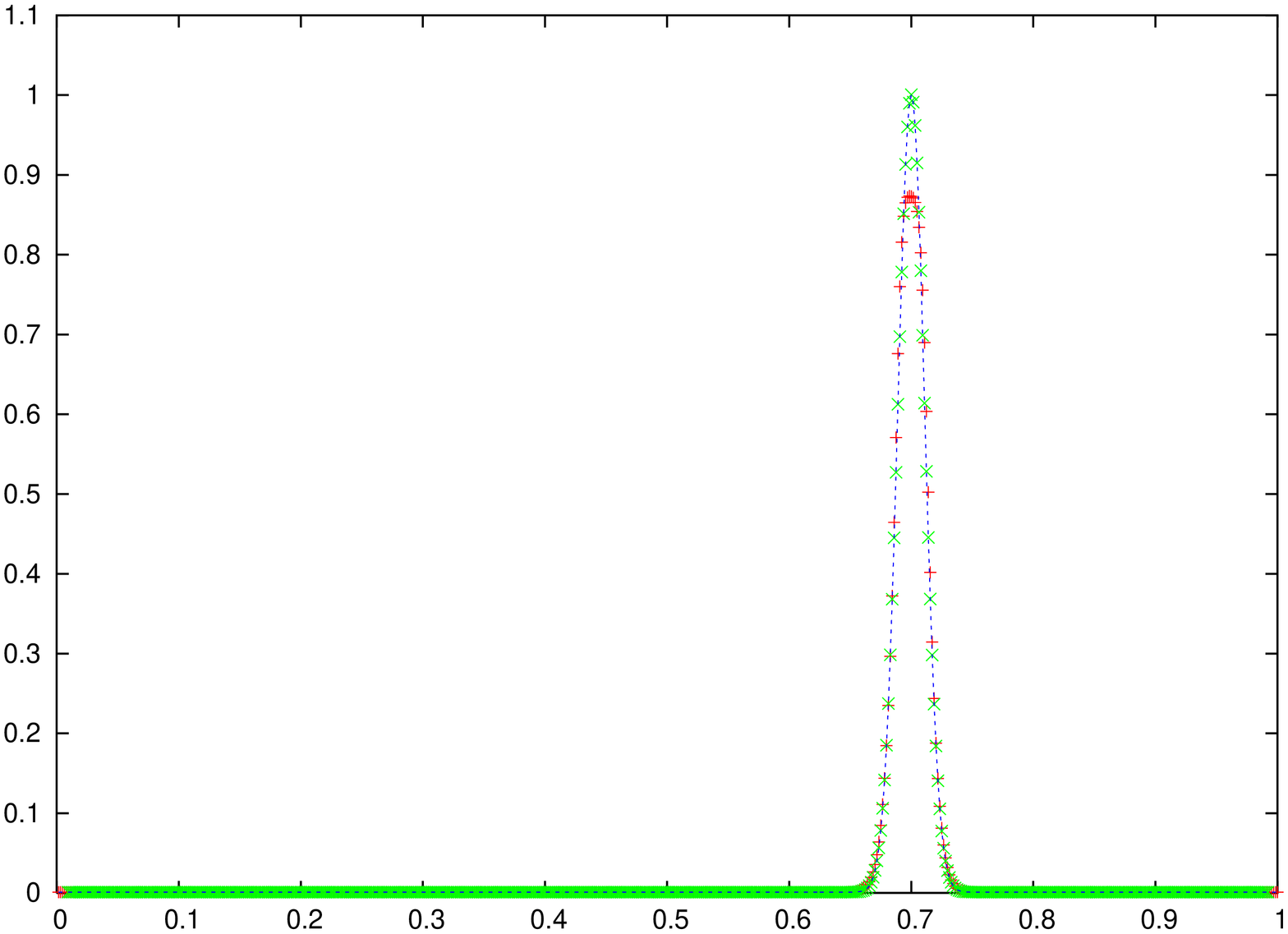}
\label{adv4}
\caption{Same as Fig. 3 for n=320 (top), and n=640(bottom)}
\end{figure}

\begin{figure}[h]
\centering
  \includegraphics[angle=0,width=0.8\textwidth]{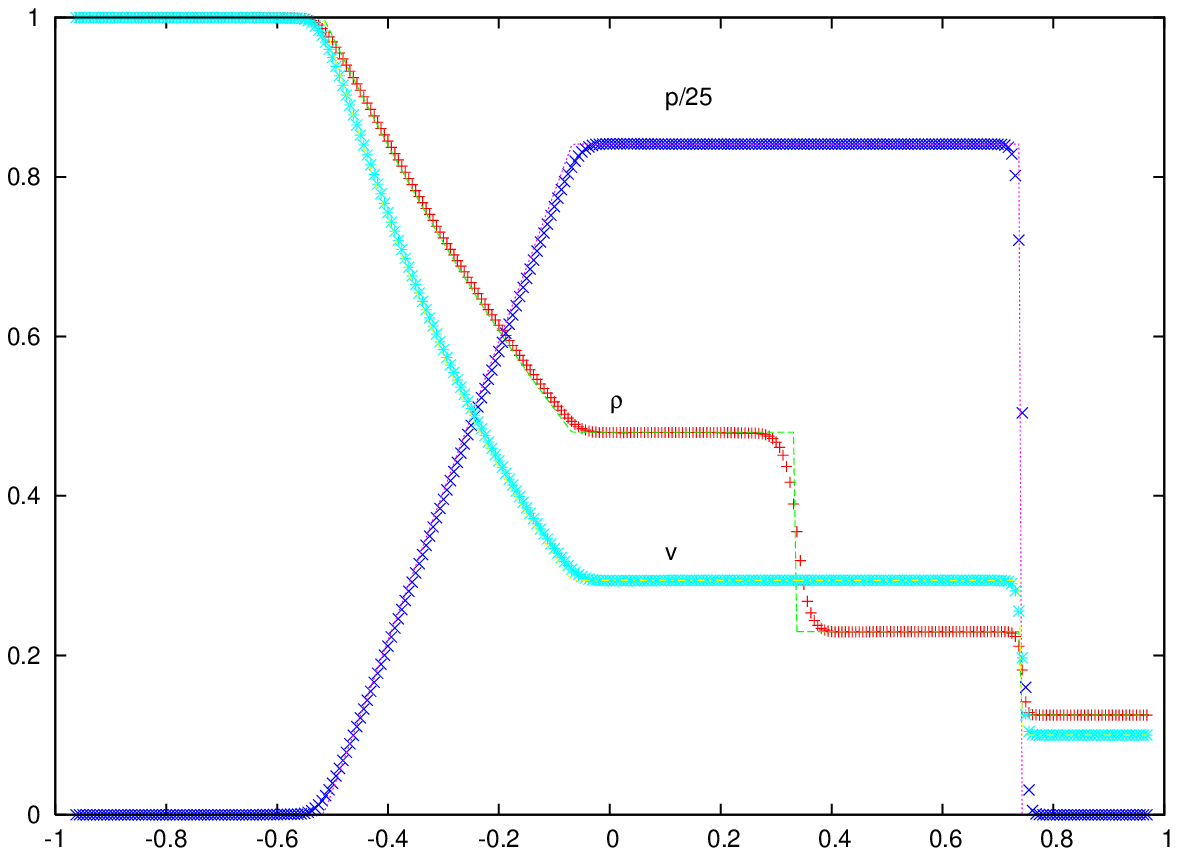}
  \includegraphics[angle=0,width=0.8\textwidth]{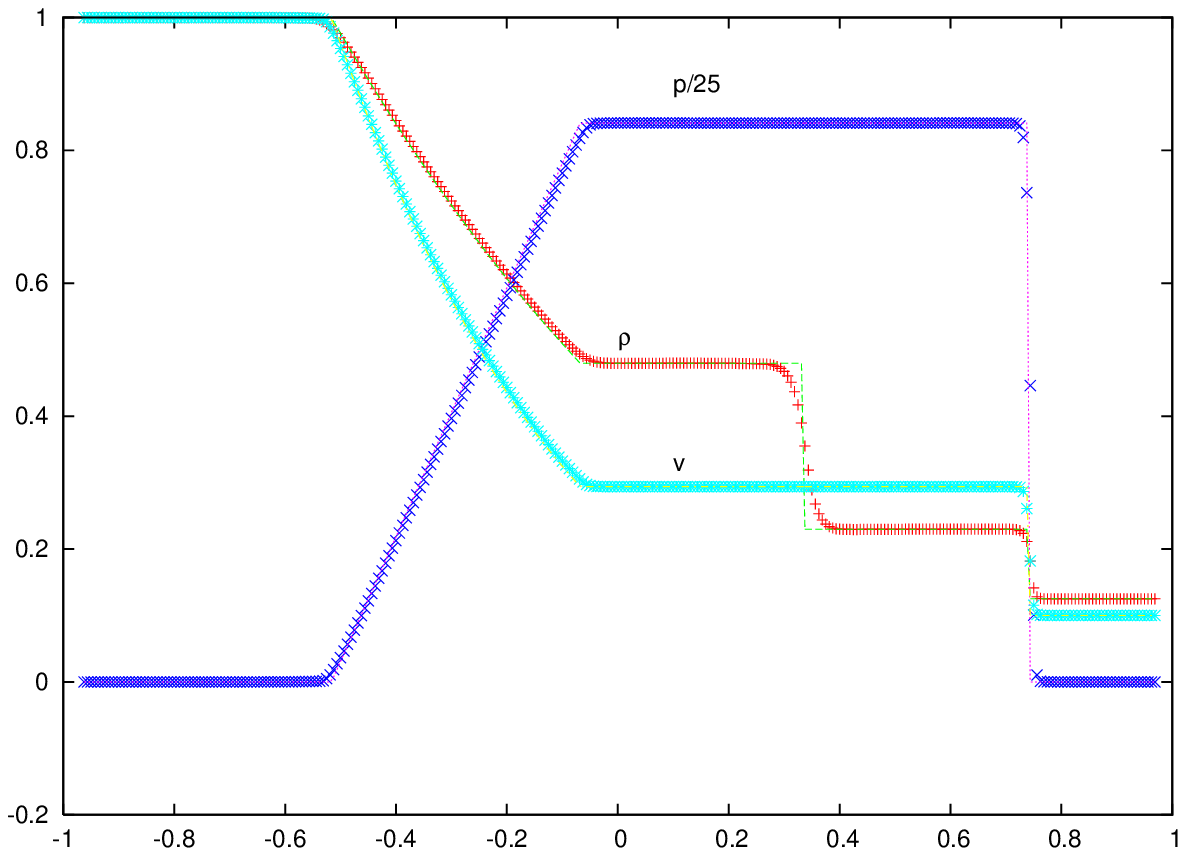} 
\label{Sod}
\caption{Sod shock Capturing test of the scheme (Test 1, Sec. 5.1). 
Solutions shown at t=0.4, n=320. 3$^{rd.}$ order 
WENO reconstruction (top).  4$^{th.}$ order 
WENO reconstruction (bottom).}  
\end{figure}

\begin{figure}[h]
\centering
  \includegraphics[angle=0,width=0.8\textwidth]{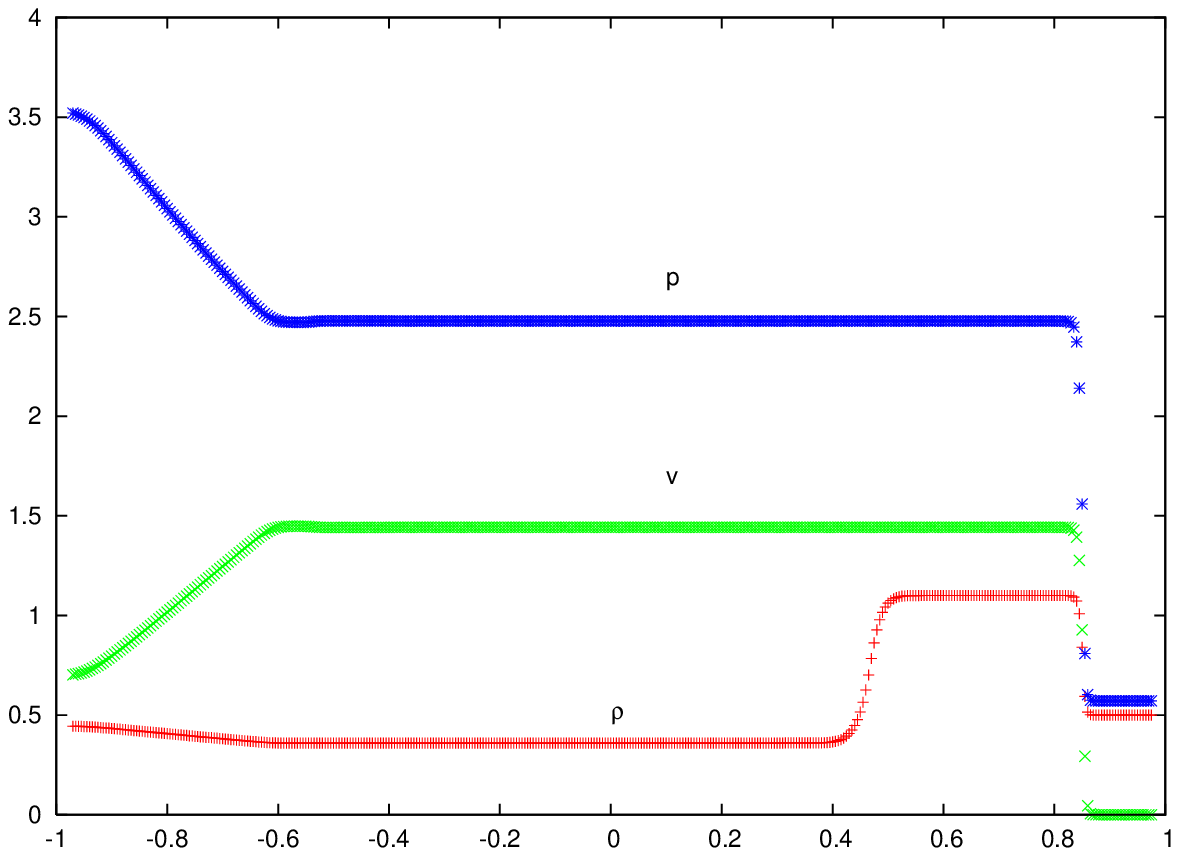}
  \includegraphics[angle=0,width=0.8\textwidth]{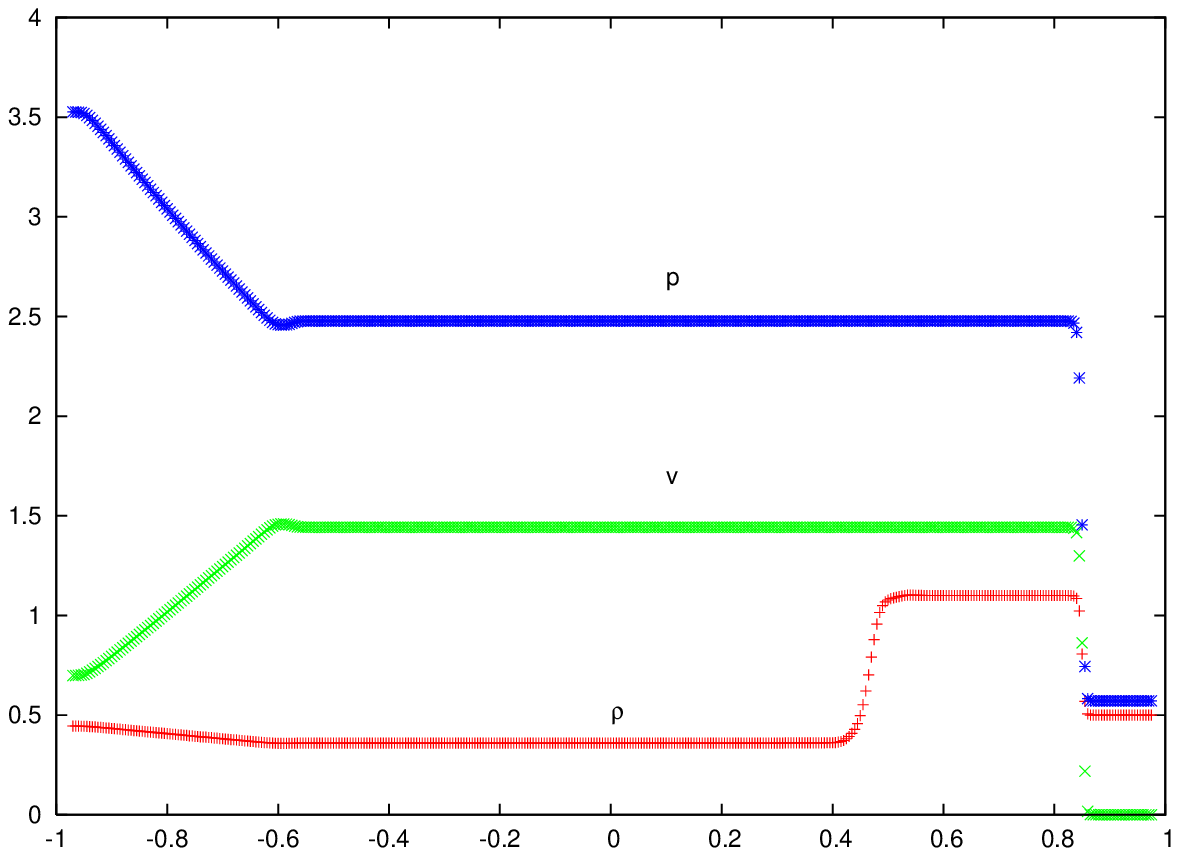}
\label{Lax}
\caption{Lax's shock Capturing test (test 2 of Sec. 5.1). 
Solutions shown at t=0.4, n=320.   3$^{rd.}$ order 
WENO reconstruction (top).  4$^{th.}$ order 
WENO reconstruction (bottom).}  
\end{figure}

\newpage

\begin{figure}[h]
\centering
  \includegraphics[angle=-90,width=0.9\textwidth]{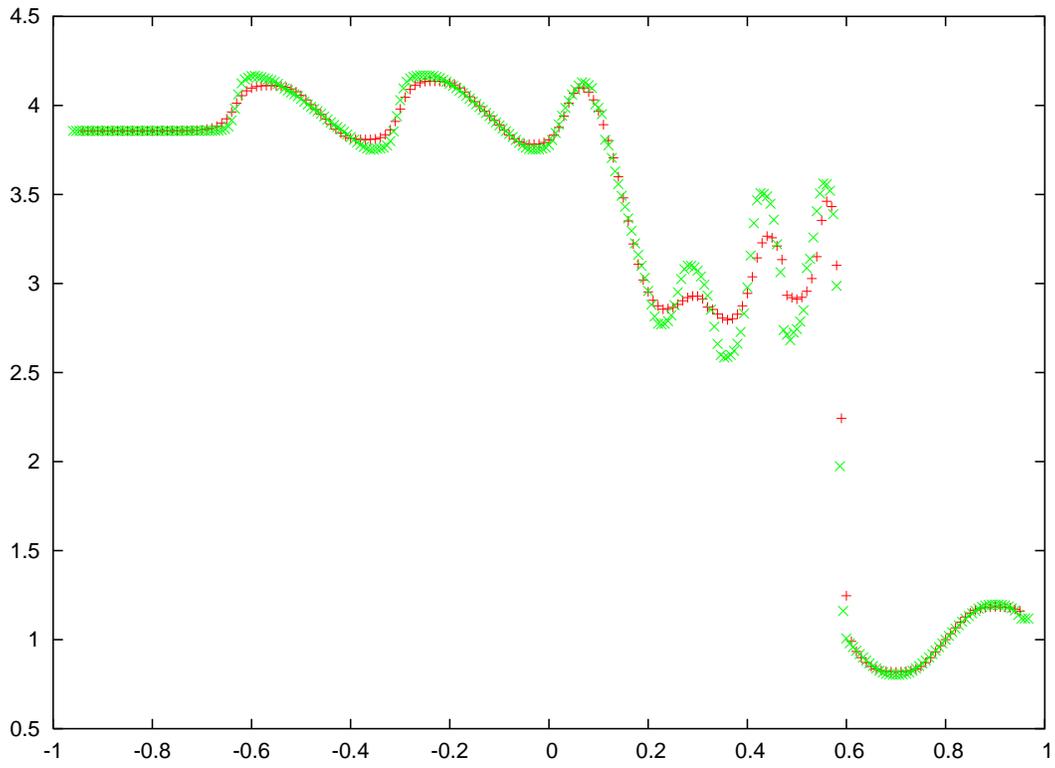}
\label{Shu}
\caption{Shu's shock Capturing test (test 3, Sec. 5.1). 
Solutions shown at t=0.4. 3$^{rd.}$ order 
WENO reconstruction (red).  4$^{th.}$ order 
WENO reconstruction (green).} 
\end{figure}

\newpage

\begin{figure}[h]
\centering
  \includegraphics[angle=0,width=0.6\textwidth]{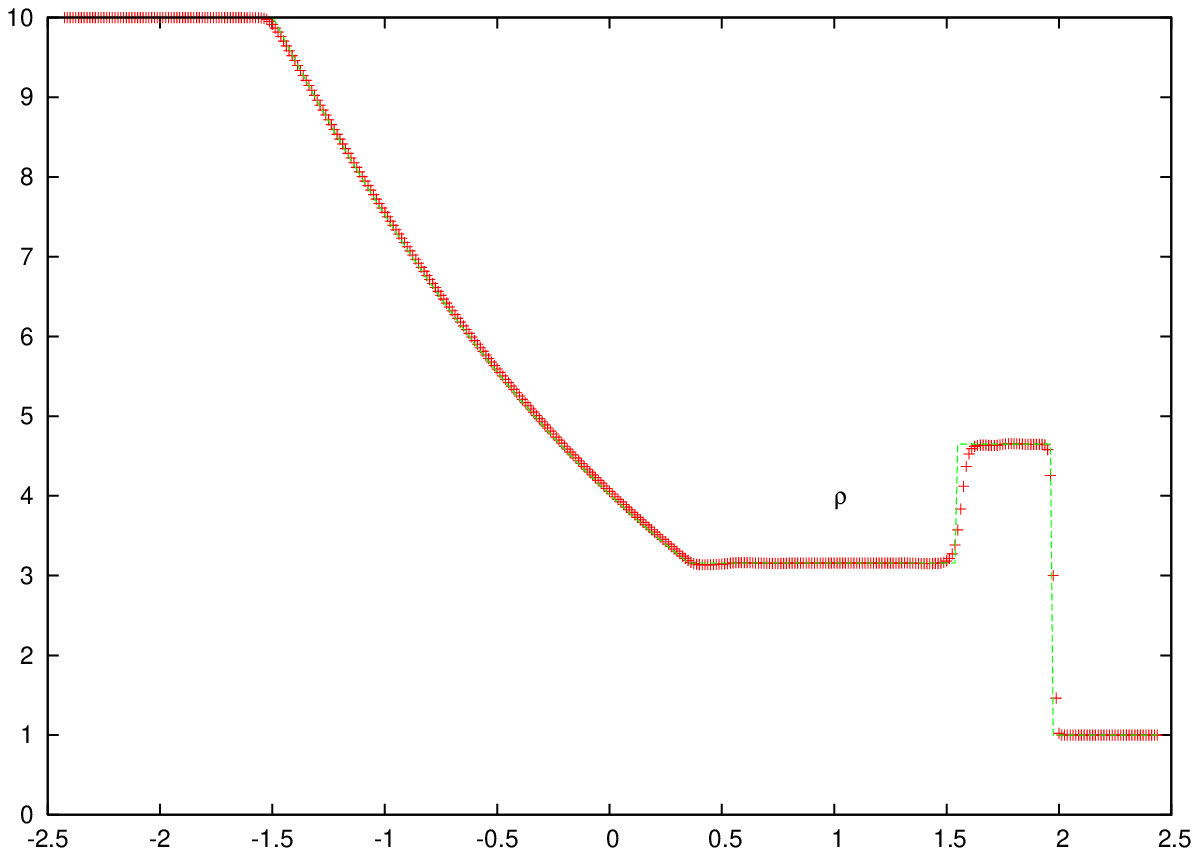}
  \includegraphics[angle=0,width=0.6\textwidth]{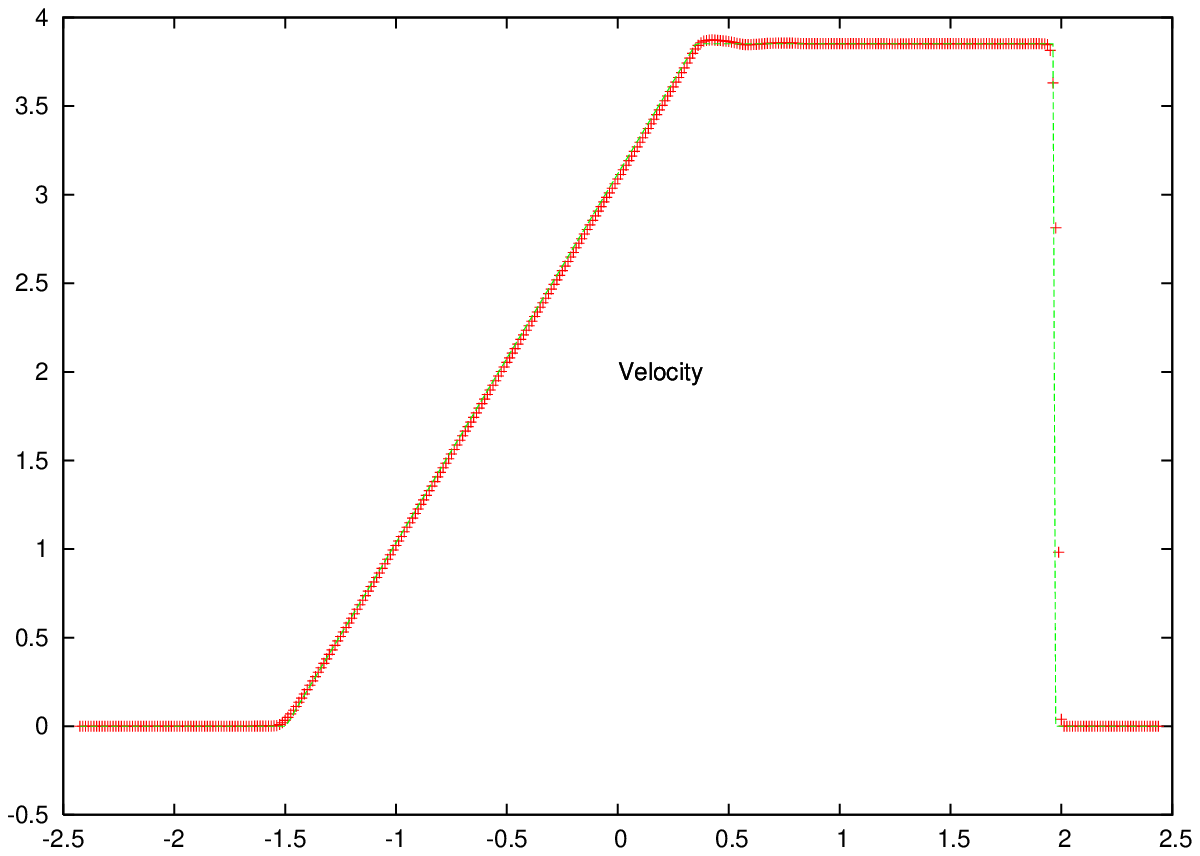}
 \includegraphics[angle=0,width=0.6\textwidth]{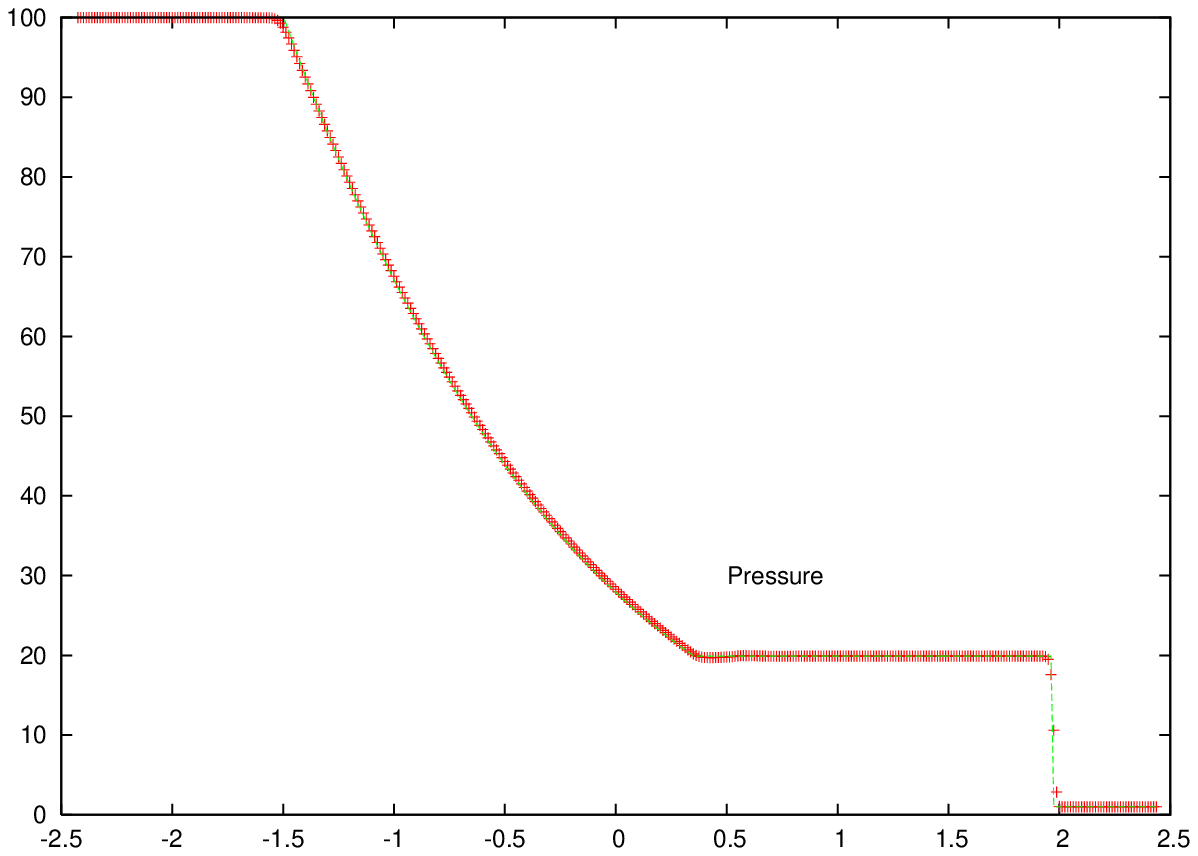}
\label{strongsod}
\caption{Strong Sod shock Capturing test (test 4, Sec. 5.1). 
Solutions shown at t=0.4 using 4th. order WENO.} 
\end{figure}

\clearpage

\begin{figure}[h!]
\centering
 \includegraphics[angle=0,width=0.5\textwidth]{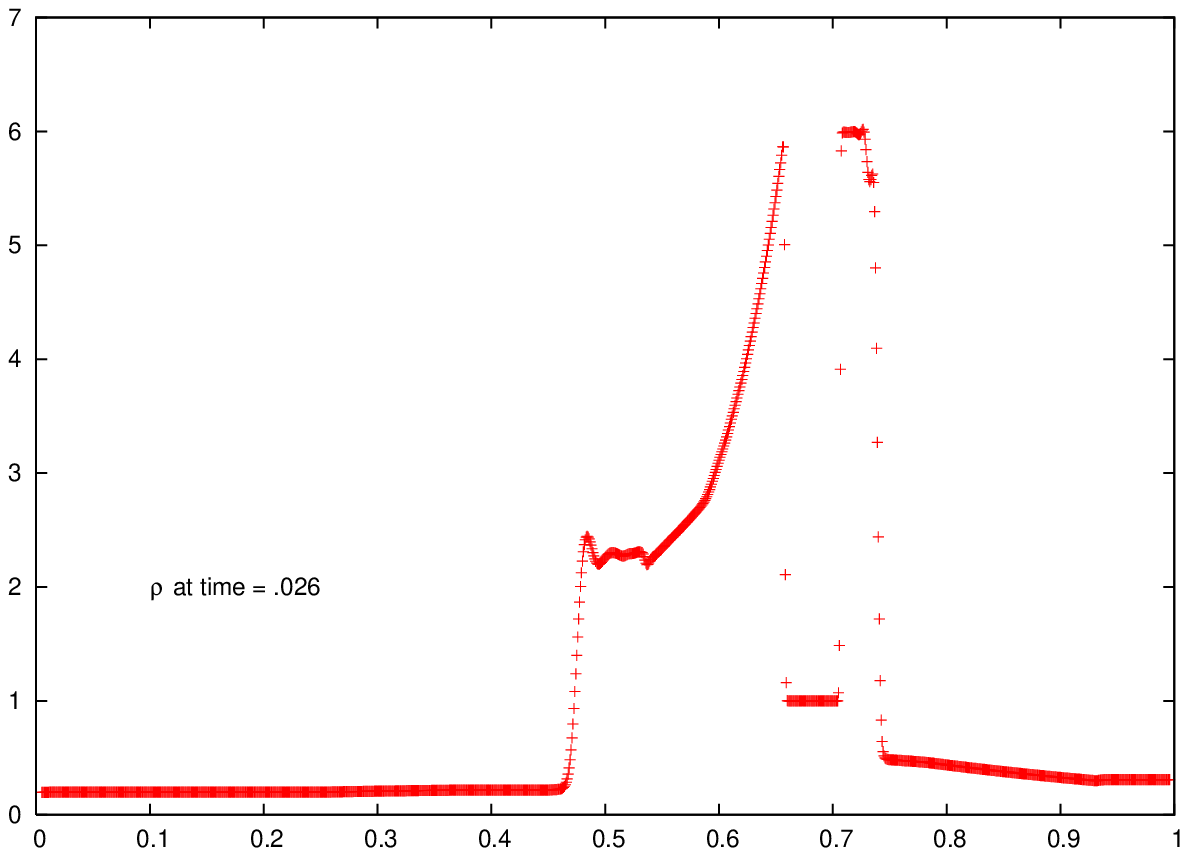}
 \includegraphics[angle=0,width=0.5\textwidth]{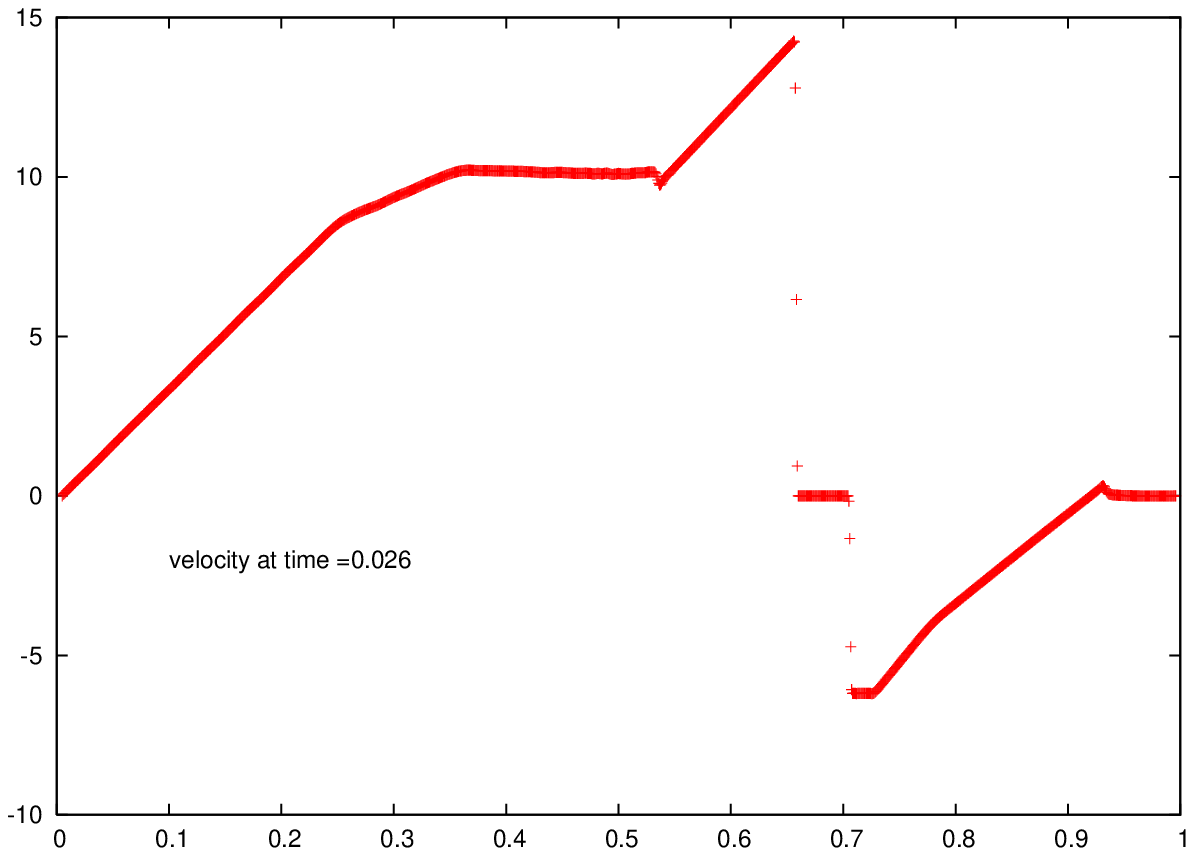}
 \includegraphics[angle=0,width=0.5\textwidth]{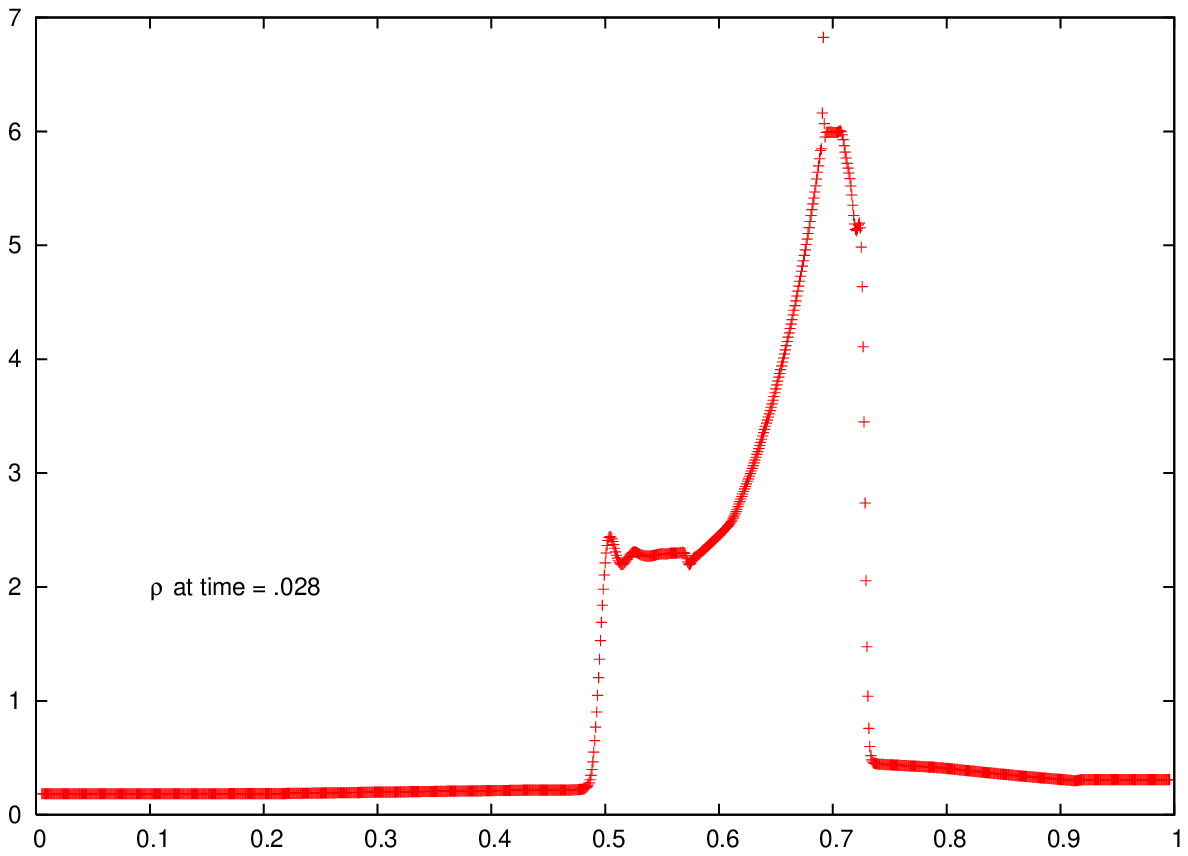}
 \includegraphics[angle=0,width=0.5\textwidth]{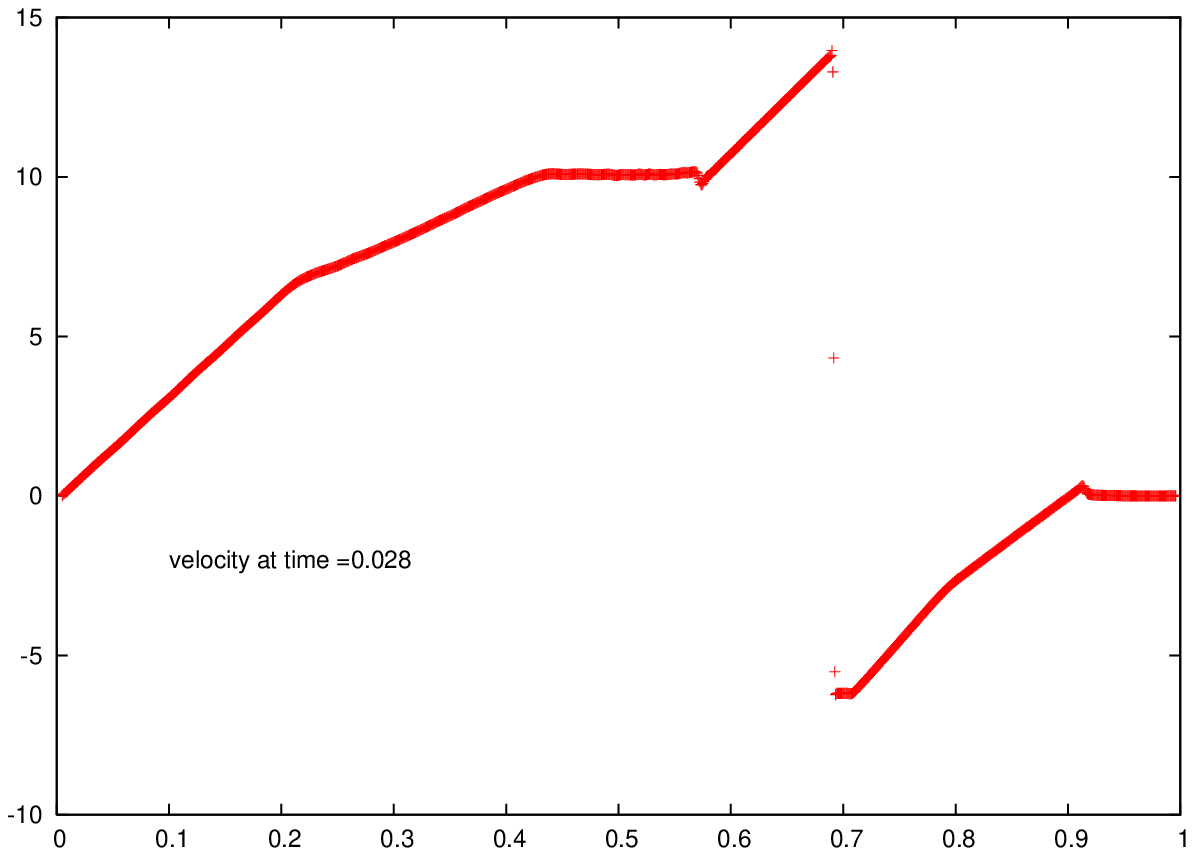}
\caption{Interaction between two blast waves (test 5 Sec. 5.1): 
Density and velocity are shown at t=0.026 and t= 0.028, for n=1200.} 
\end{figure}

\begin{figure}[h]
\centering
 \includegraphics[angle=0,width=0.5\textwidth]{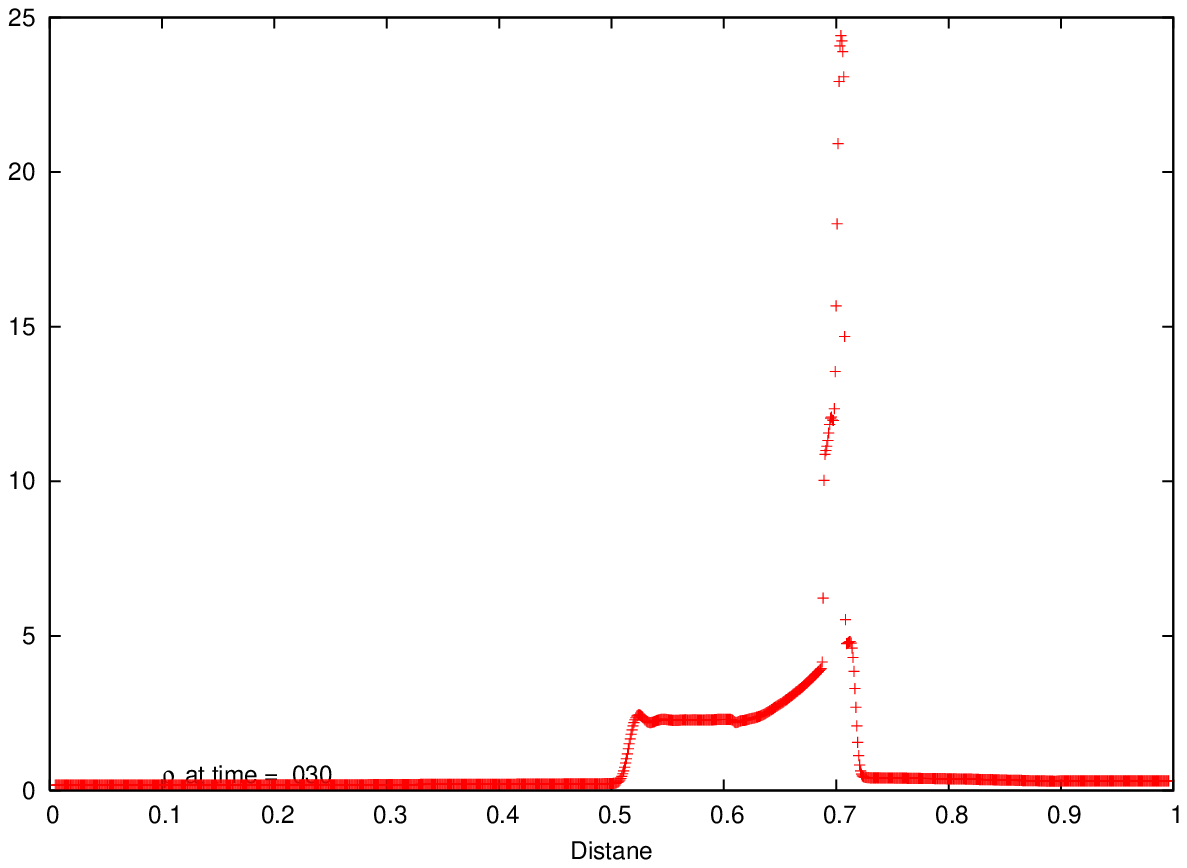}
 \includegraphics[angle=0,width=0.5\textwidth]{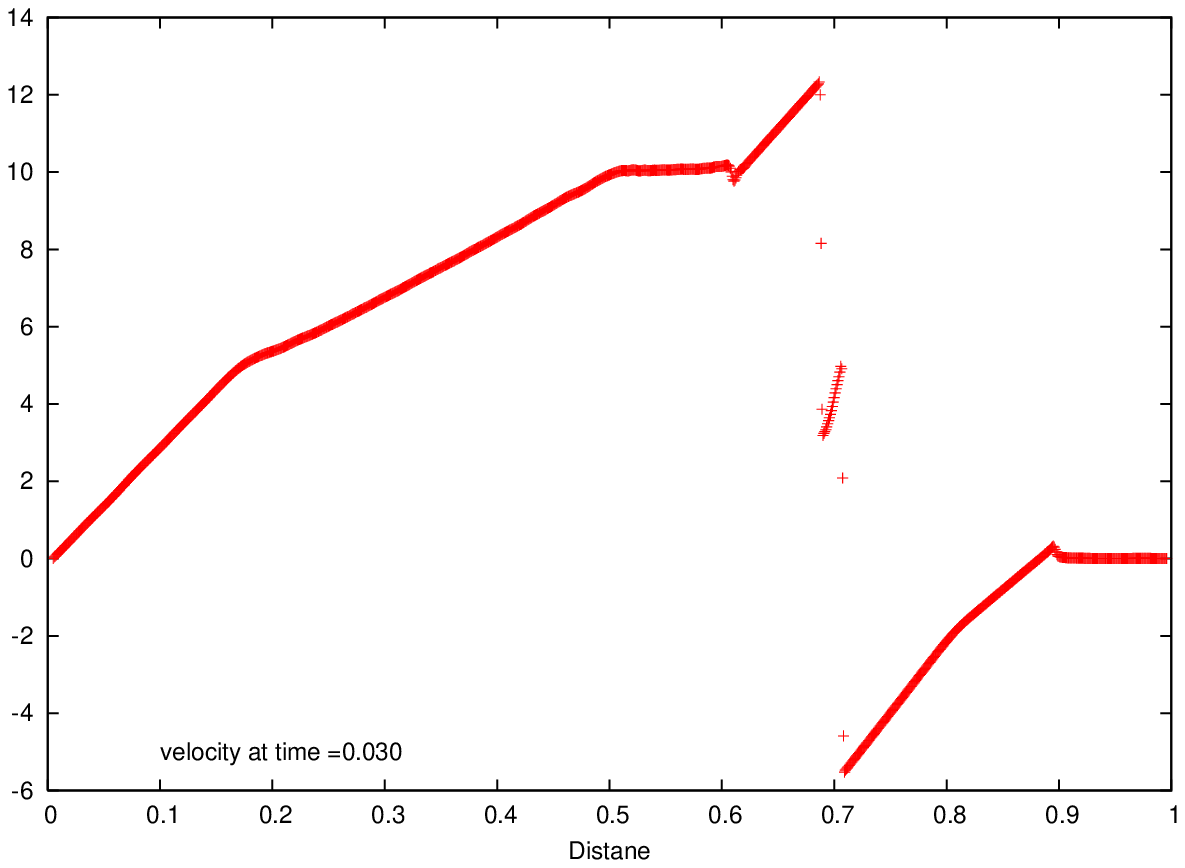}
 \includegraphics[angle=0,width=0.5\textwidth]{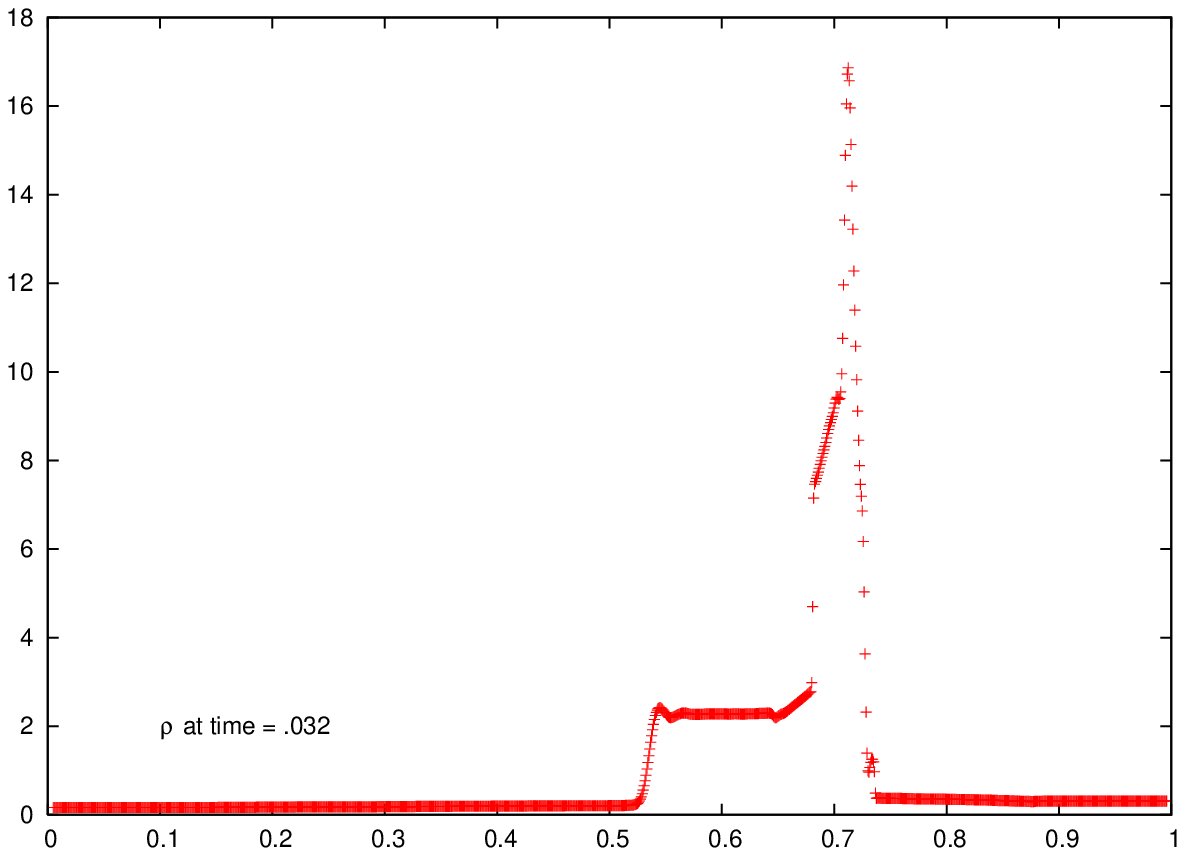}
 \includegraphics[angle=0,width=0.5\textwidth]{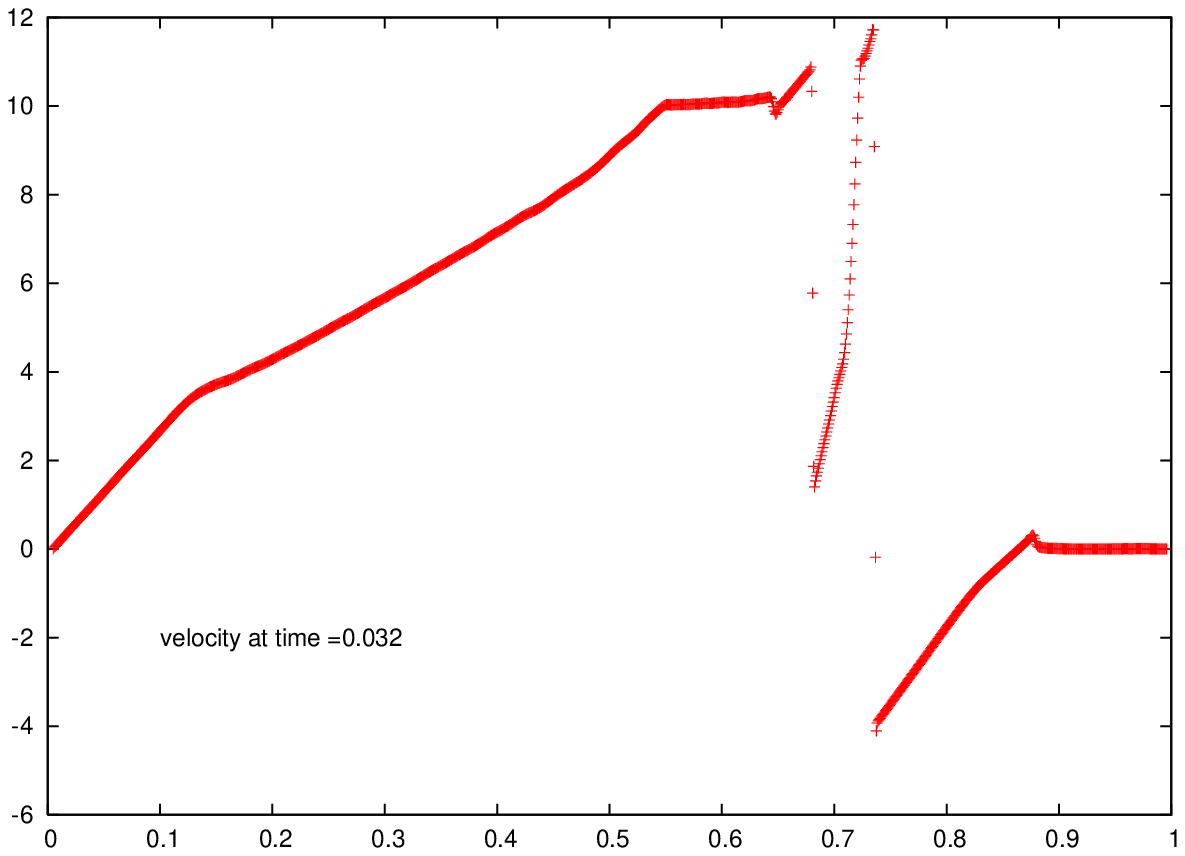}
\label{inter2}
\caption{Interaction between two blast waves (test 5 Sec. 5.1): 
Density and velocity are shown at t=0.030 and t= 0.032, for n=1200.} 
\end{figure}

\begin{figure}[h]
\centering
 \includegraphics[angle=0,width=0.5\textwidth]{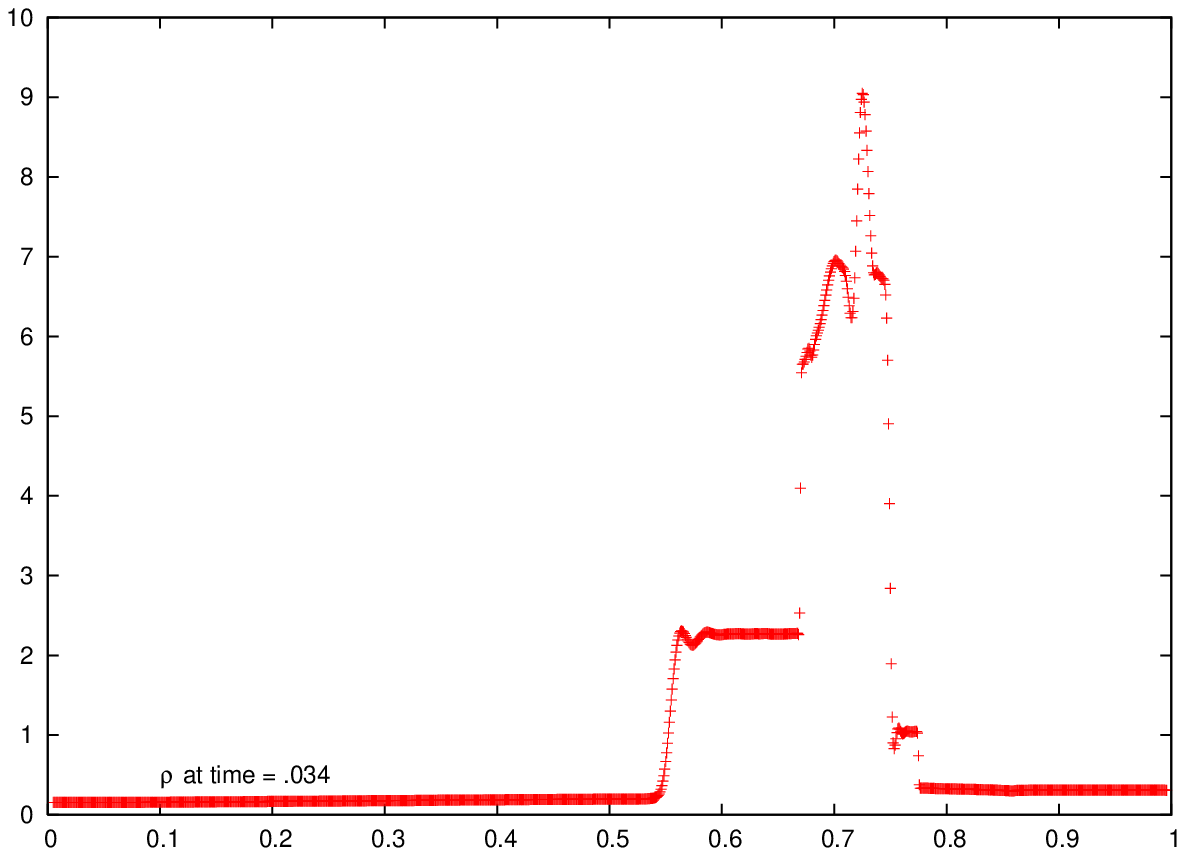}
 \includegraphics[angle=0,width=0.5\textwidth]{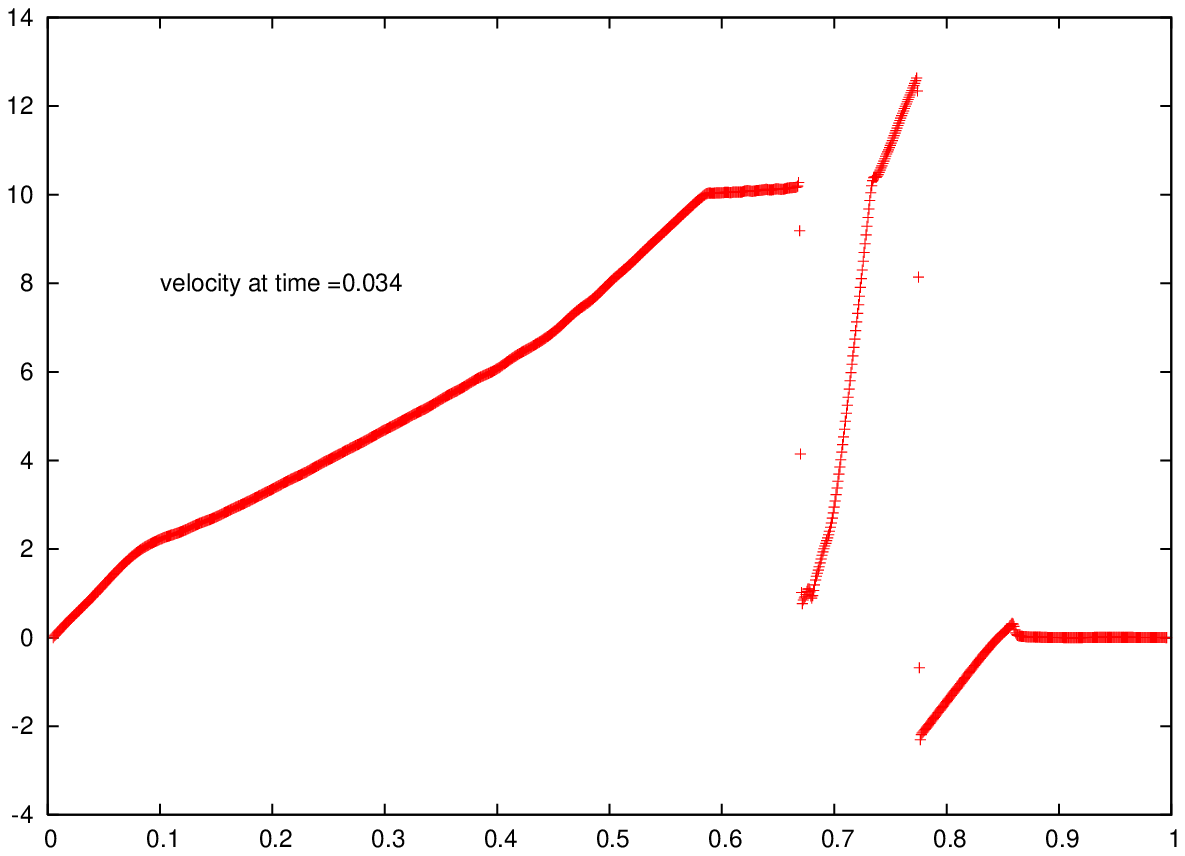}
 \includegraphics[angle=0,width=0.5\textwidth]{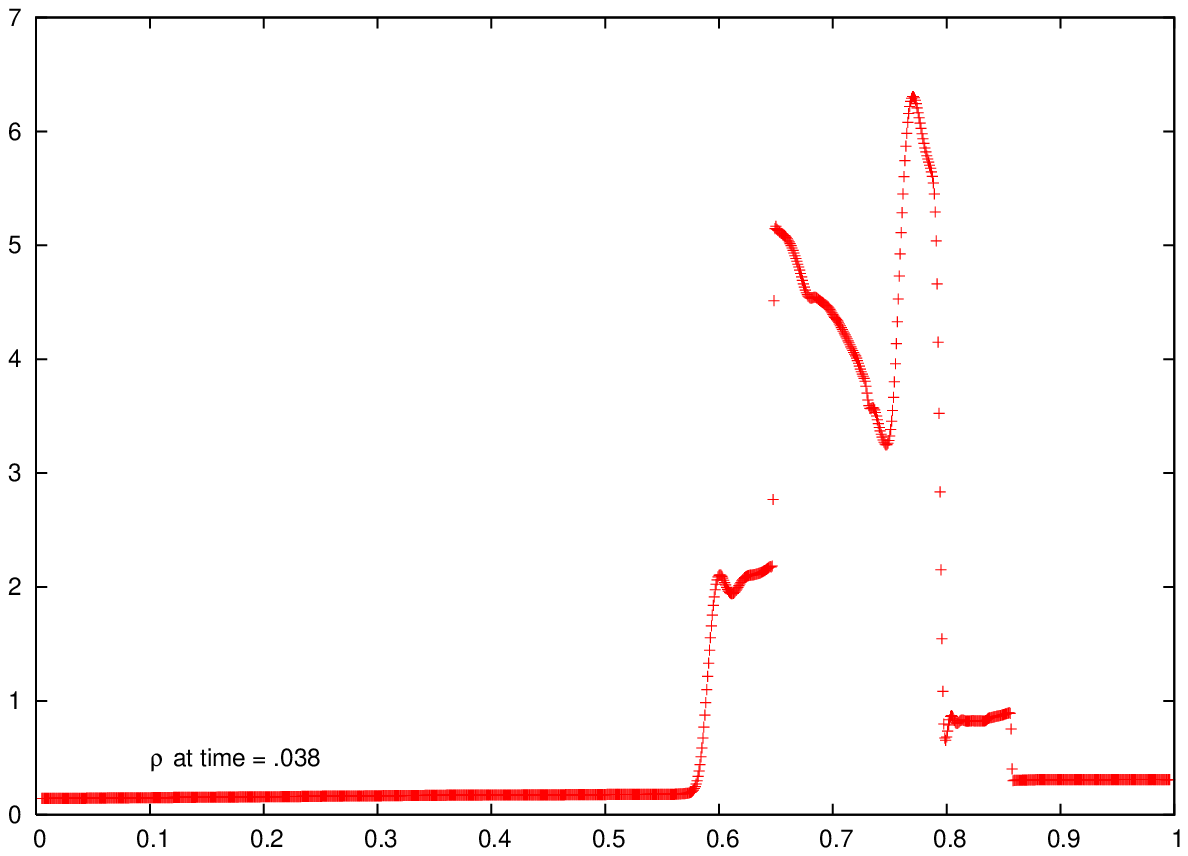}
 \includegraphics[angle=0,width=0.5\textwidth]{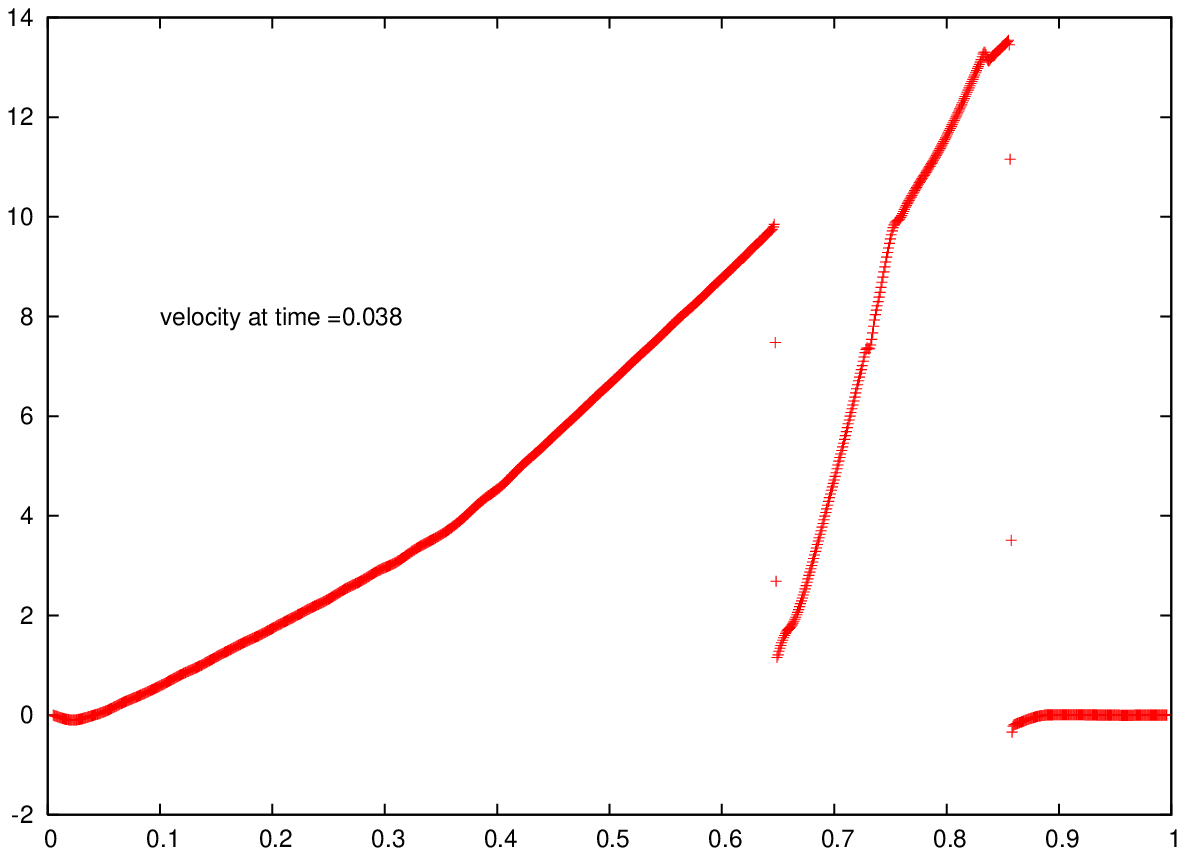}
\label{inter2}
\caption{Interaction between two blast waves (test 5 Sec. 5.1): 
Density and velocity are shown at t=0.034 and t= 0.038, for n=1200.} 
\end{figure}

\clearpage

\begin{figure}[h]
\centering
  \includegraphics[angle=-90,width=0.5\textwidth]{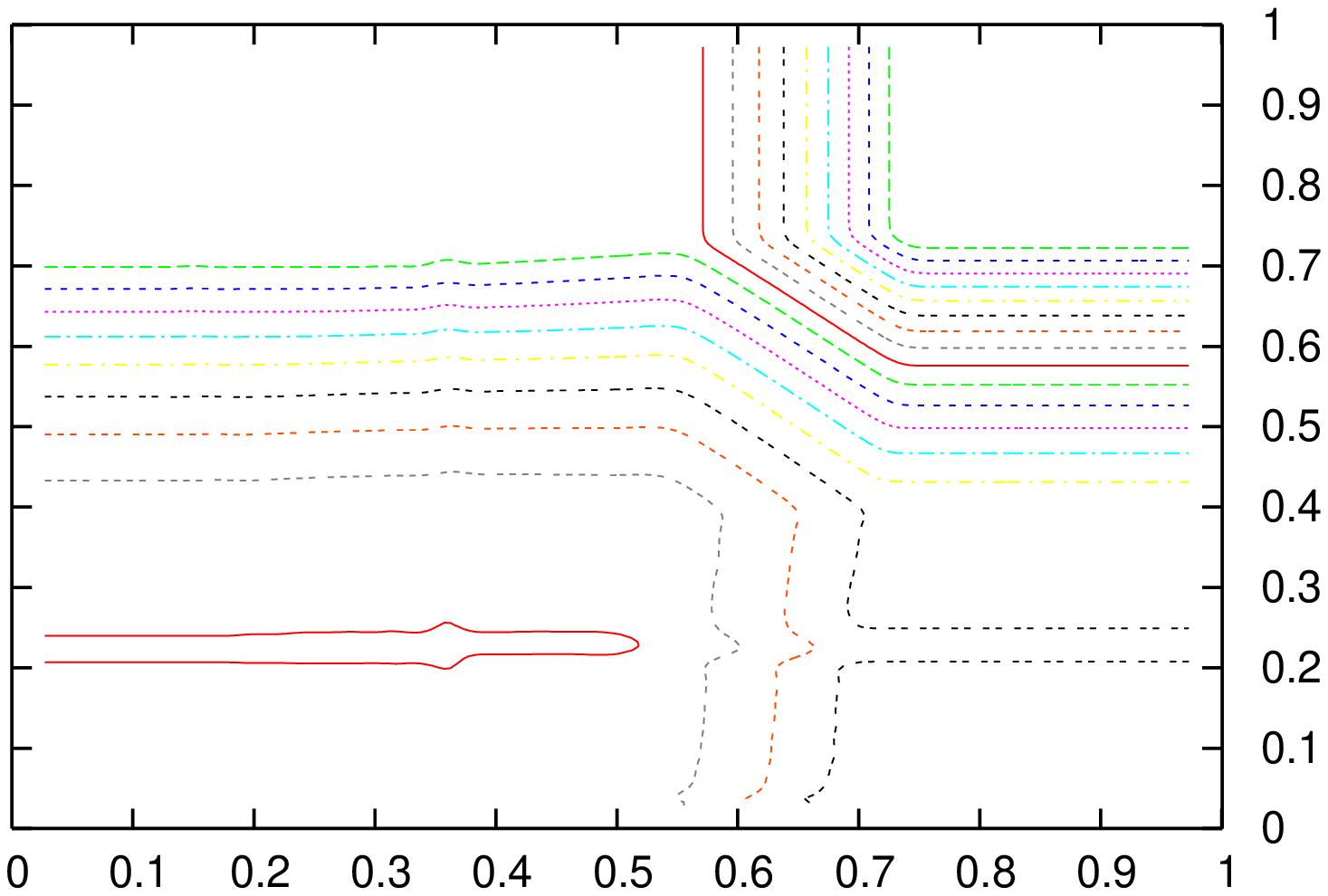}
  \includegraphics[angle=-90,width=0.5\textwidth]{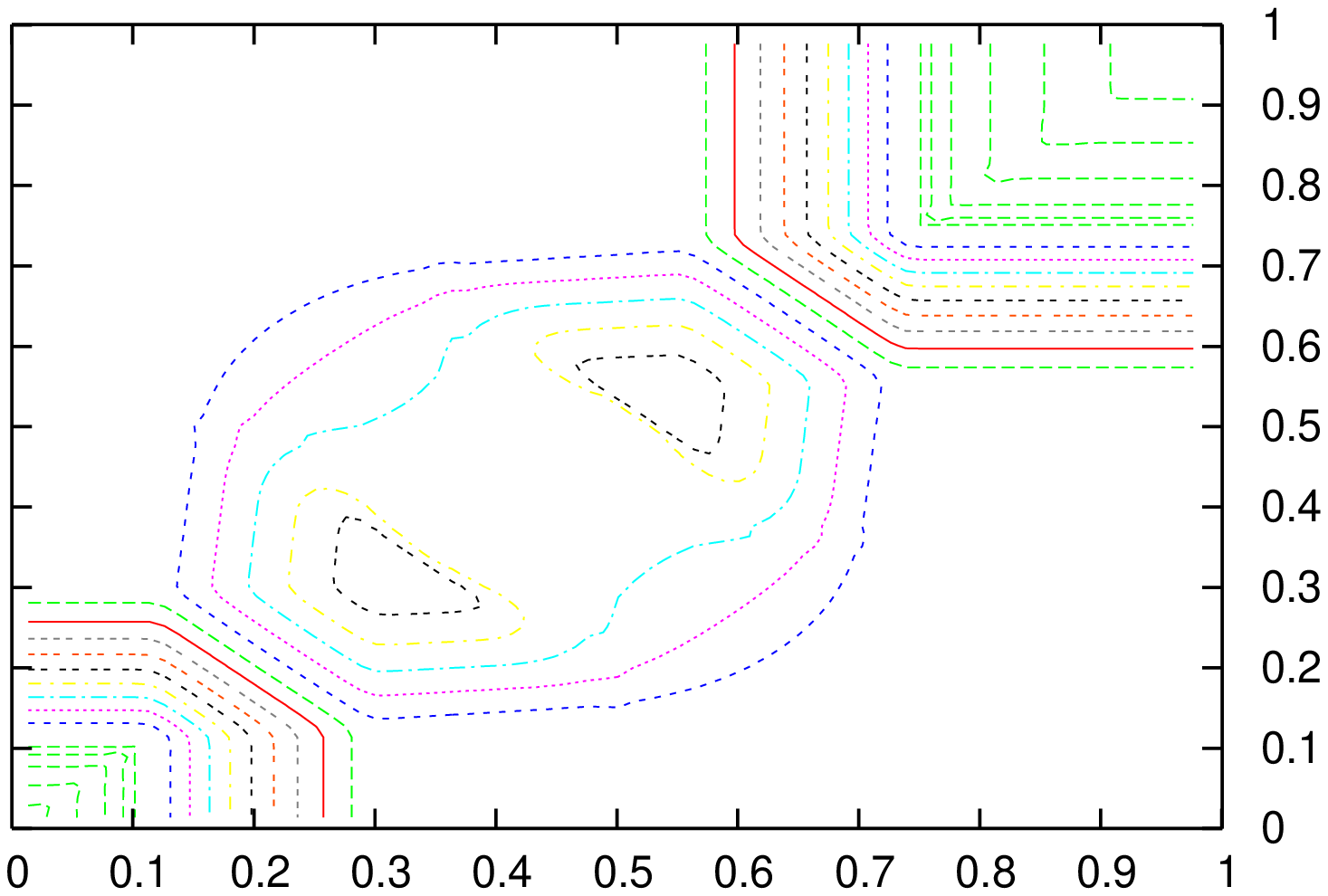}
  \includegraphics[angle=-90,width=0.5\textwidth]{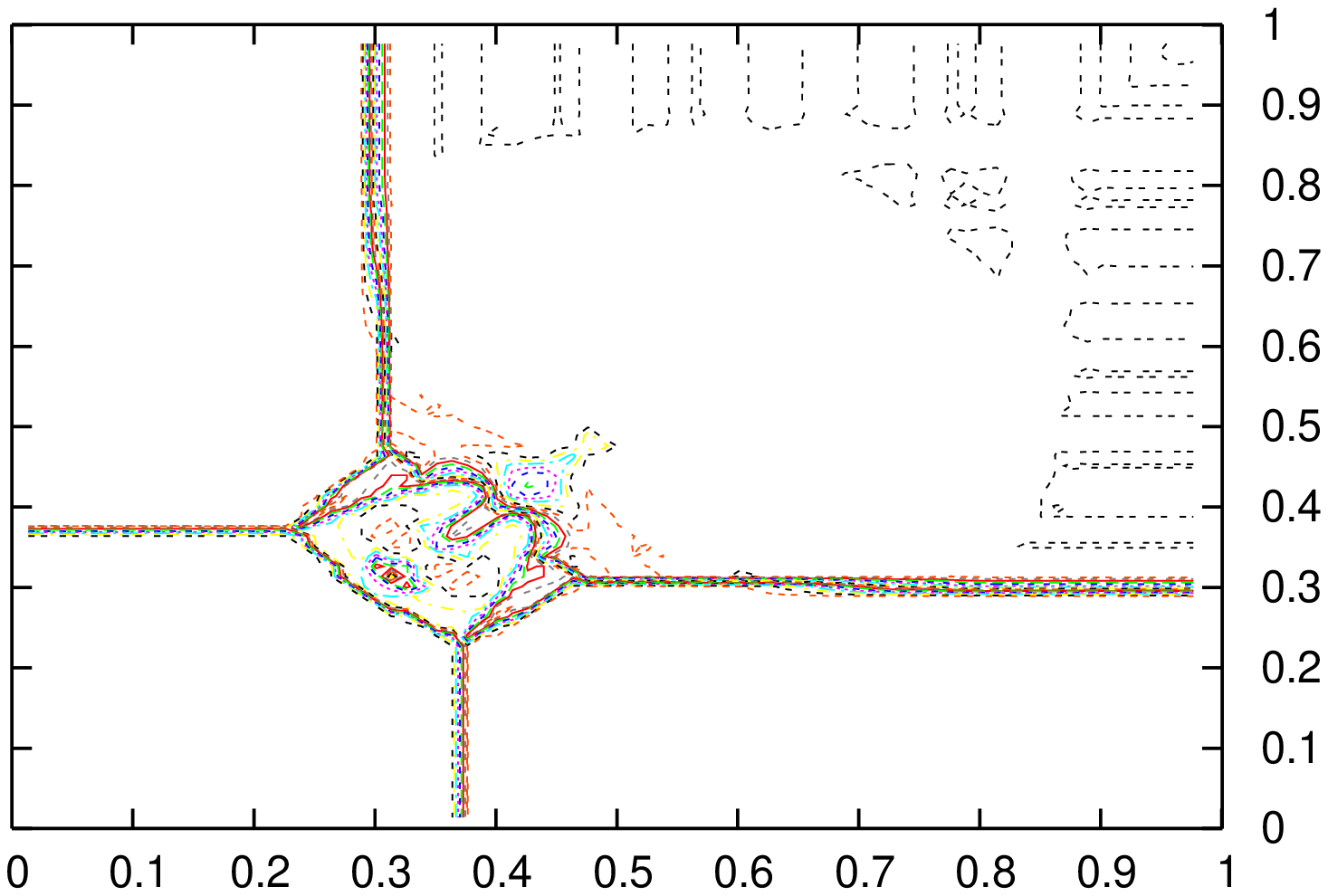}
  \includegraphics[angle=-90,width=0.5\textwidth]{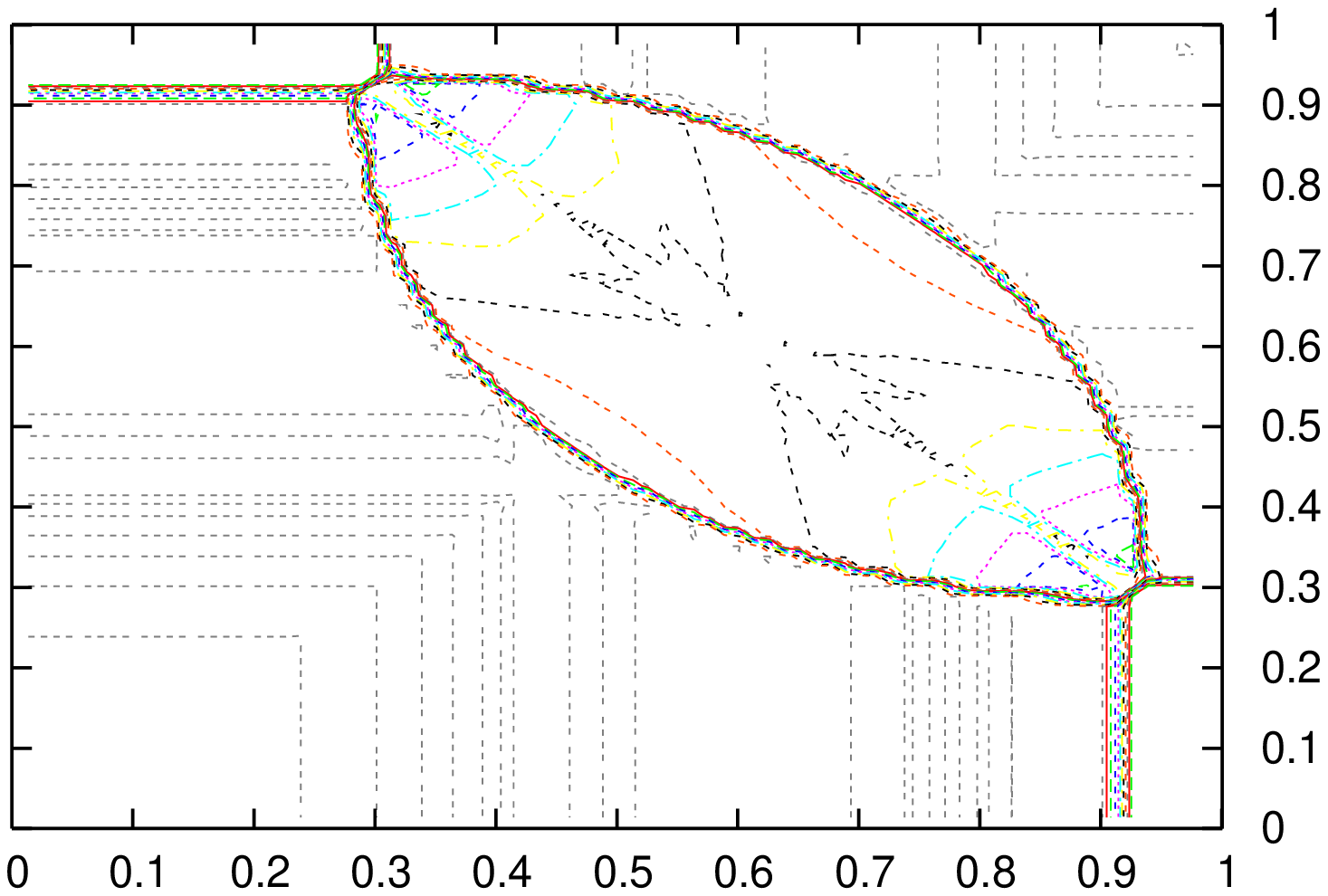}
\caption{2D Riemann problems (Sec. 5.2), 
configurations 1-4 (in ascending order from top)}
\label{2D1}
\end{figure}

\clearpage

\begin{figure}[h]
\centering
  \includegraphics[angle=-90,width=0.5\textwidth]{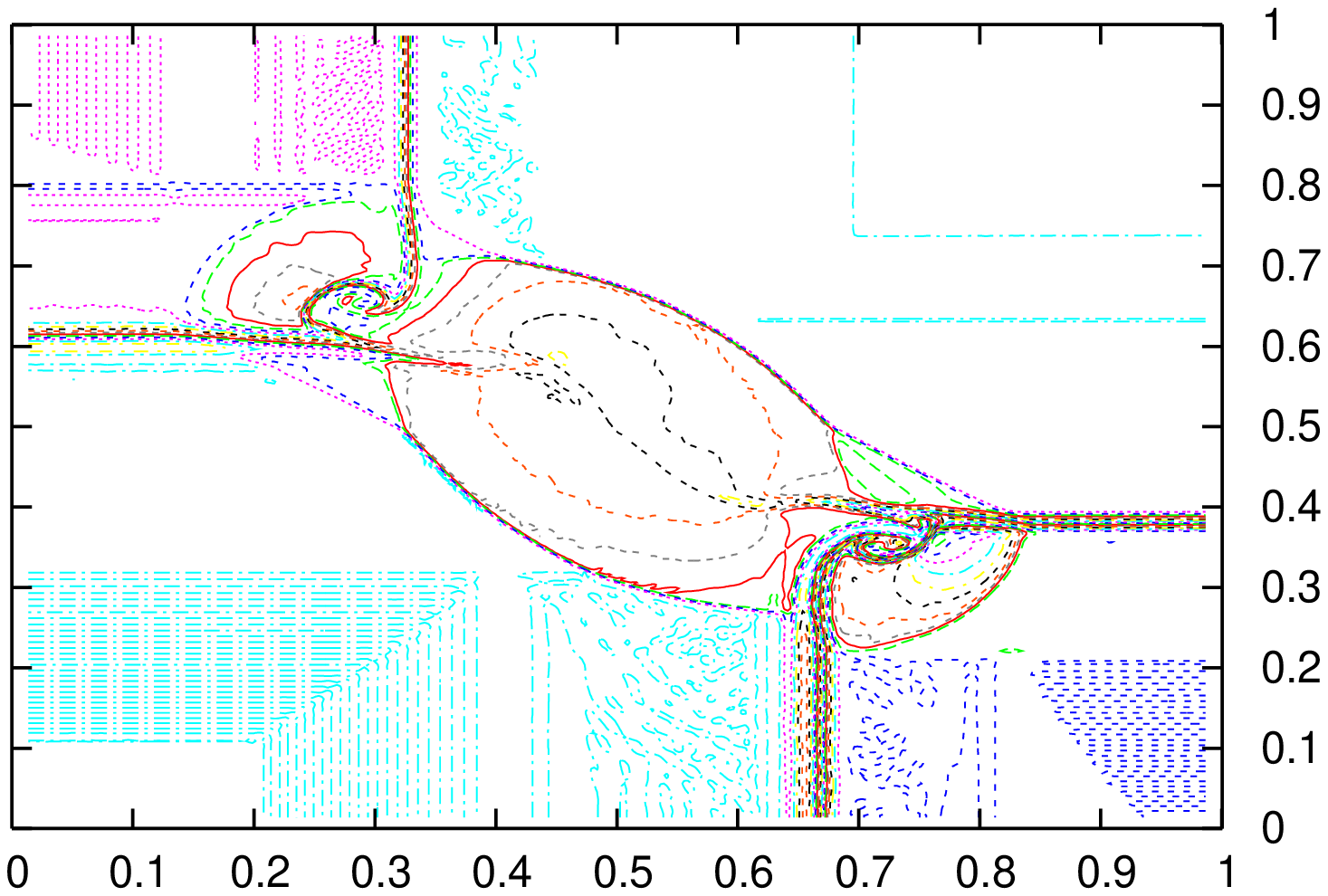}
  \includegraphics[angle=-90,width=0.5\textwidth]{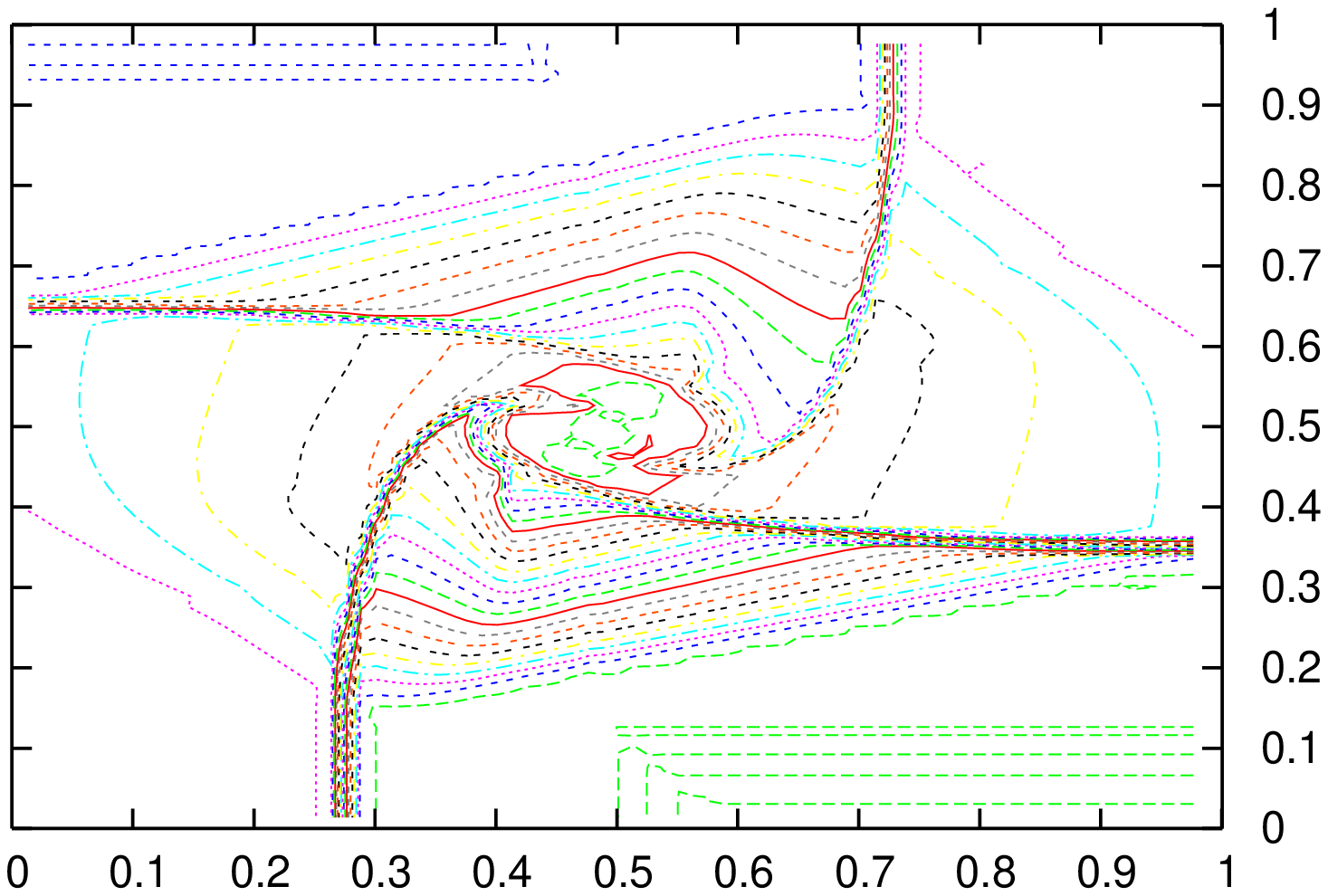}
  \includegraphics[angle=-90,width=0.5\textwidth]{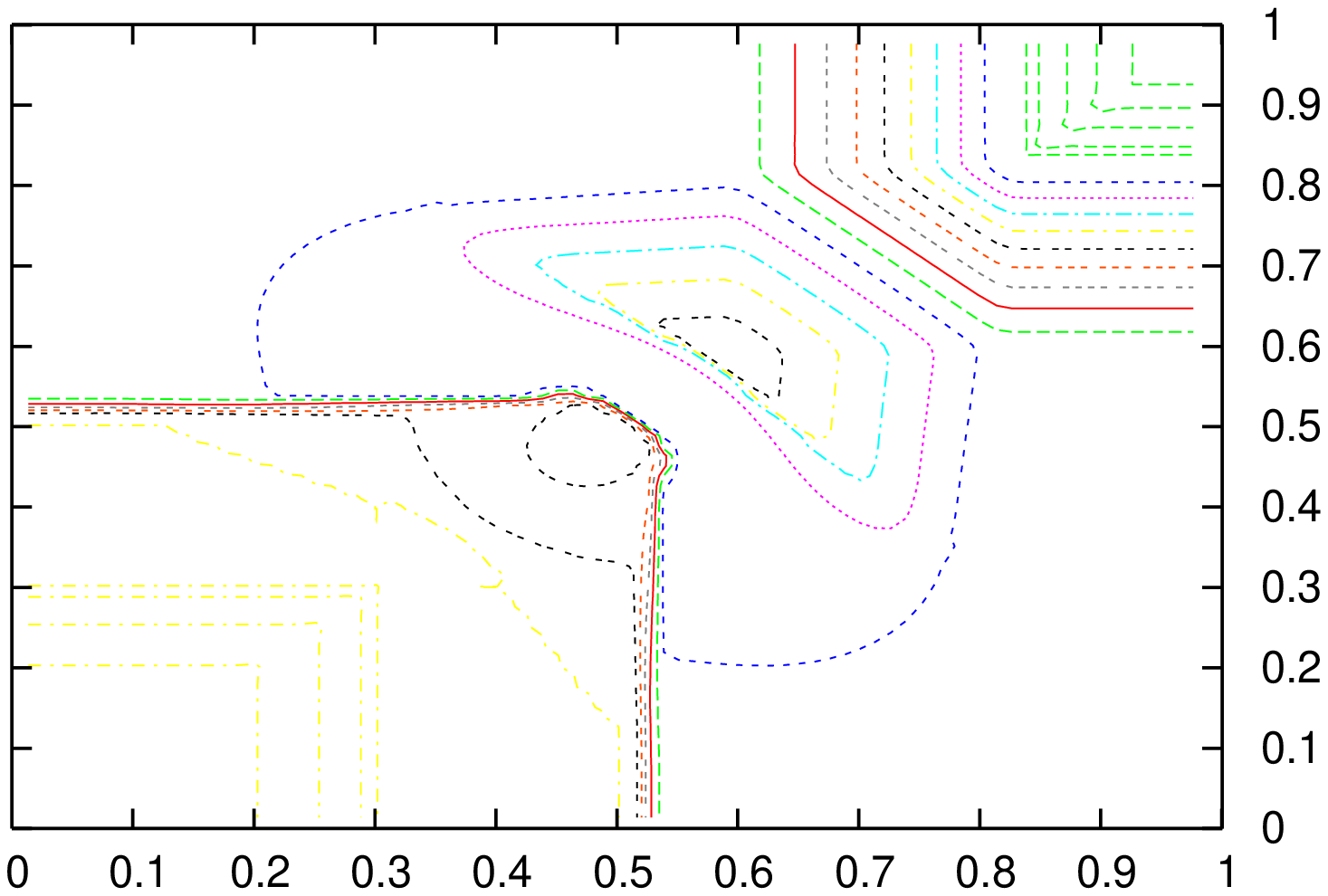}
  \includegraphics[angle=-90,width=0.5\textwidth]{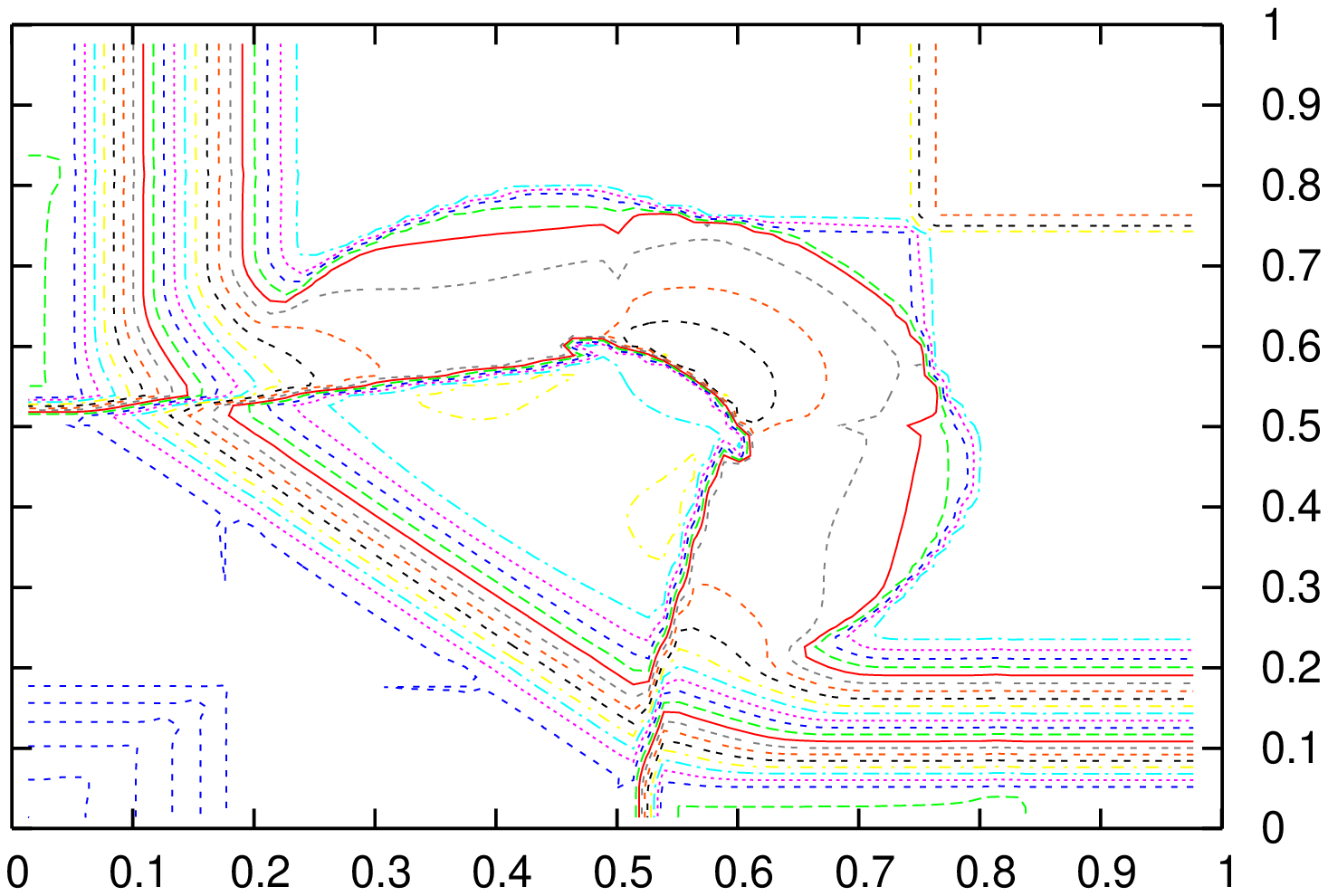}
\caption{2D Riemann problems (Sec. 5.2). Density contours for 
configurations 5-8 (in ascending order from top)}
\label{2D2}
\end{figure}

\begin{figure}[h]
\centering
  \includegraphics[angle=-90,width=0.5\textwidth]{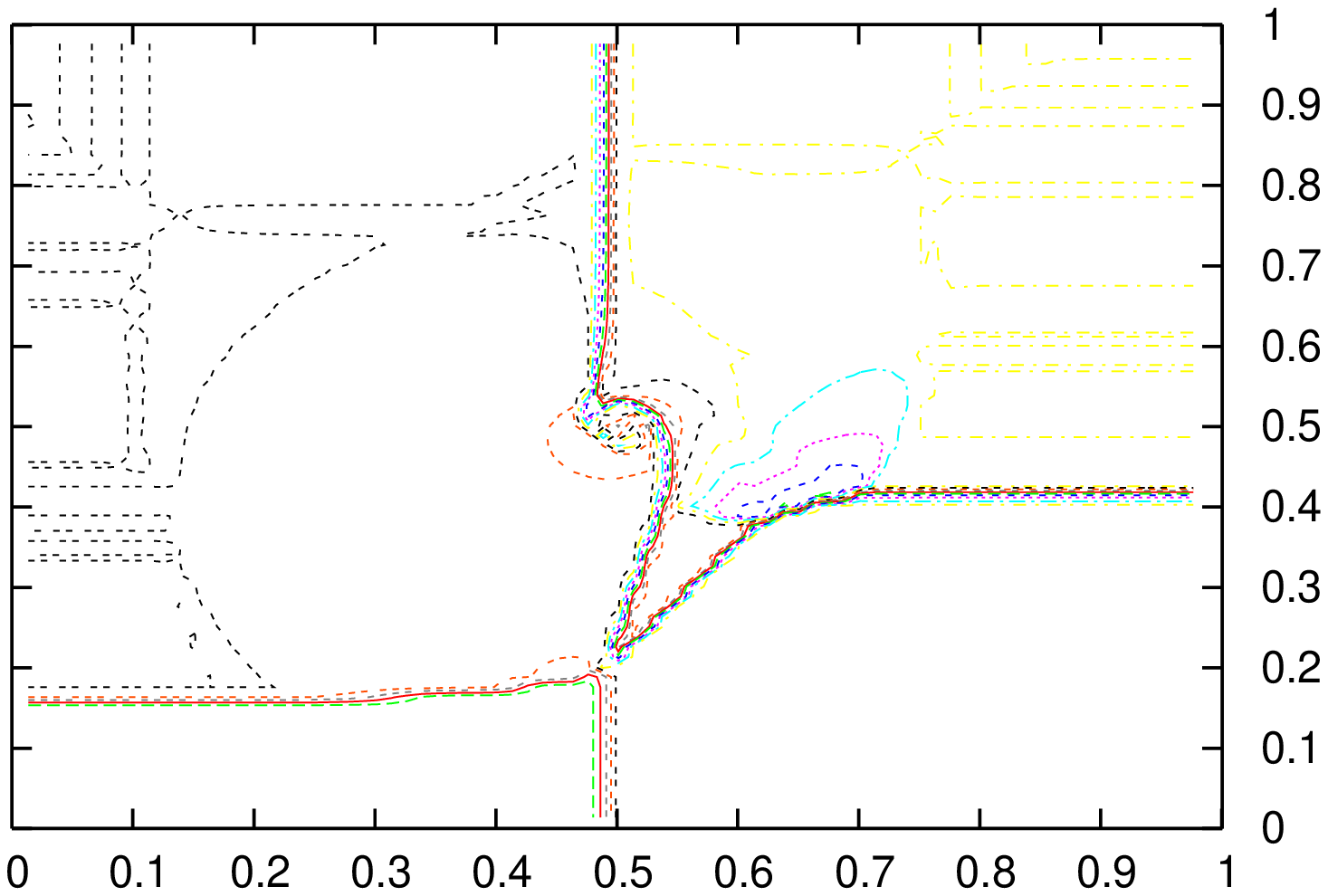}  
  \includegraphics[angle=-90,width=0.5\textwidth]{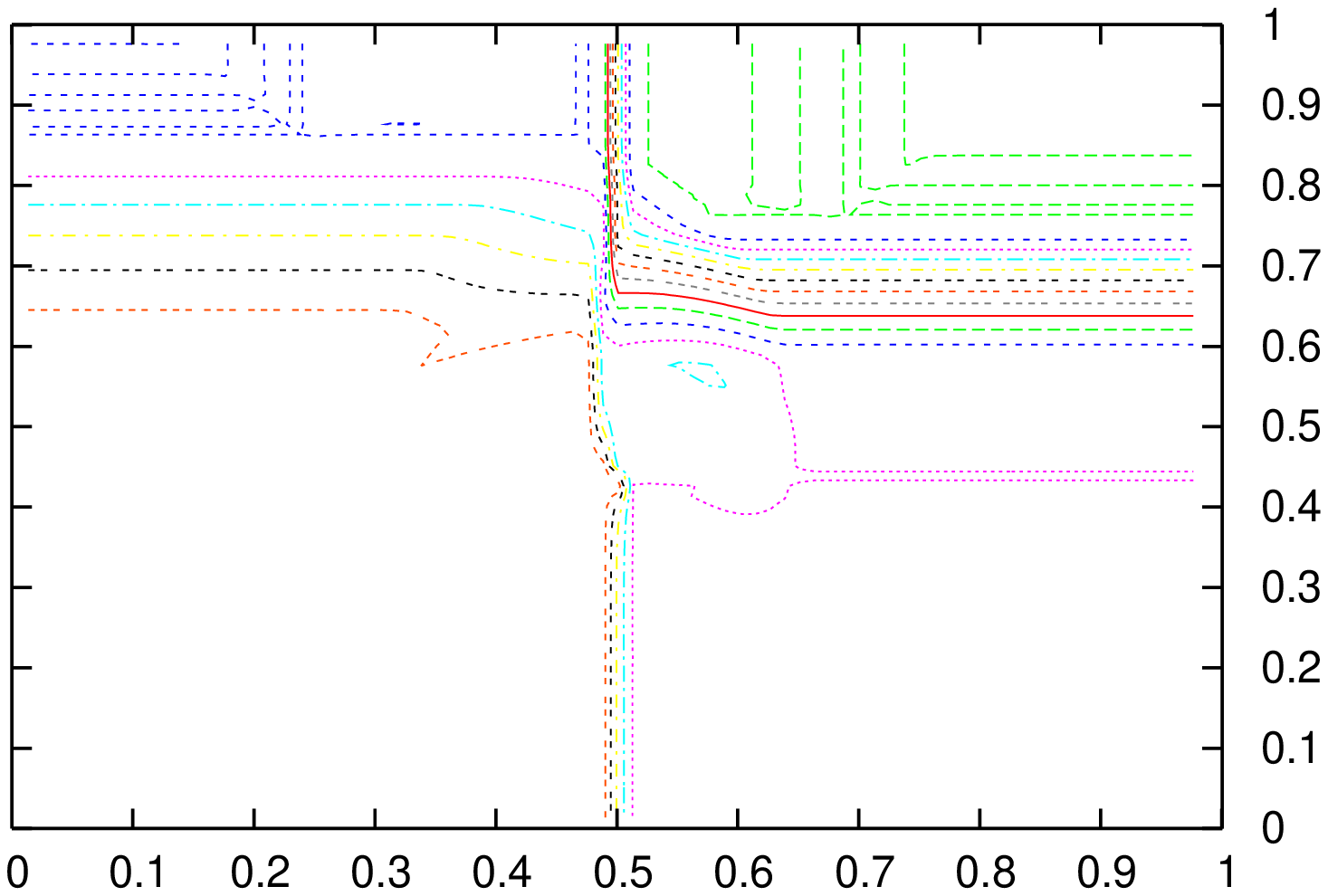}
  \includegraphics[angle=-90,width=0.5\textwidth]{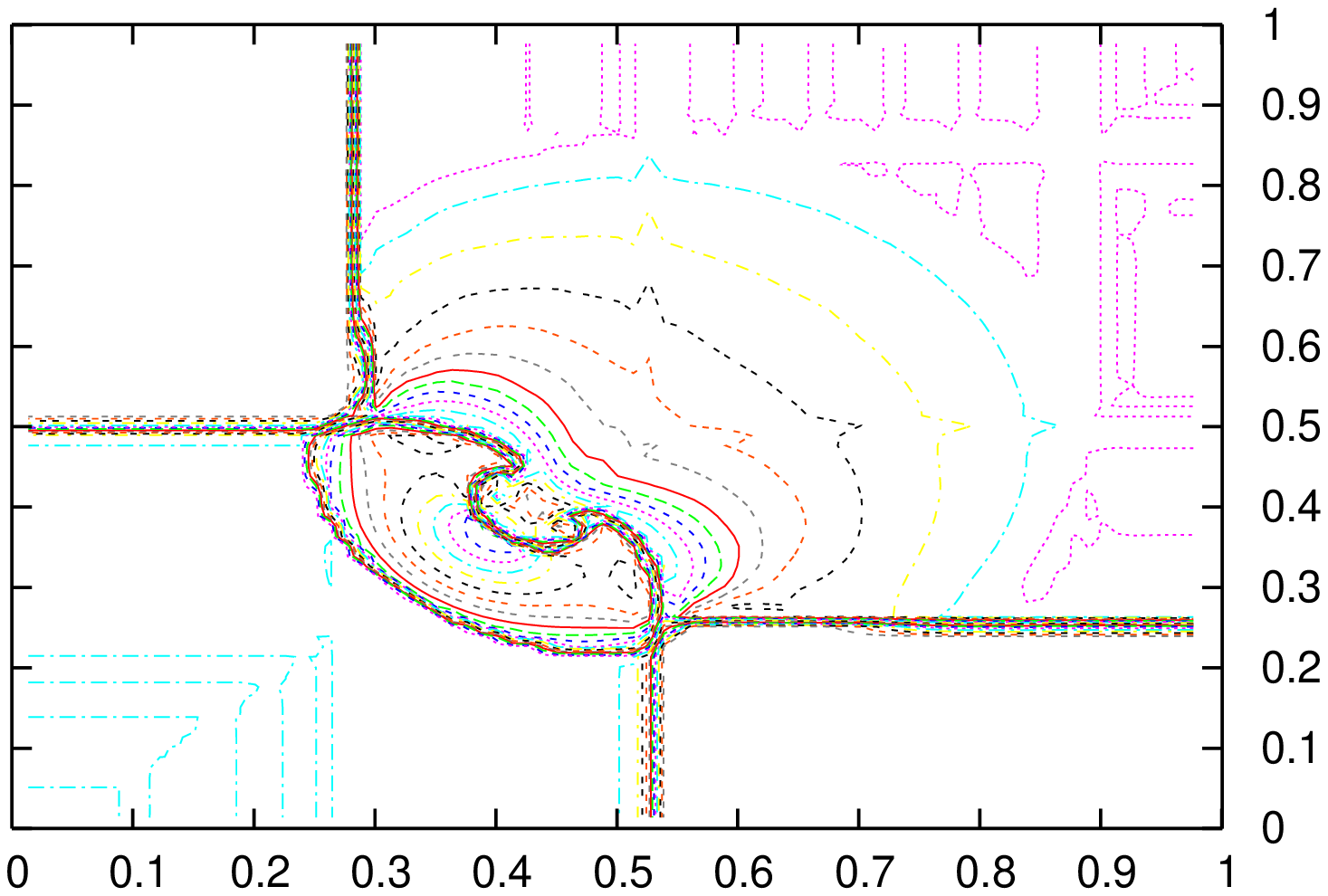}
  \includegraphics[angle=-90,width=0.5\textwidth]{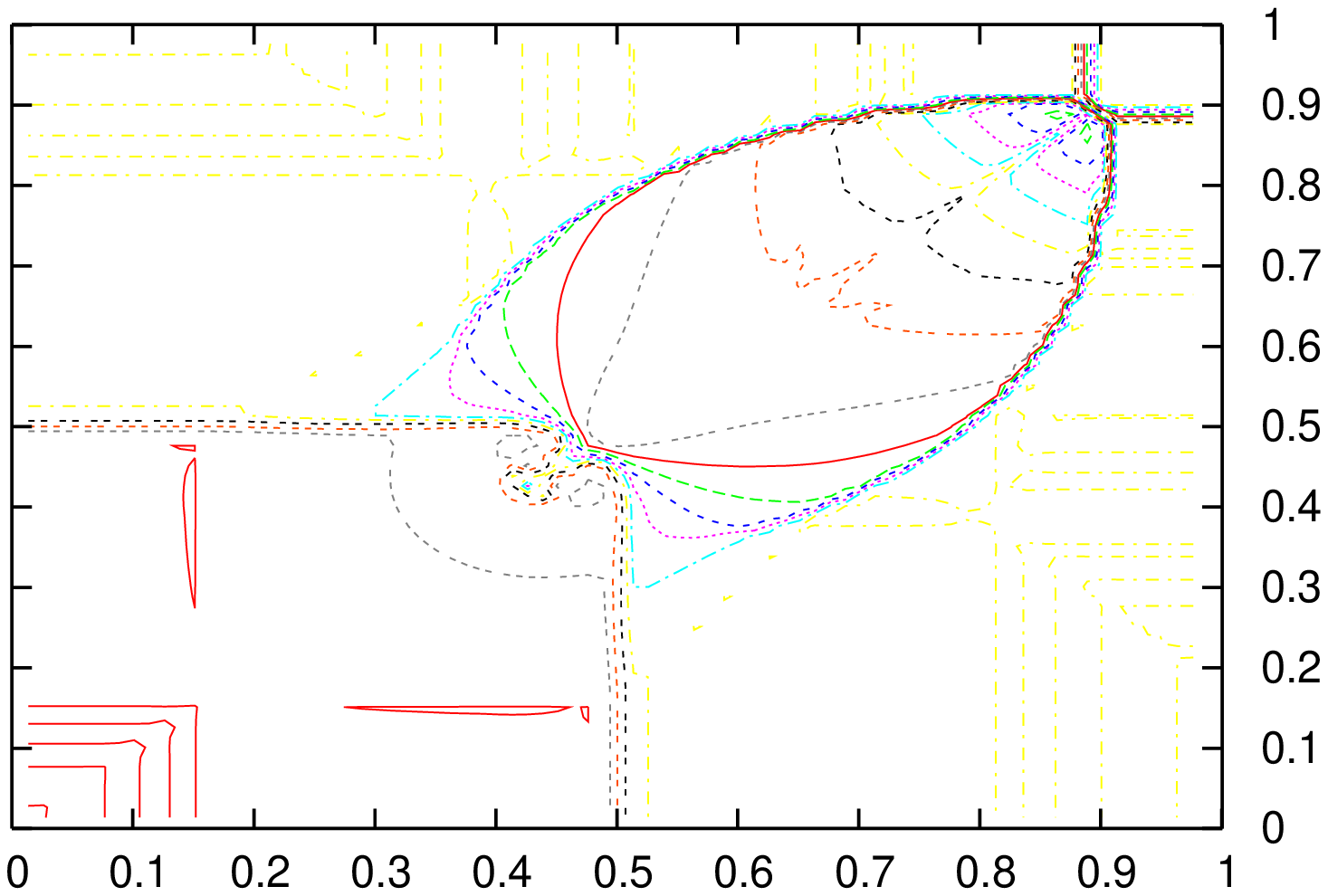}
\caption{2D Riemann problems (Sec. 5.2), Density contours for 
configurations 9-12 (in ascending order from top)}
\label{2D2}
\end{figure}

\begin{figure}[h]
\centering
  \includegraphics[angle=-90,width=0.5\textwidth]{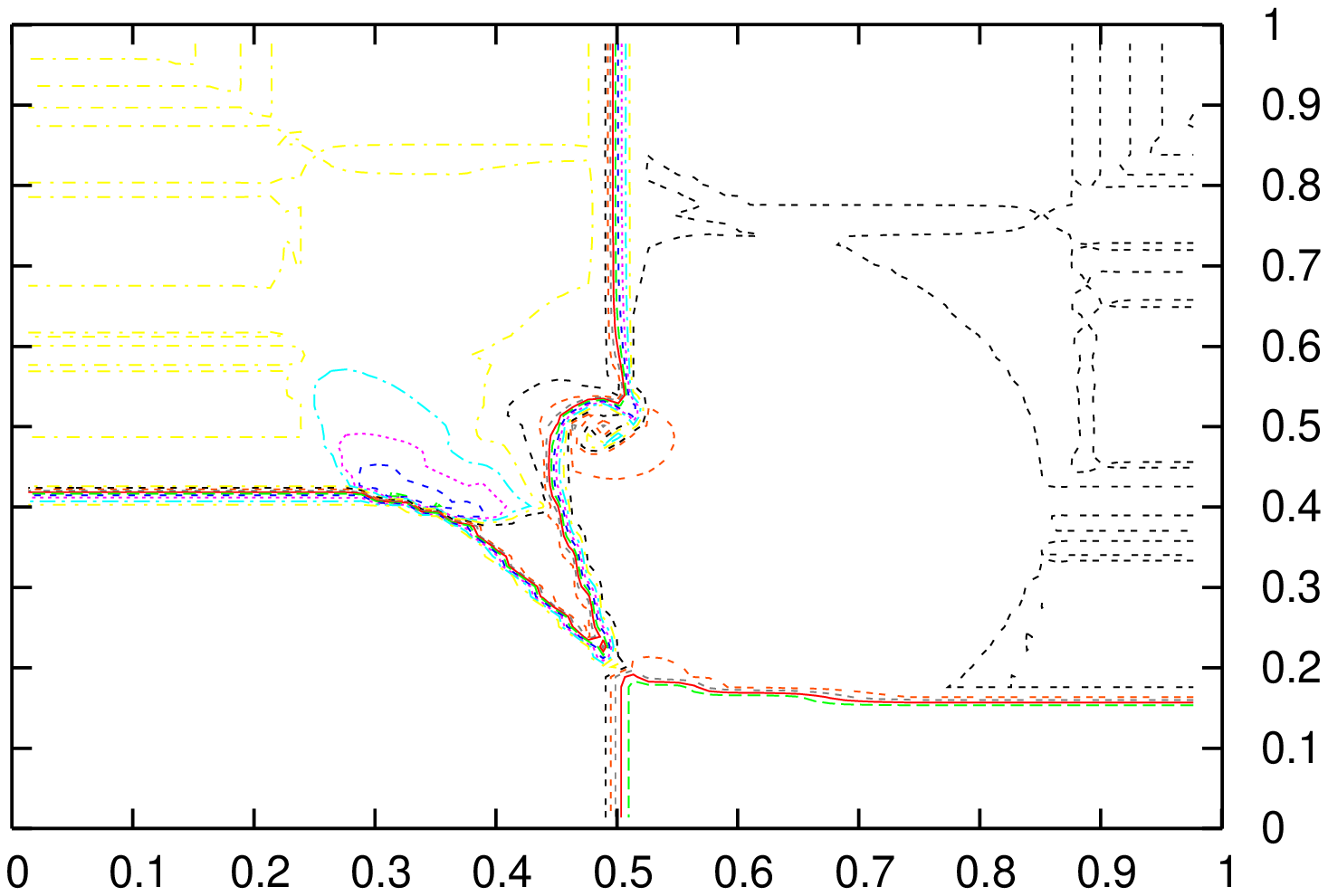}
  \includegraphics[angle=-90,width=0.5\textwidth]{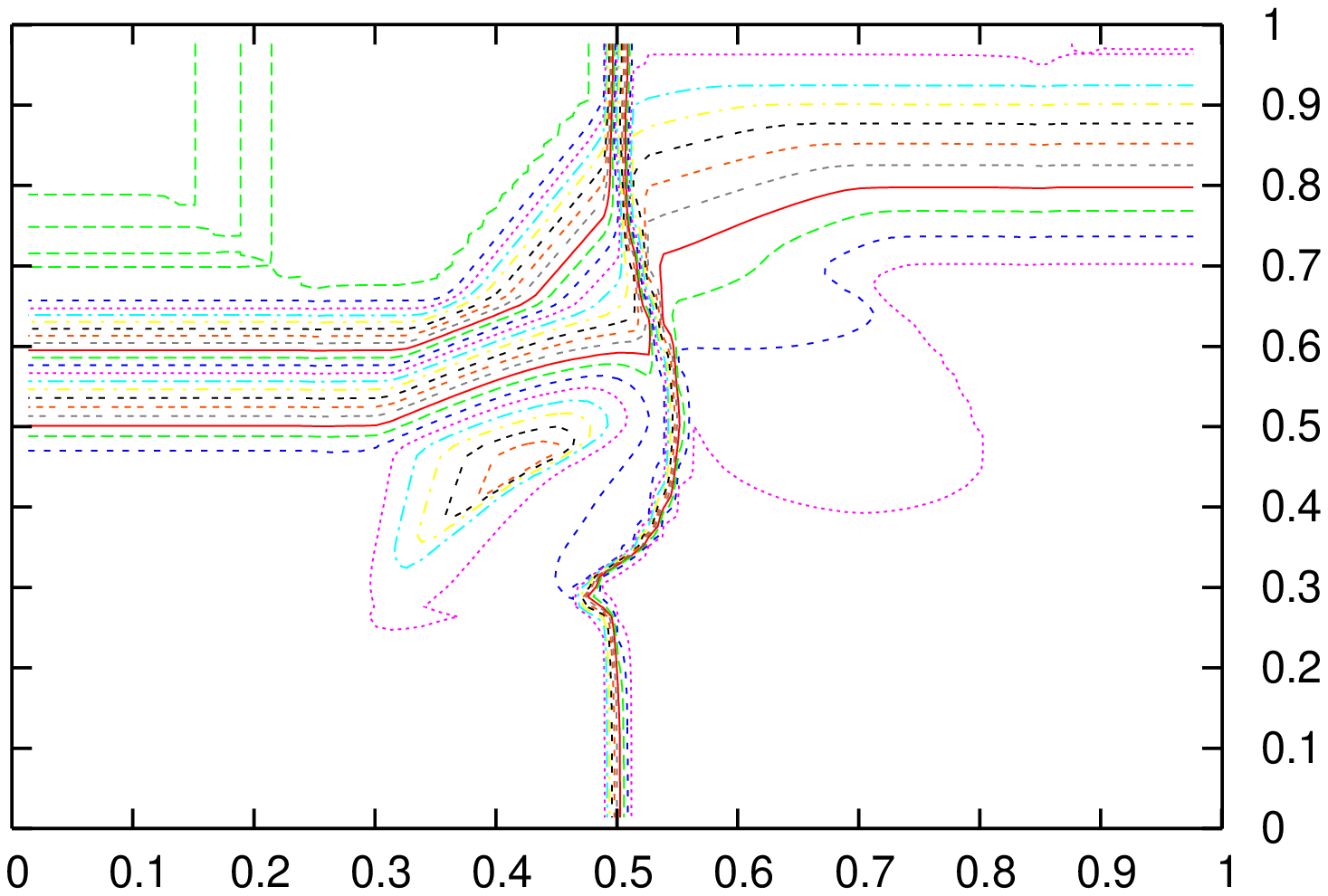}
  \includegraphics[angle=-90,width=0.5\textwidth]{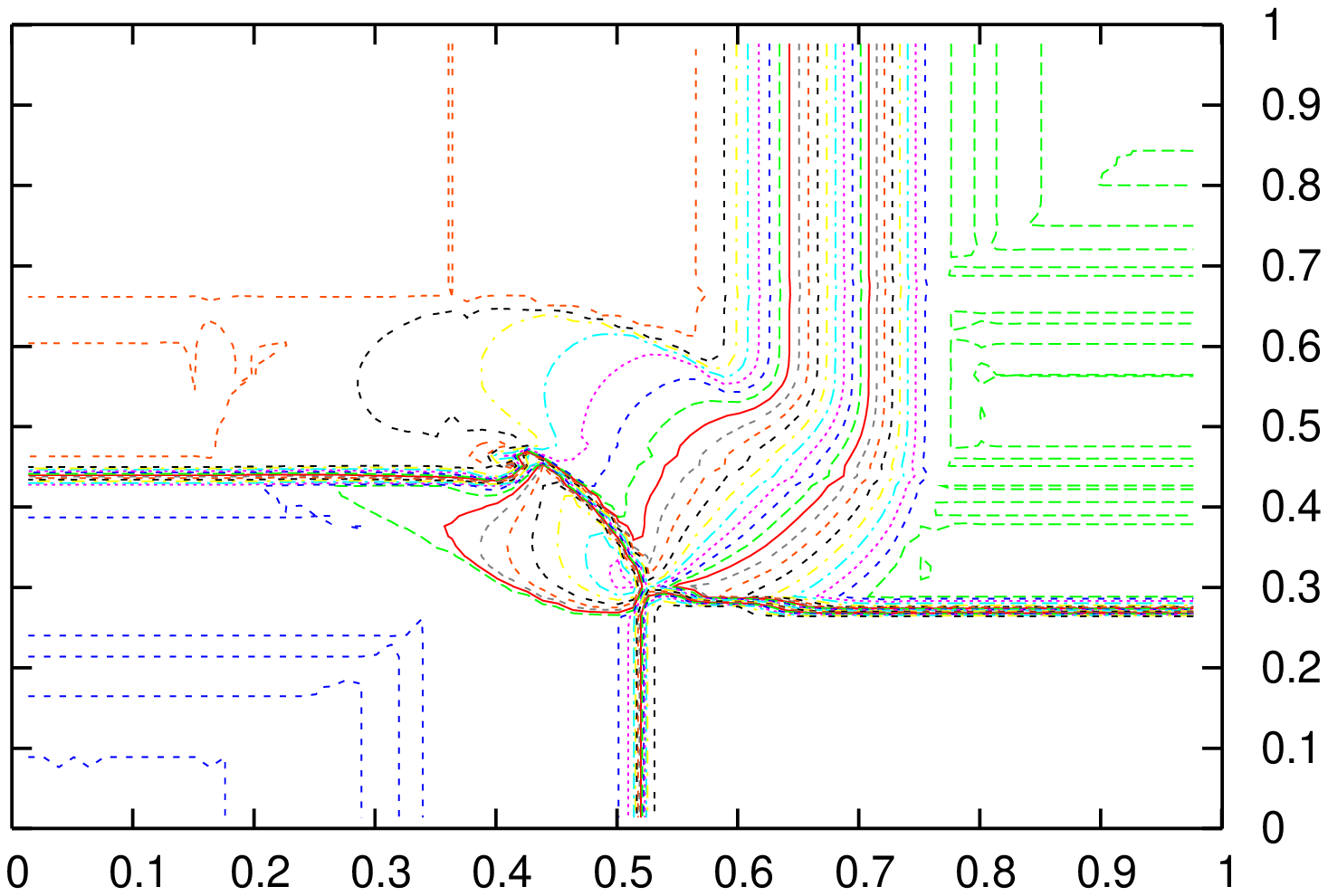}
  \includegraphics[angle=-90,width=0.5\textwidth]{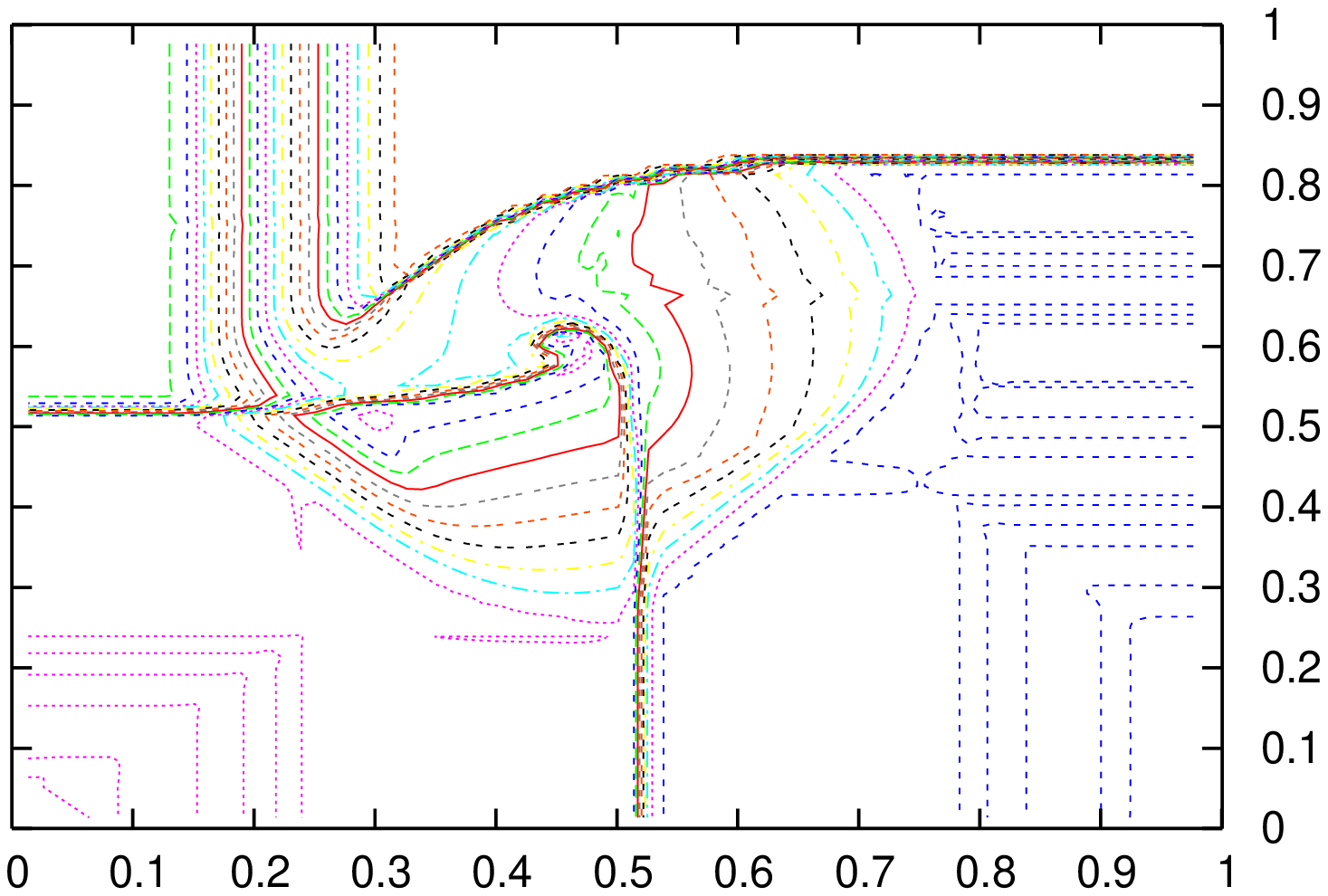}
\caption{2D Riemann problems (Sec. 5.2), Density contours for 
configurations 13-16 (in ascending order from top)}
\end{figure}

\begin{figure}[h]
\centering 
   \includegraphics[angle=-90,width=0.5\textwidth]{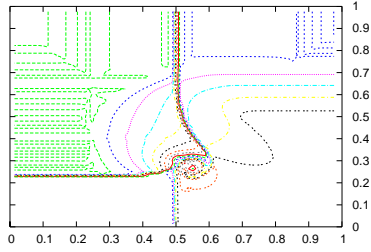}
  \includegraphics[angle=-90,width=0.5\textwidth]{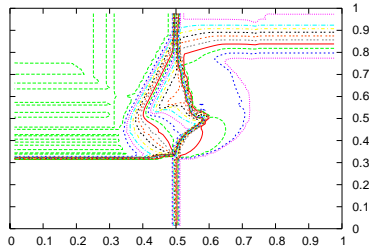}
  \includegraphics[angle=-90,width=0.5\textwidth]{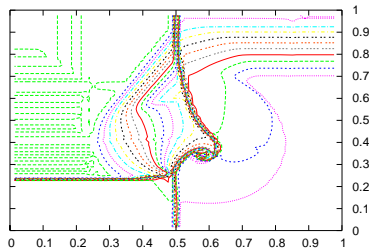}
\caption{2D Riemann problems (Sec. 5.2), Density contours for 
configurations 17-19 (in ascending order from top)}
\end{figure}

\clearpage

\begin{figure}[h]
\centering
   \includegraphics[angle=-90,width=0.8\textwidth]{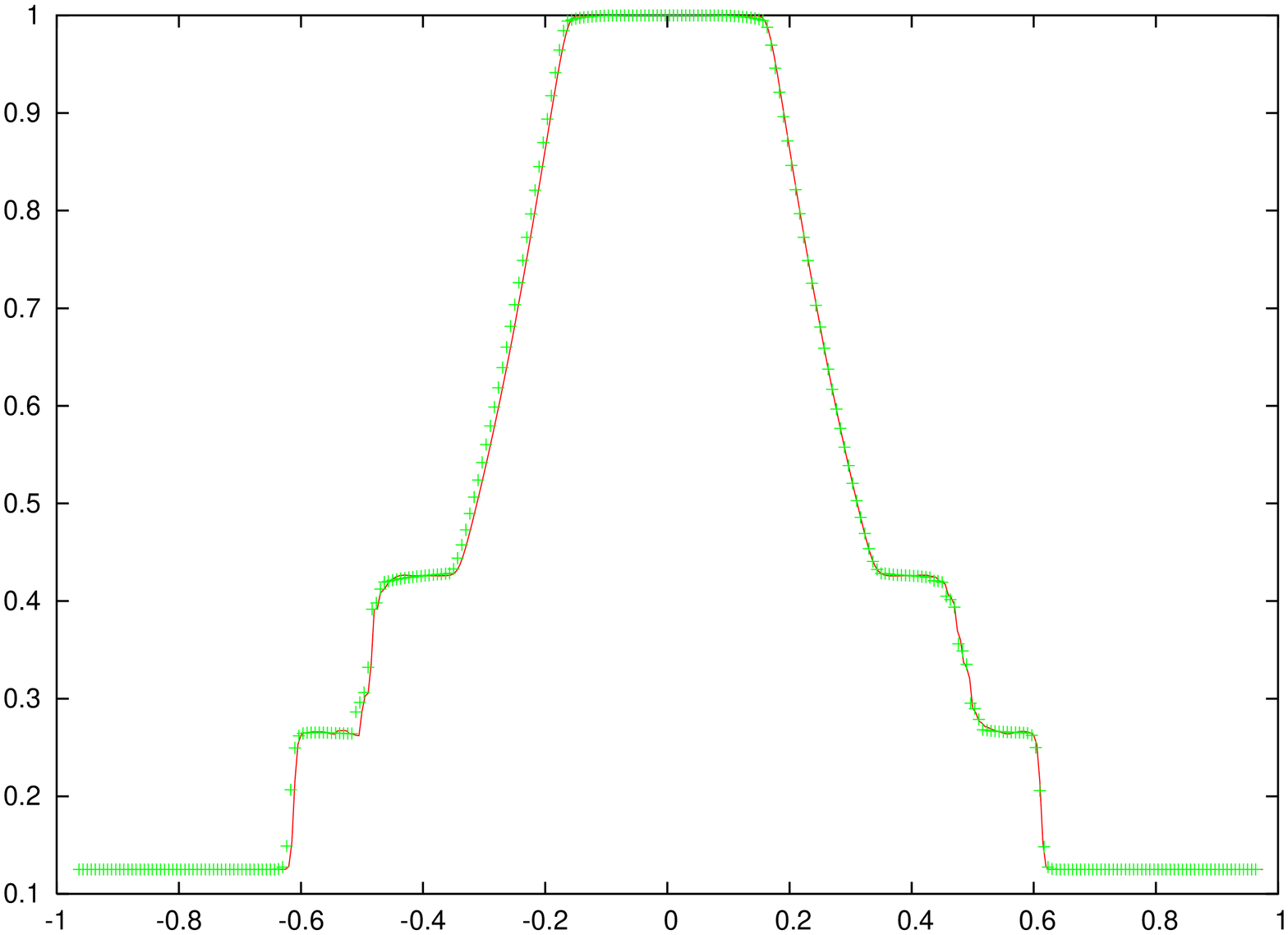}
   \includegraphics[angle=-90,width=0.8\textwidth]{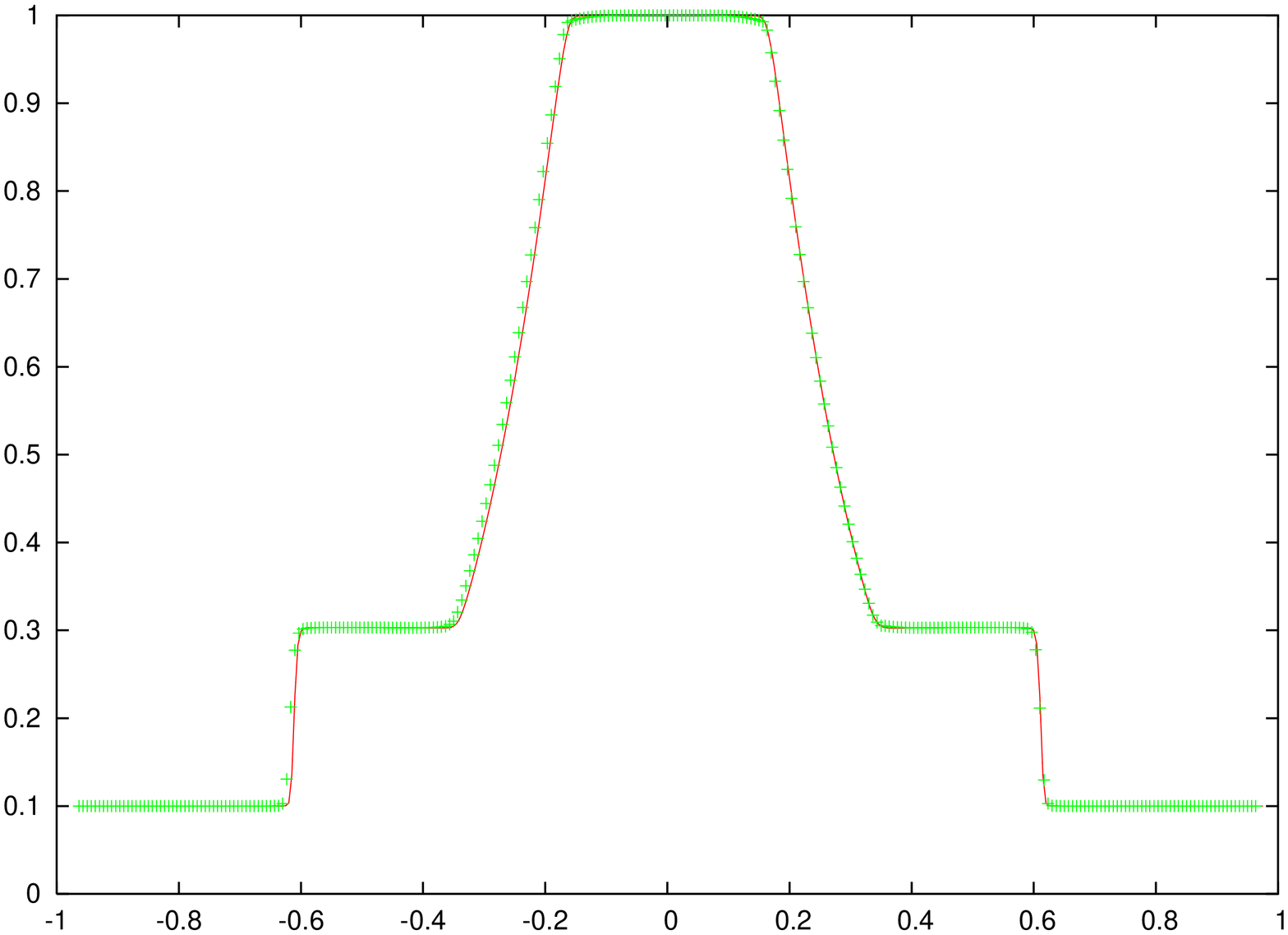}
  \label{2D6}
\caption{2D explosion test (test 1, Sec. 5.3). 
Shown are density and pressureat at t=.25, n=200 
along the x-axis. Solid lines represent 1-D 
computations for n=400.  Density (top), Pressure (bottom)}
\end{figure}

\clearpage

\begin{figure}[h]
\centering
   \includegraphics[angle=0,width=0.6\textwidth]{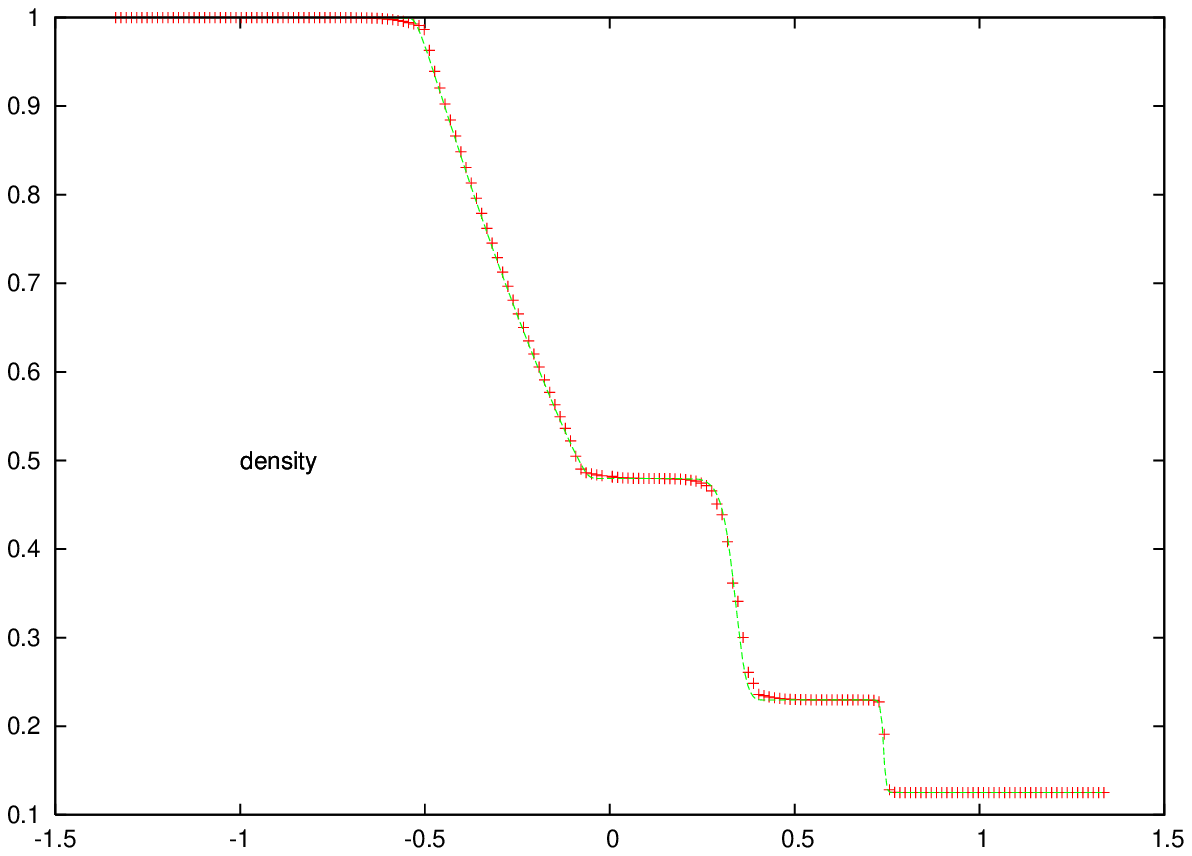}
   \includegraphics[angle=0,width=0.6\textwidth]{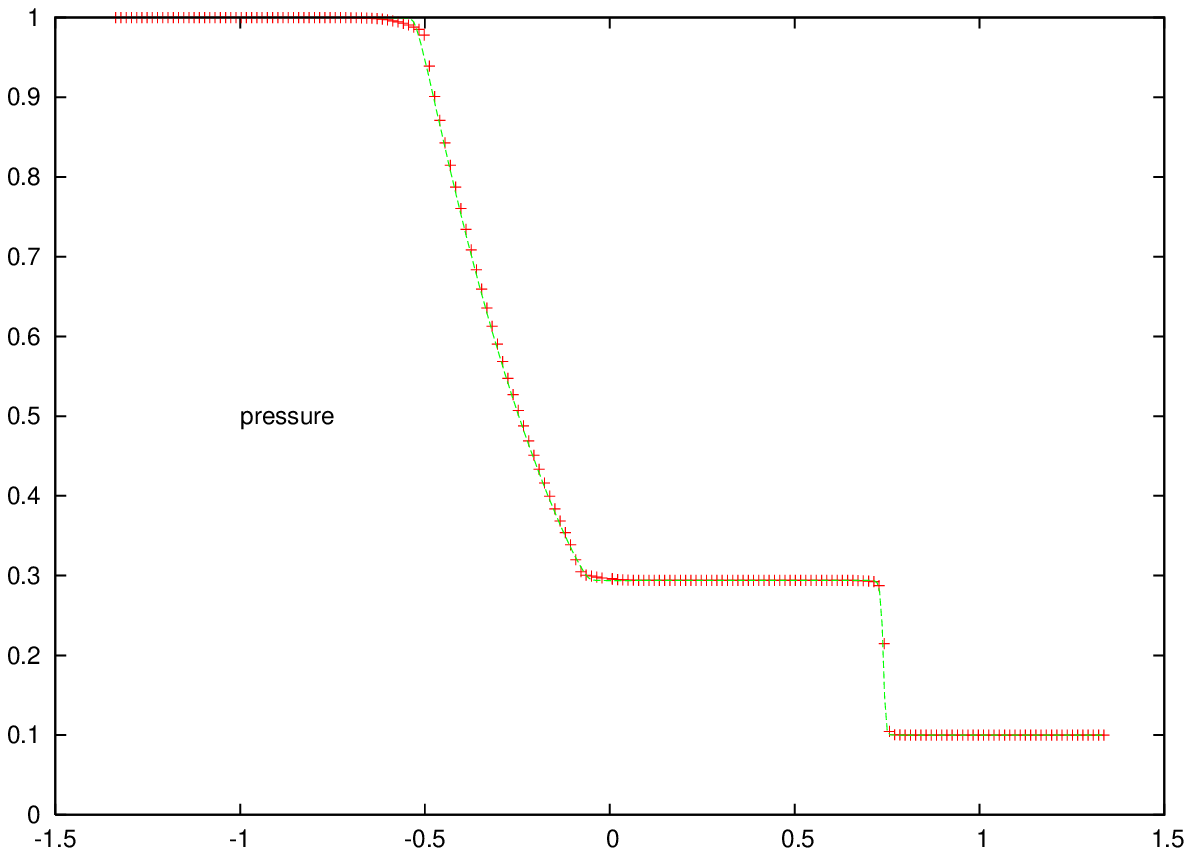}
   \includegraphics[angle=0,width=0.6\textwidth]{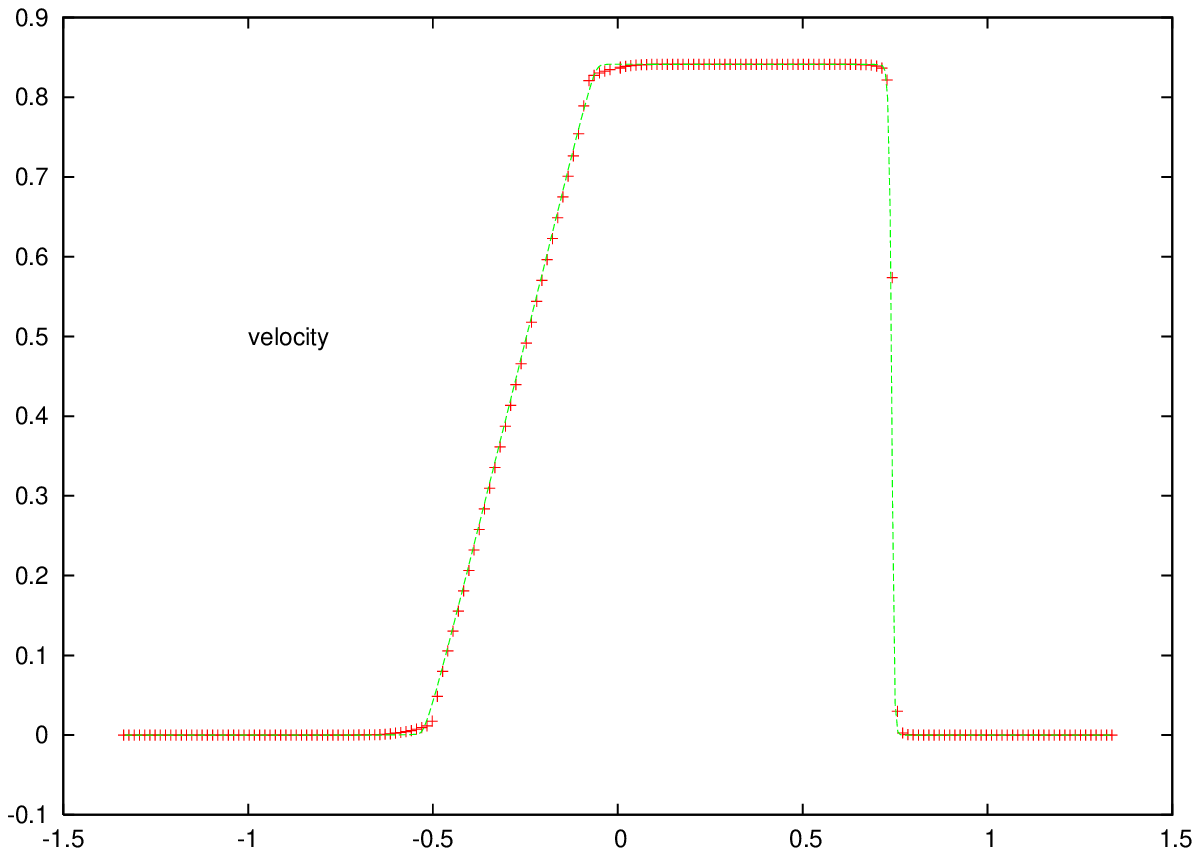}  
  \label{2D7}
\caption{The 2D Sod Shock problem (test 2, Sec. 5.3). 
Shown are the density,  pressure and velocity profiles.} 
\end{figure}

\clearpage

\begin{figure}[h]
\centering
   \includegraphics[angle=0,width=0.6\textwidth]{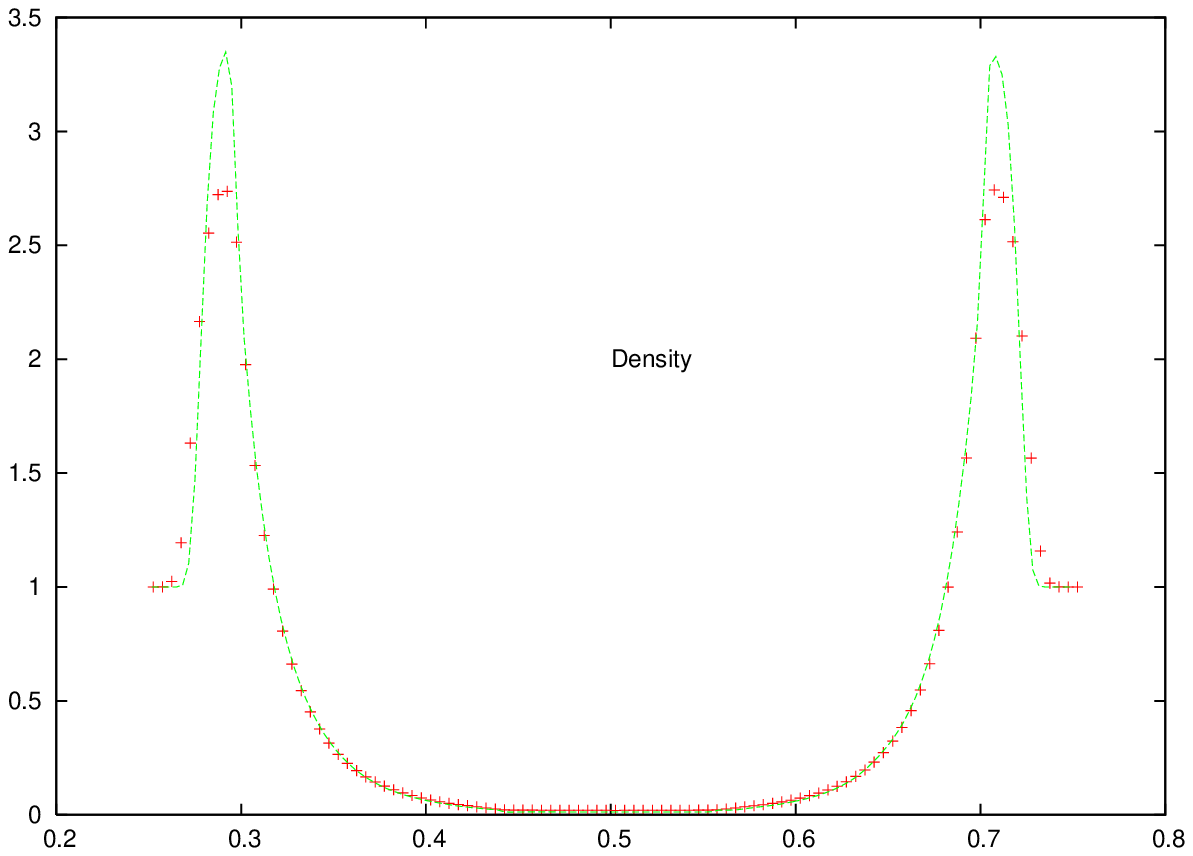}
   \includegraphics[angle=0,width=0.6\textwidth]{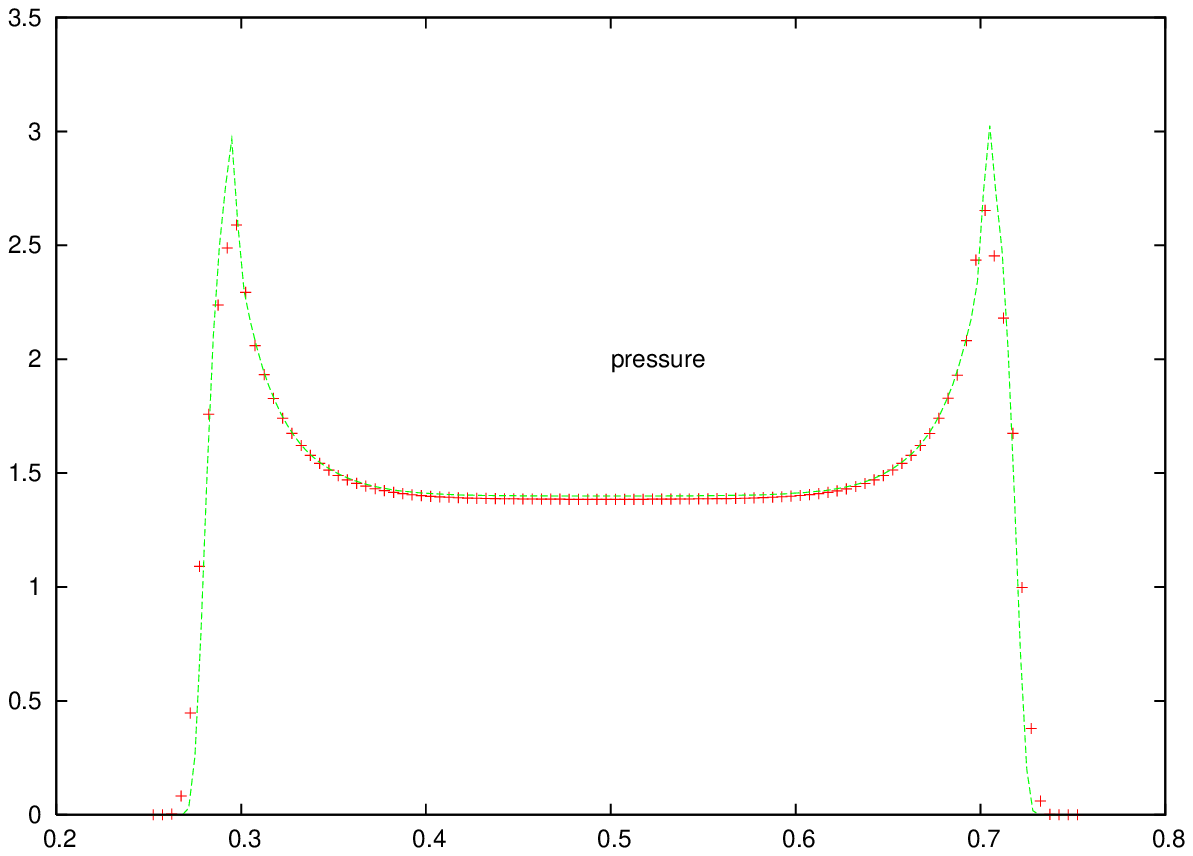}
   \includegraphics[angle=-90,width=0.6\textwidth]{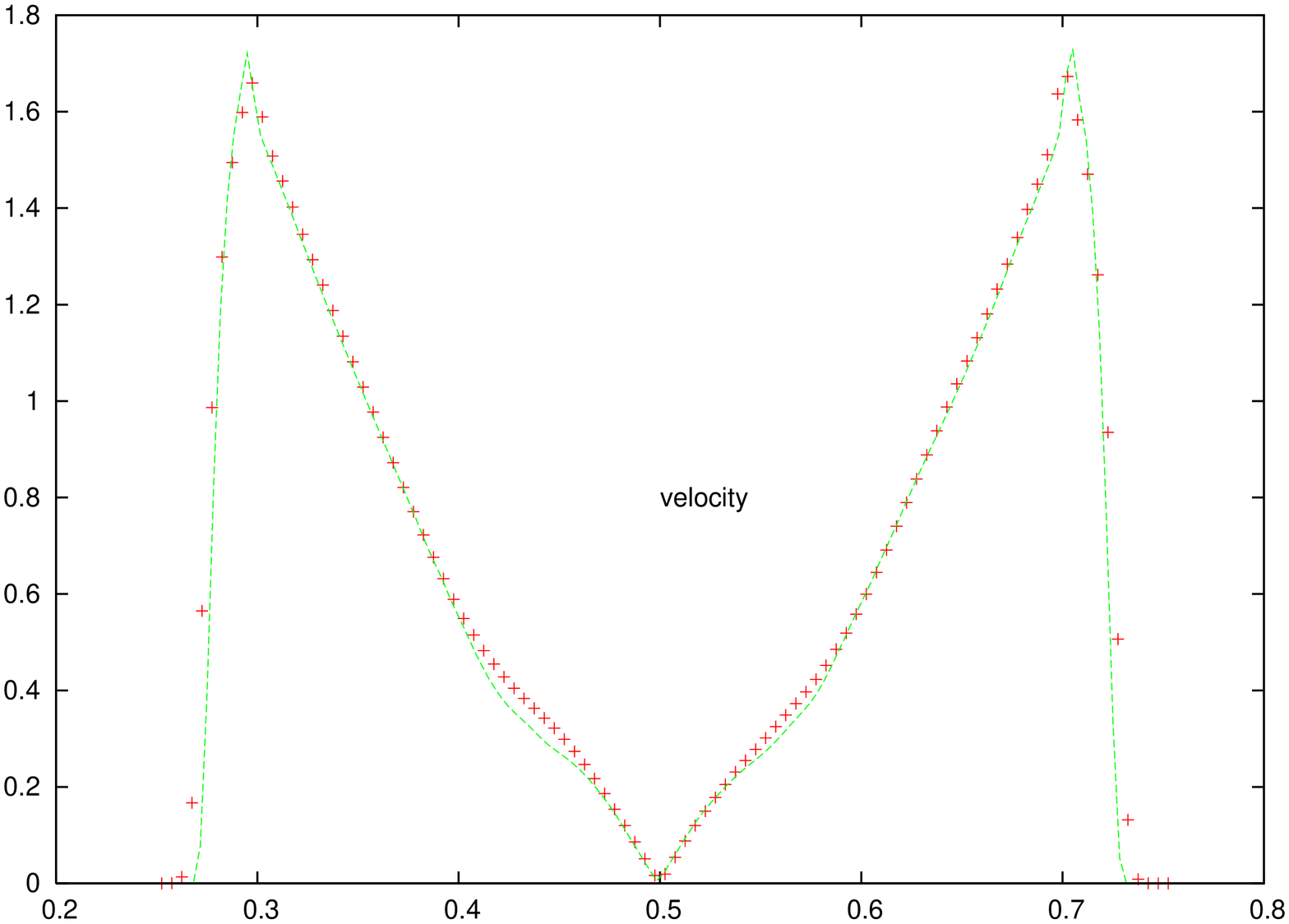}
\caption{Sedov Blast wave test (test 3, Sec. 5.3). Shown 
above are the log of density (top left), pressure (middle) and velocity (bottom) 
for n=150 at t=.05}  
  \label{2D5}
\end{figure}


\end{document}